\documentclass[pdflatex,sn-mathphys-num]{sn-jnl}

\usepackage{graphicx}%
\usepackage{multirow}%
\usepackage{amsmath,amssymb,amsfonts}%
\usepackage{amsthm}%
\usepackage{mathrsfs}%
\usepackage[title]{appendix}%
\usepackage{xcolor}%
\usepackage{textcomp}%
\usepackage{manyfoot}%
\usepackage{booktabs}%
\usepackage{algorithm}%
\usepackage{algorithmicx}%
\usepackage{algpseudocode}%
\usepackage{listings}%
\usepackage{geometry}

\newcommand{\bra}[1]{\langle {#1} |}
\newcommand{\ket}[1]{| {#1} \rangle}
\newcommand{\braket}[2]{\langle {#1} |{#2} \rangle}
\newcommand{\ketbra}[1]{| {#1} \rangle\langle {#1} |}

\newcommand{\lpnorm}[2]{\left\|#2\right\|_{#1}}

\newcommand{\tr}[1]{\text{tr}\left[#1\right]}
\newcommand{\ptr}[2]{\text{tr}_{#1}\left[#2\right]}

\newcommand{\linop}[1]{\mathbf{L}\left(#1\right)}
\newcommand{\pos}[1]{\mathbf{Pos}\left(#1\right)}

\newcommand{\dop}[1]{\mathbf{D}\left(#1\right)}
\newcommand{\puredop}[1]{\mathbf{P}\left(#1\right)}
\newcommand{\idop}{I}

\newcommand{\Schrank}[2]{{\rm Sch}_{#1}\left(#2\right)}

\newcommand{\ketMES}[2]{\ket{\phi_{#2}^+}_{#1}}
\newcommand{\ketUMES}[2]{\ket{I_{#2}}_{#1}}
\newcommand{\braUMES}[2]{\bra{I_{#2}}_{#1}}
\newcommand{\Segre}[1]{\mathbb{S}\left(#1\right)}
\newcommand{\SEP}[1]{\mathbf{SEP}\left(#1\right)}
\newcommand{\PPT}[1]{\mathbf{PPT}\left(#1\right)}

\newcommand{\Csubspace}[1]{\mathcal{V}_{#1}}
\newcommand{\CVsubspace}[1]{\mathcal{V}_{#1}^\circ}
\newcommand{\CDsubspace}[1]{\mathcal{V}_{#1}^\dag}
\newcommand{\CMFLC}[1]{\hat{\mathcal{P}}(#1)}

\newcommand{\hh}{\mathcal{H}}
\newcommand{\vv}{\mathcal{V}}
\newcommand{\ww}{\mathcal{W}}
\newcommand{\pp}{\mathcal{P}}
\newcommand{\cd}{\mathbb{C}^d}
\newcommand{\cdim}[1]{\mathbb{C}^{#1}}

\newcommand{\vspan}[1]{{\rm span}\left(#1\right)}
\newcommand{\range}[1]{{\rm range}\left(#1\right)}

\newcommand{\nn}{\mathbb{N}}
\newcommand{\rr}{\mathbb{R}}
\newcommand{\cc}{\mathbb{C}}
\newcommand{\pspace}{\mathbb{P}}
\newcommand{\pspacedim}[1]{\mathbb{P}^{#1}}

\theoremstyle{thmstyleone}%
\newtheorem{theorem}{Theorem}
\newtheorem{lemma}{Lemma}
\newtheorem{proposition}{Proposition}%
\newtheorem{corollary}{Corollary}%

\theoremstyle{thmstyletwo}%

\theoremstyle{thmstylethree}%
\newtheorem{definition}{Definition}%

\begin{document}

\title[Article Title]{Optimizing Entanglement Manipulation via Algebraic-Geometric Decompositions and Semidefinite Programming Hierarchies}


\author*[1]{\fnm{Seiseki} \sur{Akibue}}\email{seiseki.akibue@ntt.com}

\author[2]{\fnm{Jisho} \sur{Miyazaki}}

\author[3]{\fnm{Hiroyuki} \sur{Osaka}}

\affil*[1]{\orgname{Communication Science Laboratories, NTT, Inc.,}\\
\orgname{NTT Research Center for Theoretical Quantum Information,}\\
\orgname{NTT Institute for Fundamental Mathematics,}\\
\street{3--1 Morinosato Wakamiya}, \city{Atsugi}, \state{Kanagawa} \postcode{243-0198}, \country{Japan}}

\affil[2]{\orgdiv{Graduate School of Science, The University of Tokyo}, \street{7--3--1, Hongo}, \city{Bunkyo-ku}, \state{Tokyo} \postcode{113-0033}, \country{Japan}\\
\orgdiv{Ritsumeikan University BKC Research Organization of Social Sciences},
\street{1--1--1, Noji-Higashi}, \city{Kusatsu}, \state{Shiga} \postcode{525-8577}, \country{Japan}}

\affil[3]{\orgdiv{Department of Mathematical Sciences, Ritsumeikan University,}
\street{1--1--1, Noji-Higashi}, \city{Kusatsu}, \state{Shiga} \postcode{525-8577}, \country{Japan}}


\abstract{In the study of distributed quantum information processing, it is a fundamental problem to optimize local operations in the implementation of non-local quantum operations assisted by limited entanglement.
We develop an algebraic-geometric framework that systematically simplifies optimization over separable (SEP) channels---widely used as approximations of local operations---and strengthens the Doherty--Parrilo--Spedalieri (DPS) hierarchy for solving such problems.
We apply this framework to computing maximum success probability for exactly implementing a broad range of different non-local operations under SEP channels.
First, we present a unified generalization of previous analytical results on the entanglement cost. 
Via the generalization, we resolve an open problem posed by Yu et al. regarding the entanglement cost of local state discrimination.
Second, we numerically determine the trade-off between the strength of entanglement and the success probability of implementing various operations---such as entanglement distillation, non-local unitary channels, measurements, and state verification.
}




\maketitle

\section{Introduction}
In distributed quantum computation, entanglement serves as a pivotal resource, enabling the execution of quantum operations among distant parties via local operations and classical communication (LOCC)~\cite{HHHH09}. Although ample entanglement enables the execution of any non-local quantum operations, such as quantum communication protocols and non-local gate operations, through quantum teleportation~\cite{BBCJPW93}, more sophisticated LOCC protocols tailored to individual operations can often reduce the consumption of entanglement.
Thus, optimizing LOCC protocols in the implementation of a given non-local operation has practical relevance in designing efficient distributed quantum computation under limited or imperfect entanglement resources.
Additionally, it is essential for understanding the degree of non-locality inherent in the operation or the operational power of entanglement.

Despite its important role, the mathematical structure of LOCC is notoriously intricate~\cite{CLMOW14}. An LOCC protocol may involve an unbounded number of communication rounds in which each party’s local operation depends on the full history of classical messages~\cite{CLMO13}, making the set of LOCC channels analytically unwieldy. 
To circumvent these difficulties, much of the literature turns to separable (SEP) channels, a larger but mathematically tractable class that contains all LOCC channels.
While the inclusion relationship is known to be strict~\cite{BDCTEPSW99,KTYI08,CD09,DFXY09,CLMOW14}, the set of SEP channels is particularly important among the models related to LOCC.
Indeed, the power of SEP channels often coincides with that of LOCC when considering the distillability of entanglement, the transformation of bipartite pure states~\cite{GG08}, and the implementation of non-local operations~\cite{SG11,BCJRWY15,SM16,YDY12,YDY14}.
Moreover, SEP channels have helped to reveal fundamental limitations of LOCC, including difficulty of pure state transformation~\cite{SWGK18}, achievable entanglement distillation rates~\cite{R98} and coherence distillation rates~\cite{SRBL17}.

The optimization over SEP channels is formulated as maximizing a linear function over the separable cone $\mathbf{SEP}$:
\begin{equation}
\label{eq:generalSEPopt}
    \max\{ \tr{M(\mathcal{E},\tau) S}:S\in \mathbf{SEP},T(S)\leq \idop\},
\end{equation}
where $T$ is a partial trace mapping, $\idop$ is the identity operator, and $\mathcal{E}$ is a non-local channel to be implemented via SEP channels assisted by a resource state $\tau$. $M$ is a Hermitian operator determined by $\mathcal{E}$ and $\tau$, and $\tr{M(\mathcal{E},\tau)S}$ represents a figure of merit to be maximized, such as the success probability of implementing $\mathcal{E}$.

The algorithms for solving Eq.~\eqref{eq:generalSEPopt} have been extensively investigated due to its close relevance to several key problems in quantum information~\cite{G04,G10,HHH96,TDS20,OZNPQ24} and computer science~\cite{KMY03,BCY11-2,HM13,BKS17} beyond simply the entanglement cost.
One state-of-the-art algorithm uses the Doherty, Parrilo, and Spedalieri (DPS)~\cite{DPS04} hierarchy, a sequence of semi-definite programs (SDPs) for relaxed problems whose solutions provide converging upper bounds on Eq.~\eqref{eq:generalSEPopt}.
Since the size of SDPs grows exponentially with the level of the hierarchy, many numerical studies rely on the first level of the hierarchy, known as the positive partial transpose (PPT) relaxation, which replaces $\mathbf{SEP}$ in Eq.~\eqref{eq:generalSEPopt} with the PPT cone. However, this relaxation often yields weak bounds~\cite{EVWW01,I04,BCJRWY15}.

In parallel, there has been an independent line of research aimed at analyzing Eq.~\eqref{eq:generalSEPopt}, in which various heuristic methodologies have been developed, resulting in identifying the entanglement cost of various non-local tasks such as global unitary operations~\cite{SG11,YGC12,CY16,SM16} and state discrimination~\cite{YDY14,BCJRWY15}.
Nevertheless, these methods have been successful only for specific problems or restricted classes of nonlocal operations and lack a unified, systematic perspective.

\subsection{Summary of the main results}
\begin{figure*}[ht]
    \centering
    \includegraphics[width=14cm]{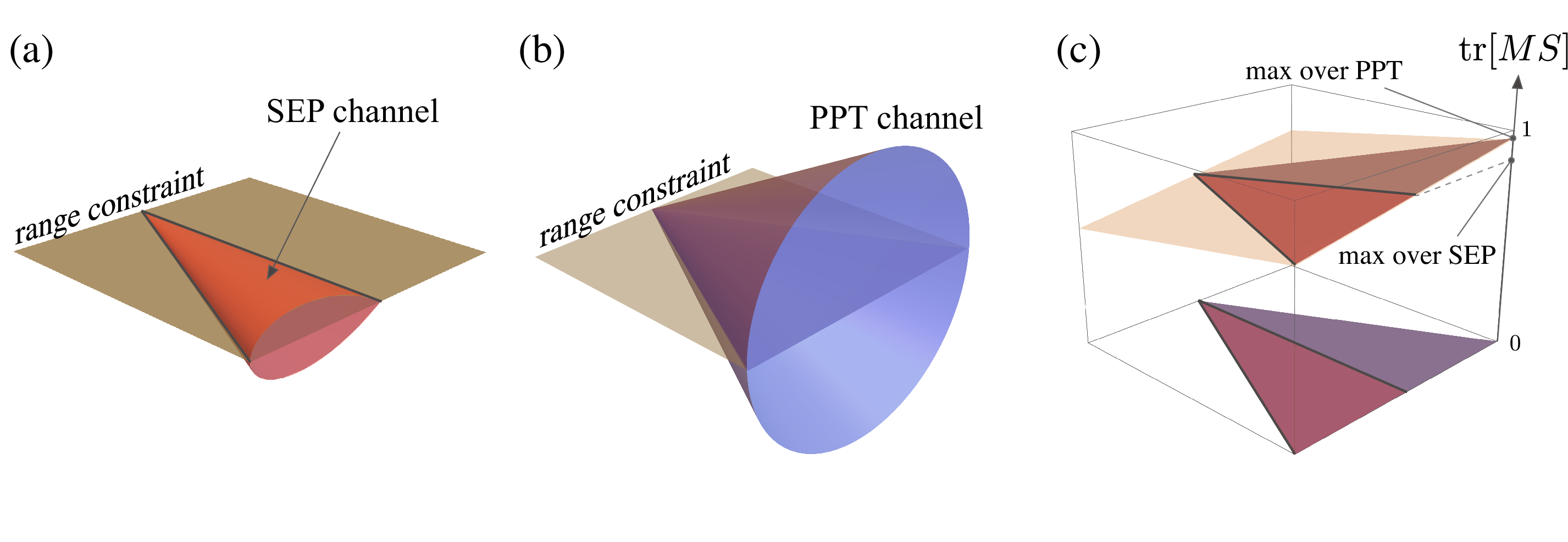}
    \caption{Feasible regions for operator $S$ in the range-constrained SEP optimization problem and of its relaxed problem. 
    (a) The SEP cone is represented by the red convex cone, with the feasible region being the intersection with the horizontal plane (range constraint). (b) In the relaxed problem, the PPT cone (blue convex cone) replaces the SEP cone, enlarging the feasible region to a large triangular area on the plane. (c) Due to the difference in the feasible region between the original problem (red triangle) and the relaxed problem (purple triangle), the solution to the relaxed problem is often strictly larger than the original one.}
    \label{fig:diagram}
\end{figure*}

$\mathcal{E}$ is deterministically implementable under SEP channels if and only if Eq.~\eqref{eq:generalSEPopt} is 1 when the figure of merit $\tr{M(\mathcal{E},\tau)S}$ is the success probability of implementing $\mathcal{E}$. Despite the importance of testing deterministic implementability, the relevant computational problems indicate that this test is also computationally hard in general~\cite{BKS17,BHKN19}.

To extend the analysis beyond the deterministic regime, we consider a probabilistic and zero-error implementation scenarios. In this setting, the corresponding optimization problem can be formulated in a more constrained form as follows:
\begin{equation}
\label{eq:SEPopt}
    \max\{ \tr{M(\mathcal{E},\tau) S}:S\in \mathbf{SEP},T(S)\leq \idop,\range{S}\subseteq\vv\},
\end{equation}
where $\vv$ is a subspace in the composite system, which arises when we require zero-error implementation.
For example, consider the case where $\mathcal{E}$ represents the preparation of a pure entangled state $\phi$, and the success probability is used as the figure of merit. In this setting, $M(\mathcal{E},\tau)=\phi\otimes\overline{\tau}$ and $S$ is the Choi operator of a (probabilistic) SEP channel that inputs a resource state $\tau$ and outputs a state $\rho\propto\ptr{R}{(\idop\otimes\overline{\tau})S}$, where $\ptr{R}{\cdot}$ denotes the partial trace of the resource system.
The zero-error constraint imposes $\rho=\phi$. This condition allows us to set
\begin{equation}
    \vv=\{\ket{\Xi}:\forall\ket{\eta}\in\range{\overline{\tau}},\braket{\eta}{\Xi}\propto\ket{\phi}\}.
\end{equation}

Problems of the form given in Eq.~\eqref{eq:SEPopt} are often referred to as \textit{range-constrained SEP optimization problems} in this paper.
Even in this problem, low levels of the DPS hierarchy fail to yield tight bound because the relaxed problem significantly enlarges the feasible region in many cases (see Fig.~\ref{fig:diagram}).

\medskip
\noindent
Our main contributions are as follows.

\begin{enumerate}
\item \textbf{Minimum algebraic coverings and MFLC.}
We formulate an algebraic approximation of the extreme rays
compatible with the range constraint.  In projective space, the rank-one extreme
rays of $\mathbf{SEP}$ correspond to the Segre variety $\mathbb{S}$, and the
range constraint restricts them to an intersection of $\mathbb{S}$ with the
projectivization of $\vv$.  

We prove that for any subset $\mathbb{E}$ of projective space and
any degree parameter $m\in\mathbb{N}$ there exists a unique \emph{minimum degree-$m$
covering} of $\mathbb{E}$ by algebraic sets of generating degree at most $m$
(Theorem~\ref{thm:minext}), and we give an explicit construction in terms of the irreducible
components of the Zariski closure of $\mathbb{E}$.  The minimum degree-$1$ covering yields the
\emph{minimum finite linear covering} (MFLC), i.e.\ the smallest union of linear
subspaces that covers $\mathbb{E}$.
Moreover, we explicitly compute the MFLC of $\mathbb{S}\cap\vv$ for several classes of $\vv$.

\item \textbf{Improved SDP hierarchy for cone optimization.}
Using the minimum covering, we refine the SDP hierarchy for cone
optimization over $\pos{\mathbb{A}}:={\rm cone}\left(\{\ketbra{\phi}:\ket{\phi}\in\mathbb{A}\}\right)$, where $\mathbb{A}$ is an algebraic set.
By extending the framework of the DPS hierarchy and its generalization proposed by Derksen \emph{et al.}~\cite{HNB24}, we introduce a decomposed symmetric-extension condition (``prime $\mathbb{A}$-tension'') that reflects the structural constraints arising from the minimum degree-$m$ covering of $\mathbb{A}$ and yields a
convergent hierarchy (Theorem~\ref{thm:hierarchy}).  A key point is that the \emph{first} level of the
resulting hierarchy already imposes nontrivial linear constraints coming from the
MFLC, in contrast to the standard DPS hierarchy and its generalization~\cite{HNB24}.

\item \textbf{Universal irreversibility for entanglement manipulation.}
As an analytical application, we obtain a unified extension of several
irreversibility results for entanglement-assisted implementations of non-local
operations.  In particular, we show (Theorem~\ref{thm:EntforQI}) that if a non-local instrument $\{\mathcal{E}_m\}_m$ is
deterministically implementable by SEP instruments assisted by a pure entangled
state of minimal Schmidt rank, and if one branch of the instrument is an inverse-isometry
(i.e.\ a projection followed by a unitary on the projected subspace), then the
resource state must be maximally entangled, regardless of how little entanglement $\{\mathcal{E}_m\}_m$ itself can generate.  This subsumes the known unitary~\cite{SG11} and
entangled-measurement~\cite{BHN18} special cases and resolves, as a corollary, the open problem
posed by Yu \emph{et al.}~\cite{YDY14} on the entanglement cost of local state discrimination.

The proof of Theorem \ref{thm:EntforQI} relies on the observation that the MFLC constraint was implicitly employed in previous studies~\cite{SG11,BHN18}. 
We substantially extend the proof strategy of Ref.~\cite{SG11} by exploiting the universality of the Choi operator as a representation of quantum instruments, together with the MFLC constraint.

\item \textbf{Practical numerical method for the range-constrained SEP optimization.}
For several benchmark tasks---including entanglement distillation, non-local unitary
channels, joint measurements, and state verification---the MFLC constraints
substantially tighten the PPT relaxation of Eq.~\eqref{eq:SEPopt} and
yield nearly tight trade-off curves between the strength of the resource
entanglement and the achievable figure of merit (see Fig.~\ref{fig:Entcost}.)
Moreover, the runtime required to solve the ``PPT+MFLC'' relaxation is typically shorter than that of the PPT relaxation alone. In many instances, the resulting ``PPT+MFLC'' bounds even outperform those obtained from the second level of the DPS hierarchy.

\end{enumerate}

\smallskip
\noindent
Finally, we stress that a strict improvement over DPS-type relaxations and their generalization~\cite{HNB24}, when approximating $\pos{\mathbb{A}}$, requires $\mathbb{A}$ to be \emph{reducible} in the Zariski topology. Here, $\mathbb{A}$ is often taken to be the intersection $\mathbb{S}\cap\vv$ in this paper. Otherwise, the MFLC typically collapses to $\vspan{\mathbb{A}}$, the first level of the hierarchy becomes trivial, and higher-level prime $\mathbb{A}$-tensions coincide with those proposed by Derksen \emph{et al}.~\cite{HNB24}.

\subsection{Related work}
Recently, the algebraic-geometric approach has gained importance in quantum information research because it provides useful concepts for capturing the complex mathematical structures of entanglement~\cite{MBDM13,HNW17,GMO20,GM21,LJ22,G24,HNB24} and circuit complexity~\cite{HFKEY22}.
For example, the size of $\mathbb{S}\cap\vv$ has been extensively studied in the context of entanglement criteria~\cite{H97,BGR05} and completely entangled subspaces~\cite{P04}. The subspace $\vv$ is called a completely entangled subspace if $\mathbb{S}\cap\vv=\emptyset$. Various algebraic-geometric methods have been developed to determine whether $\vv$ is completely entangled~\cite{P04,WS08,JLV22,HNB24} and to provide explicit constructions~\cite{GM25}.
In contrast, we consider the case when $\mathbb{S}\cap\vv\neq\emptyset$ and demonstrate that an algebraic-geometric approach remains effective. Thus, from a theoretical perspective, our research establishes a new direction that complements studies on completely entangled subspaces.

\section{Notations}
Let us briefly introduce the notation and concepts of quantum information in this subsection. Readers can find a more comprehensive introduction to quantum information and semi-definite programming in~\cite{WBook, MHBook}.

We denote the multiplicative group of non-zero complex numbers by $\cc^\times:=\cc\setminus\{0\}$.
The complex conjugate of $x\in\cc$ is denoted by $\overline{x}$.
We only consider finite-dimensional Hilbert spaces.
A pure state is represented by a unit vector $\ket{\phi}\in\hh$ in a Hilbert space $\hh$. Its density operator, denoted by $\phi:=\ketbra{\phi}$, is also often referred to as a pure state. $\puredop{\hh}$ represents the set of (density operators of) pure states $\phi$.

Vectors that are not necessarily normalized are denoted with capital letters such as $\ket{A}$ and $\ket{\Pi}$. 
$\linop{\hh_A:\hh_B}$ represents the set of linear operators mapping from a Hilbert space $\hh_A$ into a Hilbert space $\hh_B$.
We sometimes use a subscript or superscript to emphasize the Hilbert space where the vector lies or the operator acts, respectively.
For $A\in\linop{\hh_A:\hh_B}$, we sometimes define its corresponding vector $\ket{A}\in\hh_A\otimes\hh_B$ by $\ket{A}:=(\idop\otimes A)(\sum_{i}\ket{i}_A\ket{i}_A)$.
This notation is used in~\cite{MHBook}.
$\pos{\hh}$ represents the set of positive semi-definite operators acting on a Hilbert space $\hh$.
We sometimes denote the condition $E\in\pos{\hh}$ as $E\geq0$.
$\dop{\hh}$ represents the set of density operators $\rho$, which satisfies $\rho\in\pos{\hh}$ and $\tr{\rho}=1$.
We define normalized and unnormalized maximally entangled vectors in $\hh_A\otimes\hh_B$ as $\ketMES{AB}{d}=\frac{1}{\sqrt{d}}\ketUMES{AB}{d}$ and $\ketUMES{AB}{d}:=\sum_{i=0}^{d-1}\ket{i}_{A}\ket{i}_{B}$, respectively, where $\{\ket{i}_A\}_i$ and $\{\ket{i}_B\}_i$ are computational bases in $\hh_A$ and $\hh_B$, respectively.
For vectors $\ket{X}\in\hh_A\otimes\hh_B$ and $\ket{Y}\in\hh_A$, we often use an abuse of notation for a vector $\bra{Y}_A\ket{X}_{AB}$ in $\hh_B$ to represent $(\bra{Y}\otimes\idop)\ket{X}=\sum_{ij}(\alpha_{ij}\braket{Y}{i})\ket{j}$, where $\ket{X}=\sum_{ij}\alpha_{ij}\ket{i}_A\ket{j}_B$.

We define the set of product vectors as follows:
\begin{equation}
    \Segre{\hh_1:\cdots:\hh_N}:=\{\ket{A_1}\otimes\cdots\otimes\ket{A_N}:\ket{A_n}\in\hh_n\}.
\end{equation}
Moreover, we define the separable cone and the PPT cone as follows:
\begin{equation}
\SEP{\hh_1:\cdots:\hh_N}\nonumber:=\left\{\sum_x\ketbra{\Pi_x}:\ket{\Pi_{x}}\in\Segre{\hh_1:\hh_2:\cdots:\hh_N}\right\},
\end{equation}

\begin{equation}
\PPT{\hh_1:\cdots:\hh_N}\nonumber:=\left\{P\in\pos{\otimes_{n=1}^N\hh_n}:\forall \Sigma\subseteq \{1,\cdots,N\},P^{T_\Sigma}\geq0\right\},\nonumber\\
\end{equation}
where $T_\Sigma$ represents the partial transpose that acts as the transpose on systems in $\Sigma$ and the identity on the others.
It is easy to show that $\SEP{\hh_1:\cdots:\hh_N}\subseteq\PPT{\hh_1:\cdots:\hh_N}$.

A quantum channel is represented by a linear completely positive and trace-preserving (CPTP) map $\mathcal{E}:\linop{\hh_1}\rightarrow\linop{\hh_2}$. The Choi-Jamio\l kowski isomorphism defines its Choi operator $E=\sum_{i,j}\ket{i}\bra{j}\otimes\mathcal{E}(\ket{i}\bra{j})\in\linop{\hh_1\otimes\hh_2}$. The condition for a linear map $\mathcal{E}$ to be CPTP is equivalent to $E\in\pos{\hh_1\otimes\hh_2}$ and $\ptr{2}{E}=\idop$, where $\ptr{2}{E}$ represents the partial trace of the second (output) system where $E$ acts.
A quantum instrument is represented by a labeled set $\{\mathcal{E}_m:\linop{\hh_1}\rightarrow\linop{\hh_2}\}_m$ of CP maps such that $\sum_m\mathcal{E}_m$ is TP. This instrument represents the process such that we obtain a measurement outcome labeled by $m$ with probability $\tr{\mathcal{E}_m(\rho)}$ and an input $\rho\in\dop{\hh_1}$ is transformed into $\mathcal{E}_m(\rho)/\tr{\mathcal{E}_m(\rho)}$. We regard a quantum channel as a special instance of a quantum instrument.
A separable instrument is a quantum instrument $\{\mathcal{E}_m:\linop{\hh_{A_1}\otimes\hh_{B_1}}\rightarrow\linop{\hh_{A_2}\otimes\hh_{B_2}}\}_m$ each of which Choi operator $E_m$ is in the separable cone, i.e., $E_m\in\SEP{\hh_{A_1}\otimes\hh_{A_2}:\hh_{B_1}\otimes\hh_{B_2}}$.

\section{Results}
\subsection{Algebraic covering}
This section relies on several basic facts from algebraic geometry, which are summarized in Supplementary Information~\ref{sec:alggmt}.
Since our analysis focuses on optimization over a convex cone $\mathbf{C}$, we work in the associated projective space $\mathbb{P}^{d-1}$, which provides a natural representation of extreme rays of $\mathbf{C}$.
Note that there are two equivalent ways to interpret the set $\Segre{\cdim{d_1}:\cdots:\cdim{d_N}}$ of product vectors in a projective space. First, $\Segre{\cdim{d_1}:\cdots:\cdim{d_N}}$ can be viewed as $p(\Segre{\cdim{d_1}:\cdots:\cdim{d_N}}\setminus\{0\})$, where $p((x_1,x_2,\cdots))=[x_1:x_2:\cdots]$ is the canonical projection from $\cd\setminus\{0\}$ onto $\pspacedim{d-1}$. Second, $\Segre{\cdim{d_1}:\cdots:\cdim{d_N}}$ can be identified with the image of the Segre embedding from the product $\pspacedim{d_1-1}\times\cdots\times\pspacedim{d_N-1}$ of projective spaces into $\pspacedim{d_1\cdots d_N-1}$. These two descriptions are equivalent (see detail in Supplementary Information~\ref{sec:alggmt}).

First, we study a covering by algebraic sets in a projective space of fixed degree, defined as follows:

\begin{definition}
    For an algebraic set $\mathbb{A}$ in a projective space $\pspace^{d-1}$, we define the generating degree of $\mathbb{A}$ as the minimum integer $m$ such that $\mathbb{A}=Z(T)$ and $T\subseteq\cc[x_0,\cdots,x_{d-1}]$ is a set of homogeneous polynomials of degree at most $m$, where $Z(T):=\{x\in\pspace^{d-1}:\forall f\in T,f(x)=0\}$.
\end{definition}

Note that an algebraic set $\mathbb{A}$ is a subspace if and only if its generating degree is at most $1$. In this case, we denote it by $\vv$ to emphasize its linear structure.

\begin{definition}
    A finite family $\{\mathbb{A}_i\}_{i\in I}$ of algebraic sets in a projective space $\pspace^{d-1}$ is called an algebraic covering of a (not necessarily algebraic) set $\mathbb{E}$ if $\mathbb{E}\subseteq\cup_{i\in I}\mathbb{A}_i$.
    Its subfamily $\{\mathbb{A}_i\}_{i\in I'}$ with $I'\subseteq I$ is called a subcovering if $\mathbb{E}\subseteq\cup_{i\in I'}\mathbb{A}_i$.
    Moreover, $\{\mathbb{A}_i\}_{i\in I}$ is called irredundant if $\mathbb{A}_i\not\subseteq\mathbb{A}_j$ for any $i\neq j$ and $\mathbb{E}\not\subseteq\cup_{i\in I'}\mathbb{A}_i$ for any proper subset $I'\subsetneq I$.
\end{definition}

    Note that, for any algebraic covering $\{\mathbb{A}_i\}_{i\in I}$, one can obtain an irredundant subcovering $\{\mathbb{A}_i\}_{i\in I'}$ by repeatedly removing a member $\mathbb{A}_i$ such that $\mathbb{E}\subseteq \bigcup_{j\in I\setminus\{i\}}\mathbb{A}_j$
    and updating $I$ to $I\setminus\{i\}$. Since an algebraic covering is finite, this procedure terminates. However, such an irredundant subcovering is not unique in general.
    For example, let $\mathbb{E}=\{p_1,p_2,p_3\}\subset\pspace^{d-1}$ be a set of three distinct points, and define $\mathbb{A}_i=\mathbb{E}\setminus\{p_i\}$. Then
    \[
        \{\mathbb{A}_1,\mathbb{A}_2\},\quad
        \{\mathbb{A}_1,\mathbb{A}_3\},\quad
        \{\mathbb{A}_2,\mathbb{A}_3\}
    \]
    are all irredundant algebraic coverings of $\mathbb{E}$.

\begin{proposition}
    \label{prop:algcov}
    
    Suppose that a finite family $\{\mathbb{A}_i\}_{i\in I}$ of algebraic sets is an algebraic covering of an algebraic set $\mathbb{A}(\subseteq\pspace^{d-1})$.
    Then, 
    \[ 
        \forall j\in J,\exists i\in I,\mathbb{I}_j\subseteq \mathbb{A}_i,
    \]
    where $\{\mathbb{I}_j\}_{j\in J}$ is the irredundant irreducible components of $\mathbb{A}$. 
    Moreover, if $\{\mathbb{A}_i\}_{i\in I}$ is irredundant,
    \[ 
        \forall i\in I,\exists j\in J,\mathbb{I}_j\subseteq \mathbb{A}_i.
    \]
\end{proposition}
\begin{proof}
    Assume, for contradiction, that there exists $j\in J$ such that $\mathbb{I}_j\not\subseteq\mathbb{A}_i$ for every $i\in I$. Since $\{\mathbb{A}_i\}_{i\in I}$ is an algebraic covering, there exists a minimal subset $I'\subseteq I$ such that $\mathbb{I}_j\subseteq\cup_{i\in I'}\mathbb{A}_i$. Take any $i\in I'$. By the minimality of $I'$, we find that $\mathbb{I}_j$ is a union of closed proper subsets $\mathbb{A}_i\cap\mathbb{I}_j$ and $\left(\cup_{k\in I'\setminus\{i\}}\mathbb{A}_{k}\right)\cap\mathbb{I}_j$. This contradicts the irreducibility of $\mathbb{I}_j$.

    Now we assume that $\{\mathbb{A}_i\}_{i\in I}$ is irredundant. Then, for any $i\in I$, there exists $j_i\in J$ such that $\mathbb{I}_{j_i}\not\subseteq\mathbb{A}':=\cup_{k\in I\setminus\{i\}}\mathbb{A}_{k}$ and $\mathbb{I}_{j_i}\subseteq\mathbb{A}_i\cup\mathbb{A}'$. Assume, for contradiction, that there exists $i\in I$ such that $\mathbb{I}_{j_i}\not\subseteq\mathbb{A}_i$. Then, $\mathbb{I}_{j_i}$ is a union of closed proper subsets $\mathbb{A}_i\cap\mathbb{I}_{j_i}$ and $\mathbb{A}'\cap\mathbb{I}_{j_i}$. This contradicts the irreducibility of $\mathbb{I}_{j_i}$. This implies that for any $i\in I$, $\mathbb{I}_{j_i}\subseteq\mathbb{A}_i$.
\end{proof}

\begin{definition}
    For a subset $\mathbb{E}$ of $\pspace^{d-1}$ and a non-negative integer $m\in\nn$, a finite family $\{\mathbb{A}_i\}_{i\in I}$ of algebraic sets is called a degree-$m$ covering of $\mathbb{E}$ if $\{\mathbb{A}_i\}_{i\in I}$ is an algebraic covering of $\mathbb{E}$ and the generating degree of each $\mathbb{A}_i$ is at most $m$.
\end{definition}

We refer to an irredundant degree-$1$ covering $\{\vv_i\}_{i\in I}$ of $\mathbb{E}$ as a \textit{finite linear covering} (FLC) of $\mathbb{E}$. Since each subspace $\vv_i$ is a projective variety, $\vv_i\not\subseteq\vv_j$ for any $i\neq j$, and the irreducible components of an algebraic set $\cup_{i\in I}\vv_i$ are unique, $\{\vv_i\}_{i\in I}$ is uniquely determined by its union $\cup_{i\in I}\vv_i$.
Therefore, by a slight abuse of terminology, we also refer to the
algebraic set $\cup_{i\in I}\vv_i$ as an FLC.
Before proving the existence of a minimum degree-$m$ covering, we introduce the following notions.

\begin{definition}
    For a subset $\mathbb{E}$ of $\pspace^{d-1}$ and a non-negative integer $m\in\nn$, a finite family $\{\mathbb{A}_i\}_{i\in I}$ of algebraic sets is called a minimum degree-$m$ covering of $\mathbb{E}$ if it is an irredundant degree-$m$ covering of $\mathbb{E}$ such that $\cup_{i\in I}\mathbb{A}_i\subseteq\cup_{j\in J}\mathbb{A}'_j$ for any degree-$m$ covering $\{\mathbb{A}'_j\}_{j\in J}$ of $\mathbb{E}$.
\end{definition}
Note that the existence and uniqueness of a minimum degree-$m$ covering will be established in Theorem~\ref{thm:minext}.

\begin{definition}
    For a subset $\mathbb{E}$ of $\pspace^{d-1}$, 
    we define the homogeneous ideal $I_m(\mathbb{E})$ of $\mathbb{E}$ with degree $m$ as $\langle\{f\in \cup_{n=0}^m S_n:\forall \ket{x}\in\mathbb{E},f(\ket{x})=0\}\rangle$, i.e., the ideal generated by homogeneous polynomials of degree at most $m$ that vanish everywhere in $\mathbb{E}$.
\end{definition}

Note that $I_0(\mathbb{E})\subseteq I_1(\mathbb{E})\subseteq\cdots\subsetneq I_m(\mathbb{E})= I(\mathbb{E})$ for some integer $m$.
This implies $\overline{\mathbb{E}}=Z(I(\mathbb{E}))=Z(I_m(\mathbb{E}))\subseteq\cdots\subseteq Z(I_0(\mathbb{E}))$, where $\overline{\mathbb{E}}$ denotes the closure of $\mathbb{E}$ with respect to the Zariski topology.
Consequently, increasing the degree of the algebraic coverings yields progressively sharper outer approximations of $\overline{\mathbb{E}}$.
For positive integer $m$, it is known that $\ket{x}\in Z(I_m(\mathbb{I}))$ if and only if $\ket{x}^{\otimes m}\in\vee_m\mathbb{I}:=\vspan{\{\ket{x}^{\otimes m}:\ket{x}\in\mathbb{I}\}}$~\cite{HNB24}.
For completeness, we provide a proof of this fact in the Supplementary Information~\ref{sec:alggmt}.

\begin{theorem}
\label{thm:minext}
    For a subset $\mathbb{E}$ of $\pspace^{d-1}$ and a non-negative integer $m\in\nn$, its minimum degree-$m$ covering $\{\mathbb{A}_i\}_{i\in I'}$ is uniquely given, up to relabeling, by the irredundant subcovering obtained from $\{Z(I_m(\mathbb{I}_i))\}_{i\in I}$, where $\{\mathbb{I}_i\}_{i\in I}$ is the irredundant irreducible components of $\overline{\mathbb{E}}$.
\end{theorem}

\begin{proof}
    Set
    \[
        \mathbb{A}_i:=Z(I_m(\mathbb{I}_i))
        \qquad (i\in I).
    \]
    Since $\mathbb{I}_i\subseteq\mathbb{A}_i$ for every $i\in I$, the family
    $\{\mathbb{A}_i\}_{i\in I}$ is a degree-$m$ covering of
    $\overline{\mathbb{E}}$, and hence of $\mathbb{E}$.

    We first extend Proposition~\ref{prop:algcov}, which essentially expresses the indecomposability of each irreducible component $\mathbb{I}_i$ with respect to algebraic sets, to an indecomposability of $\mathbb{A}_i$ with respect to algebraic sets of bounded degree.
    Let $\{\mathbb{A}'_j\}_{j\in J}$ be an irredundant
    degree-$m$ covering of $\mathbb{E}$. Since the family is finite and each
    $\mathbb{A}'_j$ is closed, it is also a degree-$m$ covering of
    $\overline{\mathbb{E}}$. Then, by
    Proposition~\ref{prop:algcov}, we obtain
    \[
        \forall j\in J, \exists i\in I,\mathbb{I}_i\subseteq\mathbb{A}'_j\quad\wedge\quad
        \forall i\in I, \exists j\in J,\mathbb{I}_i\subseteq\mathbb{A}'_j.
    \]
    If $\mathbb{I}_i\subseteq\mathbb{A}'_j$ and $\mathbb{A}'_j=Z(T_j)$ with a set $T_j$ of homogeneous polynomials of degree at most $m$, we find $T_j\subseteq I_m(\mathbb{I}_i)$. This implies that $\mathbb{I}_i\subseteq Z(I_m(\mathbb{I}_i))\subseteq Z(T_j)=\mathbb{A}'_j$.
    Hence
    \[
        \forall j\in J, \exists i\in I,\mathbb{A}_i\subseteq\mathbb{A}'_j\quad\wedge\quad
        \forall i\in I, \exists j\in J,\mathbb{A}_i\subseteq\mathbb{A}'_j.
    \]    
    
    By taking an irredundant subcovering $\{\mathbb{A}_i\}_{i\in I'}$ from $\{\mathbb{A}_i\}_{i\in I}$, we find that $\{\mathbb{A}_i\}_{i\in I'}$ is a minimum degree-$m$ covering of $\mathbb{E}$.

    It remains to prove uniqueness. Let
    $\{\mathbb{A}_i\}_{i\in I'}$ and $\{\mathbb{A}_j\}_{j\in J'}$ be two
    irredundant subcoverings obtained from $\{\mathbb{A}_i\}_{i\in I}$.
    By the indecomposable property shown above, for every $i\in I'$, there exists $j\in J'$ such that $\mathbb{A}_i\subseteq \mathbb{A}_j$.
    Applying the same argument again to $\mathbb{A}_j$, there exists $k\in I'$ such that $\mathbb{A}_j\subseteq \mathbb{A}_k$.
    Thus,
    \[
        \forall i\in I',\exists j\in J',\exists k\in I',\mathbb{A}_i\subseteq\mathbb{A}_j\subseteq \mathbb{A}_k.  
    \]
    Since $\{\mathbb{A}_i\}_{i\in I'}$ is irredundant, we must have
    $\mathbb{A}_i=\mathbb{A}_k$, and hence $\mathbb{A}_i=\mathbb{A}_j$. This completes the proof.
\end{proof}

This theorem establishes both the existence and an explicit construction of a minimum degree-$m$ covering. In Section~\ref{sec:application}, we focus on a specific choice of $m$, defined as follows:
\begin{definition}
\label{def:MFLC}
    The minimum degree-$1$ covering 
    obtained as the irredundant subcovering of a degree-$1$ covering
    $\{\pp_i=Z(I_1(\mathbb{I}_i))=\vspan{\mathbb{I}_{i}}\}_{i\in I}$ is called the minimum finite linear covering (MFLC) of $\mathbb{E}$, where $\{\mathbb{I}_i\}_{i\in I}$ is the irredundant irreducible components of $\overline{\mathbb{E}}$.
\end{definition}
By Theorem~\ref{thm:minext}, the irredundant subcovering is uniquely given as the irredundant irreducible components of $\cup_{i\in I}\pp_i$. Therefore, by a slight abuse of terminology, we also refer to the algebraic set $\cup_{i\in I}\pp_i$ as the MFLC.

In general, decomposing an algebraic set $\mathbb{A}$ into irreducible components is difficult since it is essentially equivalent to performing the prime decomposition of an ideal in a polynomial ring $\cc[x_0,\cdots,x_{d-1}]$, which is regarded as a difficult problem.
However, we can derive the following proposition useful for determining whether an FLC is minimum.

\begin{proposition}
    \label{prop:mintest}
    Let $\mathbb{D}\subseteq\pspace^{d-1}$ be an irreducible set with respect to the Zariski topology. For a set $T:=\{f_0(x),\cdots,f_{d'-1}(x)\}$ of homogeneous polynomials of $d$ variables of the same degree such that $Z(T)=\emptyset$, $\mathbb{E}:=f(\mathbb{D})$ is irreducible, where $f:\pspacedim{d-1}\rightarrow\pspacedim{d'-1}$ is defined by $f(x):=[f_0(x):\cdots:f_{d'-1}(x)]$. Moreover, the MFLC of $\mathbb{E}$ is $\vspan{\mathbb{E}}$.
\end{proposition}
\begin{proof}
    If we can show that $\mathbb{E}$ is irreducible, the statement after the `moreover' is a direct consequence of Theorem~\ref{thm:minext} since $\overline{\mathbb{E}}$ is irreducible if $\mathbb{E}$ is and $\vspan{\overline{\mathbb{E}}}=\vspan{\mathbb{E}}$.
    While the irreducibility of $\mathbb{E}$ is known to be an elementary property of a regular map, we provide a proof for completeness.
    If $\mathbb{E}$ is not irreducible, there exist closed sets $\mathbb{E}_1$ and $\mathbb{E}_2$ in $\pspace^{d'-1}$ such that $\mathbb{E}\subseteq \mathbb{E}_1\cup \mathbb{E}_2$ and $\mathbb{E}\not\subseteq \mathbb{E}_1,\mathbb{E}_2$.
    Then, we obtain $\mathbb{D}\subseteq f^{-1}(\mathbb{E}_1)\cup f^{-1}(\mathbb{E}_2)$, $\mathbb{D}\not\subseteq f^{-1}(\mathbb{E}_b)$, and $f^{-1}(\mathbb{E}_b)$ is closed for $b\in\{1,2\}$. This contradicts the fact that $\mathbb{D}$ is irreducible.
\end{proof}

\subsubsection*{Example I: MFLC of the symmetric product states}
The MFLC of the set $\mathbb{E}=\{\ket{\phi}^{\otimes N}:\ket{\phi}\in\pspace^{d-1}\}$ of symmetric product states is $\vspan{\mathbb{E}}$, which corresponds to the symmetric subspace $\vee_{n=1}^N\cd:=\{\ket{\Xi}\in(\cd)^{\otimes N}:\forall\pi\in S_N,P_{\pi}\ket{\Xi}=\ket{\Xi}\}$ in the ambient complex vector space, where $S_N$ is the symmetric group and $P_\pi$ is a permutation operator, defined by $P_\pi\ket{i_1\cdots i_N}=\ket{i_{\pi(1)}\cdots i_{\pi(N)}}$.

\begin{proof}
    Let $f(x)=\left[x_0:x_1:\cdots:x_{d-1}\right]^{\otimes N}$ be a vector-valued homogeneous polynomial of degree $N$ from $\pspace^{d-1}$ onto $\mathbb{E}$. By applying Proposition~\ref{prop:mintest} with $\mathbb{D}=\pspace^{d-1}$, we find the MFLC of $\mathbb{E}$ is $\vspan{\mathbb{E}}$.
\end{proof}

\subsubsection*{Example II: MFLCs in two qubits}

We provide a comprehensive characterization of MFLCs in two qubits as a pedagogical example. As demonstrated in Supplementary Information~\ref{appendix:Koashi}, they are also useful in optimizing separable measurements in a certain unambiguous local state discrimination task. We observe that $\Segre{\cdim{2}:\cdim{2}}=Z(x_{11}x_{22}-x_{12}x_{21})$, where $Z(f):=f^{-1}(0)$ is the zero set in $\pspacedim{3}$ of a homogeneous polynomial $f\in\cc[x_{11},x_{12},x_{21},x_{22}]$. 
Such a simple characterization of $\Segre{\cdim{2}:\cdim{2}}$ allows us to make a comprehensive characterization of MFLCs through the following proposition.

\begin{proposition}
\label{prop:MFLCof2Q}
    Let $\mathbb{E}=\Segre{\cdim{2}:\cdim{2}}\cap\vv$ be a nonempty subset in $\pspacedim{3}$ with a subspace $\vv\subseteq\pspacedim{3}$. The MFLC of $\mathbb{E}$ is
    \begin{itemize}
        \item $\vspan{\mathbb{E}}$ if $\mathbb{E}$ is irreducible, and
        \item $\mathbb{E}$ itself if $\mathbb{E}$ is reducible. Moreover,  $\mathbb{E}=\pp_1\cup\pp_2$ with two distinct subspaces $\pp_1$ and $\pp_2$.
    \end{itemize}
\end{proposition}
Note that $\mathbb{E}$ has a simple structure in the latter case. This simplification enables us to solve range-constrained SEP optimization problems without using the DPS hierarchy as demonstrated in Supplementary Information~\ref{appendix:Koashi}.

\begin{proof}
    If $\mathbb{E}$ is irreducible, its MFLC is $\vspan{\overline{\mathbb{E}}}=\vspan{\mathbb{E}}$ from Theorem~\ref{thm:minext}. We show that $\mathbb{E}$ can be decomposed into two distinct irreducible components as $\pp_1\cup\pp_2$ if $\mathbb{E}$ is reducible. Since $\vv\neq\emptyset$, there exist $d\geq1$ and a rank-$d$ matrix $V$ such that $\vv=V\pspacedim{d-1}$.  Thus, we can show that $\mathbb{E}=VZ(f)$ with $f(t)=\left(\sum_j V_{1j}t_j\right)\left(\sum_j V_{4j}t_j\right)-\left(\sum_j V_{2j}t_j\right)\left(\sum_j V_{3j}t_j\right)$.
    From Proposition~\ref{prop:mintest}, $Z(f)$ is reducible since $\mathbb{E}$ is reducible. Since it is known that $Z(g)$ is irreducible if $g$ is an irreducible polynomial in $\cc[t_1,t_2,\cdots,t_d]$~\cite[Exercise 2.8]{RHBook}, $f$ is a constant or non-constant reducible polynomial. If $f$ is a constant, $f(t)=0$ since $\mathbb{E}\neq\emptyset$. However, this implies $\mathbb{E}=\vv$, which contradicts the reducibility of $\mathbb{E}$. Thus, $f$ is a non-constant reducible polynomial, and it can be decomposed as
    \begin{equation}
        f(t)=\left(\sum_j\alpha_j t_j\right)\left(\sum_j\beta_j t_j\right)
    \end{equation}
    with some $\alpha_j,\beta_j\in\cc$ such that $\sum_j|\alpha_j|\neq0$ and $\sum_j|\beta_j|\neq0$ since $f$ is a homogeneous polynomial of degree $2$.
    By letting $\hat{\pp}_1=\{[t_1:\cdots:t_{d}]:\sum_j\alpha_j t_j=0\}$ and $\hat{\pp}_2=\{[t_1:\cdots:t_{d}]:\sum_j\beta_j t_j=0\}$, we can verify that the irreducible components of $Z(f)$ are $\hat{\pp}_1$ and $\hat{\pp}_2$. Since $Z(f)$ is reducible,  $\hat{\pp}_1\neq\hat{\pp}_2$. By letting $\pp_b=V\hat{\pp}_b$ for $b\in\{1,2\}$, we find that the irreducible components of $\mathbb{E}$ are $\pp_1$ and $\pp_2$ and $\pp_1\neq\pp_2$.
    
\end{proof}

\begin{figure}[ht]
    \centering
    \includegraphics[width=0.8\linewidth]{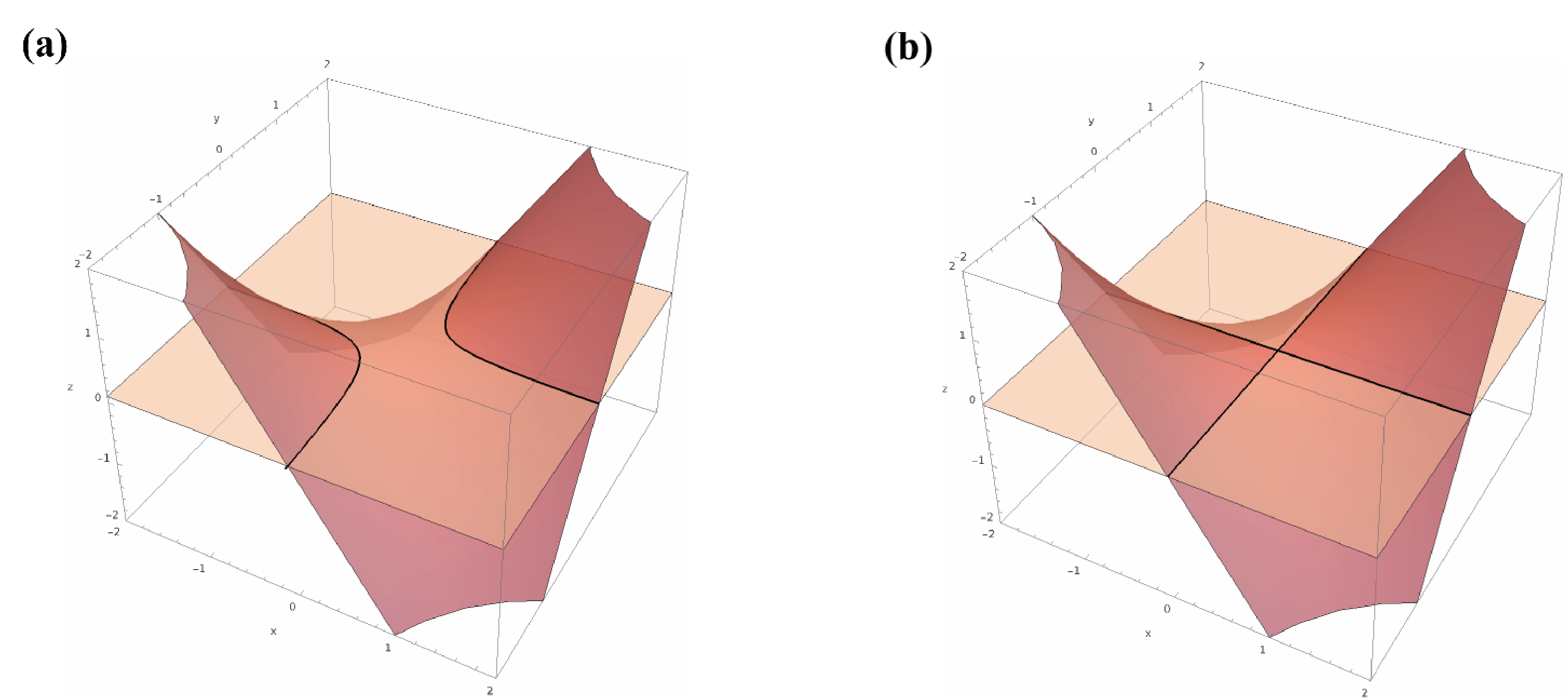}
    \caption{ Examples of MFLC and irreducible components of $\mathbb{E}=\Segre{\cdim{2}:\cdim{2}}\cap\mathcal{V}$. We plot points $\{(x,y,z):[1:x:y:z]\in\mathbb{E}\}$ in a three-dimensional space. In this space, $\Segre{\cdim{2}:\cdim{2}}$ is represented by the red surface defined by $z=xy$. $\mathbb{E}$ and $\mathcal{V}$ are depicted by black curves (or lines) and brown planes, respectively. Each of $\mathbb{E}$, $\Segre{\cdim{2}:\cdim{2}}$, and $\mathcal{V}$ is closed with respect to the Zariski topology. While $\Segre{\cdim{2}:\cdim{2}}$ and $\mathcal{V}$ are irreducible, the irreducibility of $\mathbb{E}$ differs in the two cases shown.
    (a) When $\mathcal{V}$ is defined by its normal vector $(1,0,0,-10)^T$, $\mathbb{E}$ is irreducible, as there are no polynomials whose set of zeros defines a proper subset of $\mathbb{E}$.
    In this case, the MFLC is $\vspan{\mathbb{E}}=\mathcal{V}$.
    (b) When $\mathcal{V}$ is defined by its normal vector $(0,0,0,1)^T$, $\mathbb{E}=(\pspacedim{1}\otimes\ket{0})\cup(\ket{0}\otimes\pspacedim{1})$ is reducible into two subspaces $\pspacedim{1}\otimes\ket{0}$ and $\ket{0}\otimes\pspacedim{1}$ since each is a proper closed subset of $\mathbb{E}$. In this case, the MFLC is $\mathbb{E}(\subsetneq\mathcal{V})$ itself.
    }
    \label{fig:diagramMFLC}
\end{figure}

Illustrative examples of MFLC in two qubits are shown in Fig.~\ref{fig:diagramMFLC}.

\subsubsection*{Example III: MFLCs for canonical subspace}

In this example, we calculate the MFLCs of the intersection between $\mathbb{S}$ and a certain subspace $\hat{\ww}$, which appears in almost all the optimization problems discussed in the next section. 
Considering its wide applicability, we will refer to this subspace as the \textit{canonical subspace}.

\begin{figure}[ht]
    \centering
    \includegraphics[width=5.5cm]{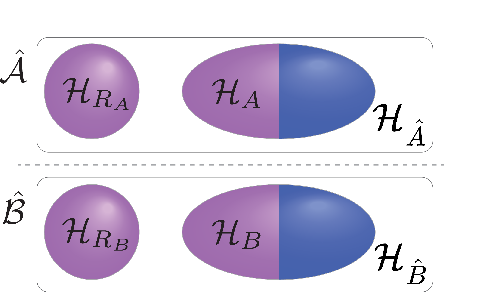}
    \caption{Hilbert spaces where the canonical subspace is defined. We consider the product vectors between $\hat{\mathcal{A}}$ and $\hat{\mathcal{B}}$.}
    \label{fig:Hspacesext0}
\end{figure}

Let $\ket{\tau}=\idop^{(R_A)}\otimes L_1^{(R_B)}\ketUMES{R_AR_B}{d}$ and $\ket{L_2}=\idop^{(A)}\otimes L_2^{(B)}\ketUMES{AB}{d}$, where $L_1^{(R_B)}\in\linop{\hh_{R_B}}$ and $L_2^{(B)}\in\linop{\hh_B}$ are full-rank operators, and $\dim\hh_A=\dim\hh_{R_A}=\dim\hh_B=\dim\hh_{R_B}=d$.
Suppose that the Hilbert spaces $\hh_A$ and $\hh_B$ are embedded in extended Hilbert spaces as $\hh_A\subseteq\hh_{\hat{A}}$ and $\hh_B\subseteq\hh_{\hat{B}}$ (see Fig.~\ref{fig:Hspacesext0}).
A canonical subspace $\hat{\ww}$ is defined by
\begin{equation}
    \hat{\ww}:=\{\ket{\Xi}\in\hat{\mathcal{A}}\otimes\hat{\mathcal{B}}:\bra{\tau}_{R_AR_B}\ket{\Xi}\in\vspan{\{\ket{L_2}\}}\},
\end{equation}
with a descriptive adjective representing the degree of freedom in the parameters, as shown in Table~\ref{table:canonicalsubspace}.

When considering the intersection $\Segre{\hat{\mathcal{A}}:\hat{\mathcal{B}}}\cap\hat{\ww}$, we identify $\hat{\ww}$ with its projectivization $p(\hat{\ww}\setminus\{0\})$ in the projective space $\pspacedim{\dim\hat{\mathcal{A}}\dim\hat{\mathcal{B}}-1}$. We adopt this slight abuse of notation throughout.

First, we show that $\pp\cup\hat{\ww}^\circ$ forms an FLC of $\Segre{\hat{\mathcal{A}}:\hat{\mathcal{B}}}\cap\hat{\ww}$, where
\begin{eqnarray}
\label{eq:MFLCofETCS}
    \pp&:=&\left(V_A\otimes\idop^{(R_A)}\otimes\left(L_1^\dag\right)^{-1}\otimes\left(V_BL_2\right)\right)\CMFLC{d},\\
    \hat{\ww}^\circ&:=&\{\ket{\Xi}\in\hat{\mathcal{A}}\otimes\hat{\mathcal{B}}:\bra{\tau}_{R_AR_B}\ket{\Xi}=0\}.
\end{eqnarray}
Here, $V_A:\hh_A\rightarrow\hh_{\hat{A}}$ and $V_B:\hh_B\rightarrow\hh_{\hat{B}}$ are isometry operators that can be represented by $V_A=V_B=\sum_{i=0}^{d-1}\ketbra{i}$ with the computational basis $\{\ket{i}\}_{i=0}^{d-1}$ of $\hh_A$ or $\hh_B$ defining the maximally entangled state in $\hh_A\otimes\hh_B$. The set $\CMFLC{d}$ is defined as
\begin{eqnarray}
         \CMFLC{d}&:=&\Csubspace{d}\cap\CDsubspace{d},\\
    \Csubspace{d}&:=&\{\ket{\Xi}\in\hh_A\otimes\hh_{R_A}\otimes\hh_{R_B}\otimes\hh_B:\braUMES{R_AR_B}{d}\ket{\Xi}\in\vspan{\{\ketUMES{AB}{d}\}}\},\\
        \CDsubspace{d}&:=&\{\ket{\Xi}\in\hh_A\otimes\hh_{R_A}\otimes\hh_{R_B}\otimes\hh_B:\braUMES{AB}{d}\ket{\Xi}\in\vspan{\{\ketUMES{R_AR_B}{d}\}}\}.
\end{eqnarray}

Second, we show that $\mathbb{V}:=\left(\Segre{\hat{\mathcal{A}}:\hat{\mathcal{B}}}\cap\hat{\ww}\right)\setminus\hat{\ww}^\circ$ is irreducible. 
Since its Zariski closure $\overline{\mathbb{V}}$ is also irreducible, Theorem \ref{thm:minext} implies that the MFLC of $\mathbb{V}$ is given by $\vspan{\overline{\mathbb{V}}}=\vspan{\mathbb{V}}$.
Since $\mathbb{V}\subseteq\Segre{\hat{\mathcal{A}}:\hat{\mathcal{B}}}\cap\hat{\ww}$, the MFLC $\cup_i\pp_i$ of $\Segre{\hat{\mathcal{A}}:\hat{\mathcal{B}}}\cap\hat{\ww}$ is an FLC of $\mathbb{V}$. From the minimality of the MFLC, this yields
\[
\vspan{\mathbb{V}}\subseteq\cup_i\pp_i\subseteq\pp\cup\hat{\ww}^\circ.
\]
Since each $\pp_i$ is irreducible, it follows that for any $i$, either $\pp_i\subseteq\pp$ or $\pp_i\subseteq\hat{\ww}^\circ$. 
Moreover, because $\vspan{\mathbb{V}}$ is irreducible, there exists $i$ such that $\vspan{\mathbb{V}}\subseteq\pp_i$.

Third, we show that $\pp\subseteq\vspan{\mathbb{V}}$. This implies that $\pp$ is an element of the MFLC of $\Segre{\hat{\mathcal{A}}:\hat{\mathcal{B}}}\cap\hat{\ww}$, and the MFLC of $\mathbb{V}$ is $\vspan{\mathbb{V}}=\pp$. Since the MFLC is irredundant, all remaining elements of the MFLC must be subspaces contained in $\hat{\ww}^{\circ}$.
As these subspaces do not contribute to the optimization problem we concern, we do not derive them explicitly.

We prove these three facts by first analyzing the simplest case and then gradually generalizing the argument, as summarized in Table~\ref{table:canonicalsubspace}.
Note that $\dim(\Csubspace{d})=d^4-(d^2-1)$, and $\dim(\CMFLC{d})=d^4-2(d^2-1)$ (in a complex vector space.)
This difference in dimension contributes to improving the bounds for numerically solving the range-constrained SEP optimization problem and reducing the size of SDP as shown in the next section. A complete proof is given in Supplementary Information~\ref{sec:MFLCCS}.

\begin{table}[ht]
 \caption{Variants of the canonical subspace we investigate and MFLCs.}
 \label{table:canonicalsubspace}
 \centering
\begin{tabular}{lll}
\hline
 Name of $\hat{\ww}$ & Conditions imposed on $\hat{\ww}$  \\
 \hline
 Canonical subspace ($\hat{\ww}=\Csubspace{d}$) & $L_1=L_2=\idop$, $\hh_{\hat{A}}=\hh_A$ and $\hh_{\hat{B}}=\hh_B$  \\
 Twisted canonical subspace & $\hh_{\hat{A}}=\hh_A$ and $\hh_{\hat{B}}=\hh_B$  \\
 Extended canonical subspace & $L_1=L_2=\idop$  \\
 Extended and twisted canonical subspace & None  \\
  \hline
\end{tabular}
\end{table}

\subsection{Applications of algebraic covering}
\label{sec:application}
\subsubsection{General framework in cone optimization}

Let $\mathbb{A}$ be an algebraic set in $\pspace^{d-1}$. We consider an optimization problem over a cone $\pos{\mathbb{A}}(\subseteq\pos{\cd})$ defined by
\begin{equation}
    \pos{\mathbb{A}}:=\left\{\sum_{j\in J}p_j\phi_j:p_j\geq0,\ket{\phi_j}\in\mathbb{A}\right\},
\end{equation}
where we regard $\mathbb{A}$ as a set of unit vectors in $\cd$. 
Note that $\pos{\mathbb{A}}$ is closed with respect to the Euclidean topology on $\linop{\cd}$ since we can assume $|J|\leq d^2+1$ by Carath\'eodory’s theorem, and the set of unit vectors in $\mathbb{A}$ is compact with respect to the Euclidean topology on $\cd$.
The optimization is computationally hard for a general algebraic set $\mathbb{A}$.
However, we can construct a convergent hierarchy by extending the one proposed by Derksen \emph{et al.}~\cite{HNB24} as follows.

\begin{definition}[Prime $\mathbb{A}$-tension]
    Let $\mathbb{A}$ be an algebraic set in $\pspacedim{d-1}$. A positive semi-definite operator $P \in \pos{\cd}$ is said to be prime $(k,\mathbb{A})$-tendable if there is a set $\{ Q^k_i \}_{i \in I} \subseteq \pos{\otimes_{n=1}^k\cd}$ of positive semi-definite operators such that
    \begin{align}
        &Q^k_i \in \pos{\vee_k\mathbb{I}_i}\quad (\forall i \in I),\\
        &\ptr{k-1}{\sum_i Q^k_i } = P,
    \end{align}
    where $\{\mathbb{I}_i\}_{i\in I}$ is the irredundant irreducible components of $\mathbb{A}$ and a subspace $\vee_m\mathbb{I}$ symmetrizing $\mathbb{I}$ is defined as $\vee_m\mathbb{I}:=\vspan{\{\ket{x}^{\otimes m}:\ket{x}\in\mathbb{I}\}}$.
    The set $\{ Q^k_i \}_{i \in I}$ satisfying the above is called a prime $(k,\mathbb{A})$-tension of $P$.
\end{definition}

Note that, if $\mathbb{A}$ is irreducible, the prime $(k,\mathbb{A})$-tension coincides with the $(k,\mathbb{A})$-tension defined by Derksen \emph{et al.}~\cite[Definition~3.1]{HNB24}.

Here, we use the term ``prime" to refer to a prime ideal, which corresponds to an irreducible component.
Since $\vee_m\mathbb{I}$ is a subspace, the maximization of a linear functional $f(P)$ over prime-$(k,\mathbb{A})$-tendable operators can be efficiently solved using semi-definite programming. The following theorem guarantees that the resulting optimal value converges from above to $\max_{P\in\pos{\mathbb{A}}}f(P)$.

\begin{theorem}[Prime-$\mathbb{A}$-tension hierarchy]
    \label{thm:hierarchy}
    For a positive semi-definite operator $P\in\pos{\cd}$ and an algebraic set $\mathbb{A}$ in $\pspacedim{d-1}$, $P\in\pos{\mathbb{A}}$ if and only if $P$ is prime $(k,\mathbb{A})$-tendable for all $k\in\nn$.
\end{theorem}
Note that we prove this theorem by modifying the argument presented in the proof of \cite[Theorem 1.3]{HNB24} and by exploiting the minimum degree-$m$ covering.
\begin{proof}
    If $P\in\pos{\mathbb{A}}$, it admits a decomposition $P=\sum_{i\in I}P_i$ with $P_i=\sum_xp_i(x)\phi_{i,x}\in\pos{\mathbb{I}_i}$, where $\{\mathbb{I}_i\}_{i\in I}$ is the irredundant irreducible components of $\mathbb{A}$.
    Defining $Q_i^k=\sum_xp_i(x)\phi_{i,x}^{\otimes k}$, we readily verify that $\{ Q^k_i \}_{i \in I}$ forms a prime $(k,\mathbb{A})$-tension of $P$.

    Conversely, suppose that $P$ is prime $(k,\mathbb{A})$-tendable for all $k\in\nn$, and let $\{ Q^k_i \}_{i \in I}$ be a corresponding a prime $(k,\mathbb{A})$-tension of $P$.
    Fix a natural number $m\in\nn$. For any $k\ge m$, define $\left\{\hat{Q}_{i,k}^m:=\ptr{k-m}{Q_i^k}\right\}_{i\in I}$. Then it constitutes a prime $(m,\mathbb{A})$-tension of $P$.
    Moreover, since $\tr{\hat{Q}_{i,k}^m}\leq\tr{\sum_i \hat{Q}_{i,k}^m}=\tr{P}$ and $\hat{Q}_{i,k}^m\in\pos{\vee_m\mathbb{I}_i}$, the sequence $\{\hat{Q}_{i,k}^m\}_{k(\geq m)}$ is contained in a compact set for each $i$.
    Consequently, there exist a subsequence $k_1\leq k_2\leq\cdots$ and a prime $(m,\mathbb{A})$-tension $\{\hat{\sigma}_i^m\}_{i\in I}$ of $P$ such that
    \begin{equation}
        \label{eq:converge1}
        \hat{\sigma}_i^m=\lim_{j\rightarrow\infty}\hat{Q}_{i,k_j}^m
    \end{equation}
    for every $i$. 
    
    On the other hand, a quantum de Finetti theorem \cite[Theorem 7.26]{WBook} implies that there exists a permutation-invariant separable operator $\sum_{x}p_{i}(x)\phi_{i,x}^{\otimes m}$ such that $\sum_xp_{i}(x)=\tr{\hat{\sigma}_i^m}$ and
    \begin{equation}
        \label{eq:converge2}
        \lim_{j\rightarrow \infty}\hat{Q}_{i,k_j}^m=\lim_{j\rightarrow \infty}\ptr{k_j-m}{Q_i^{k_j}}=\sum_{x}p_{i}(x)\phi_{i,x}^{\otimes m}.
    \end{equation}
    Combining Eq.~\eqref{eq:converge1} and \eqref{eq:converge2}, we obtain 
    \[
    \hat{\sigma}_i^m=\sum_{x}p_{i}(x)\phi_{i,x}^{\otimes m}.
    \]

    To complete the proof, it suffices to show that $P$ is included in a cone generated by the minimum degree-$m$ covering, i.e., 
    \[
    P\in\pos{\bigcup_{i\in I}Z(I_m(\mathbb{I}_i))}.
    \]
    Indeed, for sufficiently large $m$, we have $\bigcup_{i\in I}Z(I_m(\mathbb{I}_i))=\bigcup_{i\in I}\mathbb{I}_i=\mathbb{A}$.
    Since $\hat{\sigma}_i^m=\sum_{x}p_{i}(x)\phi_{i,x}^{\otimes m}\in\pos{\vee_m\mathbb{I}_i}$, we obtain $\ket{\phi_{i,x}}^{\otimes m}\in \vee_m\mathbb{I}_i$ for all $i$ and $x$. By Proposition \ref{prop:ZIm}, this implies $\ket{\phi_{i,x}}\in Z(I_m(\mathbb{I}_i))$. 
    Finally, noting that $P=\sum_{i\in I}\ptr{m-1}{\hat{\sigma}_i^m}=\sum_{i\in I}\sum_xp_{i}(x)\phi_{i,x}$, we conclude that $P\in\pos{\bigcup_{i\in I}Z(I_m(\mathbb{I}_i))}$.
\end{proof}

In the above proof, the quantum de Finetti theorem is used to guarantee the existence of separable prime $(m,\mathbb{A})$-tensions.
However, the hierarchy can be strengthened by using separability criteria, for example by incorporating PPT constraints or entanglement witnesses.
Notably, in contrast to the DPS hierarchy and its generalizations~\cite{HNB24}, the prime $(k,\mathbb{A})$-tension already provides a nontrivial characterization of $\pos{\mathbb{A}}$ even for $k=1$ as
\[
\pos{\mathbb{A}}\subseteq\pos{\bigcup_{i\in I} \vspan{\mathbb{I}_i}}
\]
when $\mathbb{A}$ is reducible.

We can also derive a quantum de Finetti theorem for prime
$(k,\mathbb{A})$-extendible states, as a quantitative version of
Theorem~\ref{thm:hierarchy}, by exploiting the quantum de Finetti theorem for their irreducible components. 

\begin{proposition}
\label{prop:prime-deFinetti}
Let $\mathbb{A}$ be an algebraic set in $\pspace^{d-1}$ and $\{\mathbb{I}_i\}_{i\in I}$ be its irredundant irreducible components.
Suppose that, for fixed $k$ and for each $i\in I$, there exists a constant $f(\mathbb{I}_i)$ such that
\begin{equation}
    \min_{\sigma\in\dop{\mathbb{I}_i}}
    \lpnorm{1}{\rho-\sigma}
    \leq f(\mathbb{I}_i)
\end{equation}
holds for any $(k,\mathbb{I}_i)$-tendable state $\rho$,
where we define
$\dop{\mathbb{B}}:=\pos{\mathbb{B}}\cap\dop{\cd}$ for an algebraic set $\mathbb{B}$.
Then, any prime $(k,\mathbb{A})$-tendable state $\rho$ satisfies
\begin{equation}
    \min_{\sigma\in\dop{\mathbb{A}}}
    \lpnorm{1}{\rho-\sigma}
    \leq
    \max_{i\in I} f(\mathbb{I}_i).
\end{equation}
\end{proposition}
The proof and an application to biseparability testing are given in Appendix~\ref{appendix:deFinetti}.

\subsubsection{Range-constrained SEP optimization}
\begin{figure}[ht]
    \centering
    \includegraphics[width=12cm]{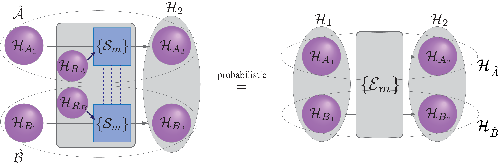}
    \caption{General setting of the implementation of a non-local instrument $\{\mathcal{E}_m\}_{m\in\Sigma}$ by using a separable instrument $\{\mathcal{S}_m\}_{m\in\Sigma}\cup\{\mathcal{S}_{\texttt{fail}}\}$ assisted by an entangled state in $\hh_{R_A}\otimes\hh_{R_B}$. For all $m\in\Sigma\cup\{\texttt{fail}\}$, the Choi operator of $\mathcal{S}_m$ is an element in $\SEP{\hat{\mathcal{A}}:\hat{\mathcal{B}}}$. We assume that we simulate the non-local instrument without error by post-selecting events corresponding to $m\in\Sigma$.
}
    \label{fig:Hspaces}
\end{figure}

In this section, we use the MFLCs to solve range-constrained SEP optimization problems for the scenario where a non-local instrument is implemented with the assistance of limited entanglement.
We depict the general setting of the entanglement-assisted implementation of a non-local instrument by using a separable instrument in Fig.~\ref{fig:Hspaces},
where $\hat{\mathcal{A}}=\hh_{\hat{A}}\otimes\hh_{R_A}$, $\hat{\mathcal{B}}=\hh_{R_B}\otimes\hh_{\hat{B}}$,
$\hh_{\hat{A}}=\hh_{A_1}\otimes\hh_{A_2}$, $\hh_{\hat{B}}=\hh_{B_1}\otimes\hh_{B_2}$
 and $\hh_b=\hh_{A_b}\otimes\hh_{B_b}$ for $b\in\{1,2\}$.
Note that the dimension of some Hilbert spaces can be $1$ for some non-local instruments.
Let us consider a separable instrument $\{\mathcal{S}_m:\linop{(\hh_{A_1}\otimes\hh_{R_A})\otimes(\hh_{B_1}\otimes\hh_{R_B})}\rightarrow\linop{\hh_{A_2}\otimes\hh_{B_2}}\}_{m\in\Sigma\cup\{\texttt{fail}\}}$ with a special measurement outcome corresponding to $m=\texttt{fail}\notin\Sigma$.
We focus on the setting where, for all input states $\rho\in\dop{\hh_{1}}$, there exists a probability $p(\rho)\in[0,1]$ such that $\mathcal{S}_{m}(\rho\otimes\tau)=p(\rho)\mathcal{E}_m(\rho)$ for any $m\in\Sigma$, where $\{\mathcal{E}_m:\linop{\hh_1}\rightarrow\linop{\hh_2}\}_{m\in\Sigma}$ is a target non-local instrument and $\tau\in\dop{\hh_{R_A}\otimes\hh_{R_B}}$ is a resource entangled state.
This constraint guarantees that we can perfectly simulate the measurement distribution and output state of the non-local instrument by post-selecting events that correspond to $m\in\Sigma$.
In this scenario, the probability $p(\rho)$ corresponds to the success probability of the post-selection.

Table~\ref{table:Nop} summarizes all classes of non-local instruments investigated in this section.
Here, we put some constraints on the dimension of the Hilbert spaces and the Schmidt rank.
Note that for any vector $\ket{\Xi}\in\hh_1\otimes\hh_2$, there exists a Schmidt decomposition $\ket{\Xi}=\sum_{k\in K}p_k\ket{\phi_k}_1\ket{\psi_k}_2$, where $p_k>0$ and $\{\ket{\phi_k}\}_{k}$ and $\{\ket{\psi_k}\}_{k}$ are orthonormal vectors in $\hh_1$ and $\hh_2$, respectively. 
The Schmidt rank of $\ket{\Xi}$, denoted by $\Schrank{1:2}{\ket{\Xi}}$, is $|K|$.

\begin{table}[ht]
 \caption{Classes of non-local instruments we investigate and the assumptions on their Schmidt rank and the dimension of Hilbert spaces,
 where $\ket{U}=\sum_{i}\ket{i}_1\otimes(U\ket{i}_1)\in\hh_1\otimes\hh_2\simeq\hh_{\hat{A}}\otimes\hh_{\hat{B}}$, $\sum_m\ketbra{M_m}=\idop$, $\phi\in\puredop{\hh_1}$, $\psi\in\puredop{\hh_2}$ and $q\in(0,1]$.
 We assume that a resource state $\tau$ is a pure state $\ketbra{\tau}$ except for distillation.
 Note that the channel representing entanglement distillation does not depend on the input state since it has no input, i.e., $\dim\hh_1=1$.
 We assume the resource state $\tau$ to be a fixed mixed state for entanglement distillation.}
 \label{table:Nop}
 \centering
\begin{tabular}{lll}
\hline
 Class & $\{\mathcal{E}_m(\rho)\}_m$ & Assumption  \\
 \hline
 State verification & $ \{\tr{(q\phi)\rho}, \tr{(\idop-q\phi)\rho}\}$ & $\Schrank{{A_1}:{B_1}}{\ket{\phi}}= \Schrank{{R_A}:{R_B}}{\ket{\tau}}$\\
Rank-1 POVM & $\{\bra{M_m}\rho\ket{M_m}\}_m$ & $\Schrank{{A_1}:{B_1}}{\ket{M_m}}= \Schrank{{R_A}:{R_B}}{\ket{\tau}}$ \\
 Unitary channel & $\{U\rho U^\dag\}$ & 
 $
 \Schrank{\hat{A}:\hat{B}}{\ket{U}}= \Schrank{{R_A}:{R_B}}{\ket{\tau}}
 $
\\
 \begin{tabular}{c}
  Entanglement\\distillation
\end{tabular}
 & $\{\psi\}$ & $\Schrank{{A_2}:{B_2}}{\ket{\psi}}=2$ \\
  \hline
\end{tabular}
\end{table}

In Supplementary Information~\ref{appendix:constsuccessprob}, we show that if $\{\mathcal{E}_m\}_{m\in\Sigma}$ shown in Table~\ref{table:Nop} can be perfectly simulated with the post-selection, the success probability $p(\rho)$ of the post-selection does not depend on the input state $\rho$.
That is, in our setting, we can assume $\exists p\in[0,1],\forall\rho\in\dop{\hh_1},\forall m\in\Sigma,\mathcal{S}_m(\rho\otimes\tau)=p\mathcal{E}_m(\rho)$ without loss of generality.
We would like to maximize the success probability $p$ except in the case of the state verification (see Section~\ref{sec:verification} for the state verification). This optimization problem can be formulated as
\begin{equation}
\label{eq:SEPoptformatpre}
p(\{\mathcal{E}_m\}_m,\tau):=\max\left\{p\in\rr:
\begin{array}{l}
\forall m\in\Sigma, S_m\in \SEP{\hat{\mathcal{A}}:\hat{\mathcal{B}}}, \ptr{R_AR_B}{S_m\overline{\tau}}= p\ketbra{E_m},     \\
\idop-\sum_{m\in\Sigma}\ptr{2}{S_m}\in \SEP{\hh_{A_1}\otimes\hh_{R_A}:\hh_{R_B}\otimes\hh_{B_1}}
\end{array}
\right\},
\end{equation}
where $S_m$ and $\ketbra{E_m}$ represent the Choi operators of $\mathcal{S}_m$ and $\mathcal{E}_m$, respectively.
 Note that we used the fact that the Choi operators of $\mathcal{E}_m$ are rank-1 operators in the classes we investigate.
Note also that $S_{\texttt{fail}}$ can be set to be $(\idop-\sum_{m\in\Sigma}\ptr{2}{S_m})\otimes\rho^{(A_2)}\otimes\rho^{(B_2)}$ for $\{\mathcal{S}_m\}_{m\in\Sigma\cup\{\texttt{fail}\}}$ to form a separable instrument.

We reformulate Eq.~\eqref{eq:SEPoptformatpre} using a range constraint to match the format of Eq.~\eqref{eq:SEPopt}.
Observe that $\exists p\in\rr,\ptr{R_AR_B}{S_m\overline{\tau}}=p\ketbra{E_m}$ is equivalent to $\range{S_m}\subseteq\ww_m$, where 
\begin{equation}
\ww_m:=\{\ket{\Xi}\in\hat{\mathcal{A}}\otimes\hat{\mathcal{B}}:\forall\ket{\eta}\in\range{\overline{\tau}},\braket{\eta}{\Xi}\in\vspan{\{\ket{E_m}\}}\}.
\end{equation}
(A full proof for this observation is provided in Lemma~\ref{lemma:range_constraint} in Supplementary Information~\ref{appendix:range_constraint}.)
Accordingly, the optimization problem can be reformulated as
\begin{equation}
\label{eq:SEPoptformat}
{\rm Eq.}~\eqref{eq:SEPoptformatpre}=\max\left\{\min_{m\in\Sigma} \frac{\tr{S_m\overline{\tau}}}{\lpnorm{2}{\ket{E_m}}^2}:
\begin{array}{l}
\forall m\in\Sigma, S_m\in \SEP{\hat{\mathcal{A}}:\hat{\mathcal{B}}},\range{S_m}\subseteq\ww_m,     \\
\idop-\sum_{m\in\Sigma}\ptr{2}{S_m}\in \SEP{\hh_{A_1}\otimes\hh_{R_A}:\hh_{R_B}\otimes\hh_{B_1}}
\end{array}
\right\}.
\end{equation}
This is because for any feasible solution $S_m$ of the optimization problem given in the right-hand side of Eq.~\eqref{eq:SEPoptformat}, $S'_m=p\frac{\lpnorm{2}{\ket{E_m}}^2}{\tr{S_m\overline{\tau}}}S_m(\leq S_m)$ with $p=\min_{m\in\Sigma} \frac{\tr{S_m\overline{\tau}}}{\lpnorm{2}{\ket{E_m}}^2}$ is a feasible solution of the optimization problem given in the right-hand side of Eq.~\eqref{eq:SEPoptformatpre}.
Note that a variable $x$ is called a feasible solution of an optimization problem $\max_{x\in X}f(x)$ if $x\in X$.

Since $S_m$ ranges over $\pos{\mathbb{A}_m}$, where $\mathbb{A}_m:=\Segre{\hat{\mathcal{A}}:\hat{\mathcal{B}}}\cap\ww_m$, our hierarchy can be applied to this cone optimization problem. 
In particular, the first level yields
\begin{equation}
\label{eq:SEPoptformatMFLC}
{\rm Eq.}~\eqref{eq:SEPoptformat}=\max\left\{\min_{m\in\Sigma} \frac{\tr{S_m\overline{\tau}}}{\lpnorm{2}{\ket{E_m}}^2}:
\begin{array}{l}
\forall m\in\Sigma,S_m\in\SEP{\hat{\mathcal{A}}:\hat{\mathcal{B}}}\cap \pos{\cup_{i\in I}\pp_i^{(m)}},\\
\idop-\sum_{m\in\Sigma}\ptr{2}{S_m}\in \SEP{\hh_{A_1}\otimes\hh_{R_A}:\hh_{R_B}\otimes\hh_{B_1}}
\end{array}
\right\},
\end{equation}
where $\cup_{i\in I}\pp_i^{(m)}$ is the MFLC of $\mathbb{A}_m$.

As shown in Appendix, $\ww_m$ is an extended and twisted canonical subspace for the classes in Table~\ref{table:Nop} except in the case of entanglement distillation.
For such classes, the MFLC of $\mathbb{A}_m$ consists of a subspace $\pp$ given by Eq.~\eqref{eq:MFLCofETCS} and subspaces contained in
\begin{equation}
\ww^\circ:=\{\ket{\Xi}\in\hat{\mathcal{A}}\otimes\hat{\mathcal{B}}:\forall\ket{\eta}\in\range{\overline{\tau}},\braket{\eta}{\Xi}=0\}.
\end{equation}
We now show that we can further assume $S_m\in\pos{\pp}$.
Let $S_m=\sum_x\ketbra{\Xi_x}$ maximize the right-hand side of Eq.~\eqref{eq:SEPoptformatMFLC}, where $\ket{\Xi_x}\in\Segre{\hat{\mathcal{A}}:\hat{\mathcal{B}}}$. For any $\ket{\Xi_{x}}\in\ww^\circ$, $S_m-\ketbra{\Xi_x}$ remains a feasible solution and attains the same optimal value.

As a first application of Eq.~\eqref{eq:SEPoptformatMFLC}, we generalize previous analytical results in a unified way in the next section.

As a second application of Eq.~\eqref{eq:SEPoptformatMFLC}, we compute an upper bound on Eq.~\eqref{eq:SEPoptformatpre} on the basis of the PPT relaxation of Eq.~\eqref{eq:SEPoptformatMFLC}. It is important to note that although Eq.~\eqref{eq:SEPoptformatMFLC} is equivalent to Eq.~\eqref{eq:SEPoptformatpre}, it includes linear constraints derived from the MFLC even after its PPT relaxation is applied. This differs from the previous way of applying the DPS hierarchy, which computes upper bounds on Eq.~\eqref{eq:SEPoptformatpre} by relaxing Eq.~\eqref{eq:SEPoptformat}.

\begin{figure}[ht]
    \centering
    \includegraphics[width=13cm]{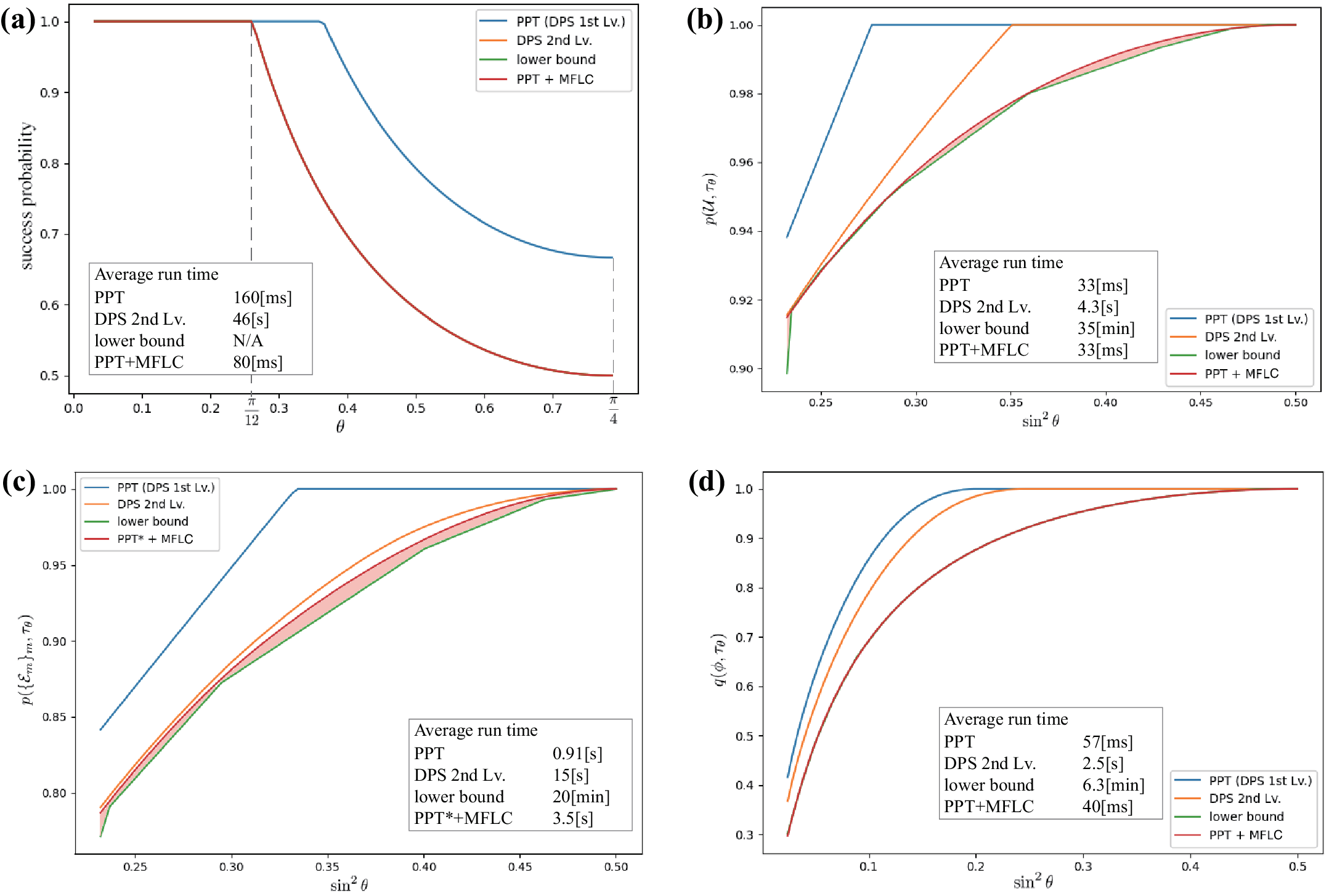}
    \caption{Solutions and run time for the relaxed problems of range-constrained SEP optimization problems across (a) $100$ different target states $\ket{\psi_\theta}=\cos\theta\ket{00}+\sin\theta\ket{11}$ or (b-d) $100$ different resource states $\ket{\tau_\theta}=\cos\theta\ket{00}+\sin\theta\ket{11}$. The relaxed problems derived from the DPS hierarchy are denoted by `PPT' or `DPS 2nd Lv.'. The relaxed problems obtained from our strengthened DPS hierarchy are denoted by `PPT+MFLC' or `PPT*+MFLC'.
    (a) Success probability of distilling the entangled state $\ket{\psi_\theta}$ from a mixed state $\frac{1}{3}\sum_{i=1}^3\tau_i$ using SEP channels, where $\ket{\tau_1}=\frac{1}{\sqrt{2}}(\ket{01}-\ket{10})$, $\ket{\tau_2}=\frac{1}{\sqrt{2}}(\ket{02}-\ket{20})$, and $\ket{\tau_3}=\frac{1}{\sqrt{2}}(\ket{12}-\ket{21})$. The solutions for `PPT+MFLC' and `DPS 2nd Lv.' coincide with the analytical lower bound $\min\left\{1,\frac{1}{2\sin2\theta}\right\}$.
    (b) Success probability of implementing the controlled $T$ gate and (c) success probability of implementing the symmetric joint POVM~\cite{CZ21,MA24,PPDG25} using SEP channels assisted by the entangled state $\ket{\tau_\theta}$. The true trade-off curve lies within the red-shaded region. The relaxed problem denoted by `PPT*' partially incorporates constraints from the second level of the DPS hierarchy.
    (d) Maximum $q$ for deterministically implementing a POVM $\{q\phi,(\idop-q\phi)\}$, where $\ket{\phi}=\frac{\sqrt{3}+1}{2\sqrt{2}}\ket{00}-\frac{\sqrt{3}-1}{2\sqrt{2}}\ket{11}$. The solution for `PPT+MFLC' coincides with the lower bound.}
    \label{fig:Entcost}
\end{figure}

Numerical experiments demonstrate that adding MFLC constraints to the PPT relaxation yields nearly optimal upper bounds on the trade-off between entanglement cost and the figure of merit for diverse tasks, as shown in Fig.~\ref{fig:Entcost}. Furthermore, since MFLC constraints reduce the SDP size, the run time for solving the PPT+MFLC relaxation is typically shorter than for the PPT relaxation alone. The specific setup for the numerical experiments is outlined in Appendix.

\subsubsection{Necessity of maximally entangled state}
\label{sec:maxent}
In Fig.~\ref{fig:Entcost}, we can observe that a maximally entangled state is necessary for deterministically implementing the optimal state verification, a PVM, and a unitary channel. The following theorem is known for the unitary case.

\noindent\textbf{Theorem}~\cite[Theorem 1]{SG11}
{\it Suppose that a unitary operator $U:\hh_{1}\rightarrow\hh_{2}$ is implemented deterministically by SEP channels that make use of the pure entangled state $\ket{\tau}\in\hh_{R_A}\otimes\hh_{R_B}$, where $\hh_b=\hh_{A_b}\otimes\hh_{B_b}$ for $b\in\{1,2\}$ (see {\rm Fig}.~\ref{fig:Hspaces}). Then
    \begin{enumerate}
        \item $\Schrank{{R_A}:{R_B}}{\ket{\tau}}\geq\Schrank{{\hat{A}}:{\hat{B}}}{\ket{U}}$, where $\ket{U}=\sum_{i}\ket{i}_1\otimes(U\ket{i}_1)\in\hh_1\otimes\hh_2\simeq\hh_{\hat{A}}\otimes\hh_{\hat{B}}$.

        \item If $\Schrank{{R_A}:{R_B}}{\ket{\tau}}=\Schrank{{\hat{A}}:{\hat{B}}}{\ket{U}}$, then $\ket{\tau}$ is maximally entangled.
    \end{enumerate}
    }
    
This theorem guarantees the optimality of the entanglement cost of several non-local unitary channels~\cite{STM11}.
From our numerical results, we expect that the constraints of the MFLC are sufficient to derive this theorem. In this subsection, we demonstrate that this is true. Moreover, we can prove a generalized theorem by exploiting the similarity of MFLCs for non-local unitary channels, non-local PVMs, and state verifications.

Before presenting the generalized theorem, we introduce two concepts.
First, a quantum instrument $\{\mathcal{E}_{m}:\linop{\hh_1}\rightarrow\linop{\hh_2}\}_{m\in\Sigma}$ is deterministically implementable by using separable instruments assisted by a pure state $\ket{\tau}\in\hh_{R_A}\otimes\hh_{R_B}$ if the success probability given in Eq.~\eqref{eq:SEPoptformatpre} satisfies $p(\{\mathcal{E}_m\}_m,\tau)=1$; more specifically, there exists a set $\left\{S_m\in\SEP{\hat{\mathcal{A}}:\hat{\mathcal{B}}}\right\}_{m\in\Sigma}$ of separable operators such that
    \begin{eqnarray}
        \label{eq:detcond}
         \forall m\in\Sigma,\ptr{R_AR_B}{S_m\overline{\tau}}=E_m,\\
        \label{eq:detcond2}
         \idop-\sum_{m\in\Sigma}\ptr{2}{S_m}\in\SEP{\hh_{A_1}\otimes\hh_{R_A}:\hh_{R_B}\otimes\hh_{B_1}},   
    \end{eqnarray}
    where $E_m$ is the Choi operator of $\mathcal{E}_m$ and the labels of Hilbert spaces are summarized in Fig.~\ref{fig:Hspaces}.
 
 The second concept is the Schmidt rank of a positive semi-definite operator, defined as follows. Note that this is sometimes called the Schmidt number~\cite{K13}.

\begin{definition}
    The Schmidt rank $\Schrank{{\hat{A}}:{\hat{B}}}{E}$ of $E\in\pos{\hh_{\hat{A}}\otimes\hh_{\hat{B}}}$ is the minimum integer $r$ such that $E$ is contained in the cone of pure states $\phi$ such that $\Schrank{{\hat{A}}:{\hat{B}}}{\ket{\phi}}\leq r$. 
\end{definition}

\begin{theorem}
\label{thm:EntforQI}
Suppose that a quantum instrument $\{\mathcal{E}_m:\linop{\hh_1}\rightarrow\linop{\hh_2}\}_{m\in\Sigma}$ is deterministically implementable by separable instruments assisted by a pure state $\ket{\tau}\in\hh_{R_A}\otimes\hh_{R_B}$, where $\hh_b=\hh_{A_b}\otimes\hh_{B_b}$ for $b\in\{1,2\}$ (see {\rm Fig}.~\ref{fig:Hspaces}). Then
\begin{enumerate}
    \item $\Schrank{{R_A}:{R_B}}{\ket{\tau}}\geq\Schrank{{\hat{A}}:{\hat{B}}}{E_m}$ for all $m$, where $E_m\in\pos{\hh_1\otimes\hh_2}\simeq\pos{\hh_{\hat{A}}\otimes\hh_{\hat{B}}}$ is the Choi operator of $\mathcal{E}_m$.

    \item If an $m$ exists such that $\Schrank{{R_A}:{R_B}}{\ket{\tau}}=\Schrank{{\hat{A}}:{\hat{B}}}{E_m}$ and $E_m=\ketbra{V^\dag}$, where $\ket{V^\dag}=\sum_i\ket{i}_1\otimes V^\dag\ket{i}_1$ and $V:\hh_2\rightarrow\hh_1$ is an isometry operator, then $\ket{\tau}$ is maximally entangled.
\end{enumerate} 
\end{theorem}

\begin{proof}
    The first statement can be proven as follows:
\begin{eqnarray}
 \Schrank{{\hat{A}}:{\hat{B}}}{E_m}=\Schrank{{\hat{A}}:{\hat{B}}}{\ptr{R_AR_B}{S_m\overline{\tau}}}
 \leq\Schrank{{R_A}:{R_B}}{\ket{\overline{\tau}}}=\Schrank{{R_A}:{R_B}}{\ket{\tau}}.
\end{eqnarray}

To prove the second statement, 
let $m$ satisfy the conditions of the theorem. By using a similar argument to the one used to derive Eq.~\eqref{eq:SEPoptformatMFLC}, Eq.~\eqref{eq:detcond} implies that we can assume that $\range{S_m}\subseteq\pp$, where
\begin{equation}
    \pp=\left(V_A\otimes\idop^{(R_A)}\otimes\left(L_1^\dag\right)^{-1}\otimes(V_BL_2)\right)\CMFLC{d},
\end{equation}
$\ket{V^\dag}=V_A\otimes(V_BL_2)\ketUMES{AB}{d}$, $\ket{\overline{\tau}}=\idop^{(R_A)}\otimes L_1^{(R_B)}\ketUMES{R_AR_B}{d}$, and $d=\Schrank{{R_A}:{R_B}}{\ket{\tau}}=\Schrank{{\hat{A}}:{\hat{B}}}{E_m}=\Schrank{{\hat{A}}:{\hat{B}}}{\ket{V^\dag}}$.
We can show that $\ket{\tau}$ is maximally entangled if and only if $\ket{\overline{\tau}}\ket{V^\dag}\in \pp$. 
First, note that $\ket{\tau}$ is maximally entangled if and only if $\sqrt{d}L_1$ is a unitary operator.
Since $\ketUMES{AB}{d}\ketUMES{R_AR_B}{d}\in\CMFLC{d}$, $\ket{\overline{\tau}}\ket{V^\dag}\in\pp$ if $\ket{\tau}$ is maximally entangled. For the converse, we can show
\begin{eqnarray}
    &&\ket{\overline{\tau}}\ket{V^\dag}\in \pp\\
    &\Rightarrow&(\idop^{(R_A)}\otimes L_1^\dag)\ket{\overline{\tau}}\braUMES{AB}{d}(V_A^\dag\otimes L_2^{-1}V_B^\dag)\ket{V^\dag}\in\vspan{\{\ketUMES{R_AR_B}{d}\}}\\
    &\Leftrightarrow&(\idop^{(R_A)}\otimes L_1^\dag L_1)\ketUMES{R_AR_B}{d}\in\vspan{\{\ketUMES{R_AR_B}{d}\}}.
\end{eqnarray}
This implies that $L_1$ is proportional to a unitary operator.
In the following, we show that $\ket{\overline{\tau}}\ket{V^\dag}\in \range{S_m}$.

Suppose that $\ptr{2}{S_m}\ket{\overline{\tau}}=\ket{\overline{\tau}}\otimes\ptr{2}{E_m}$. Let $S_m=\sum_k\ketbra{\Pi_k}$. Eq.~\eqref{eq:detcond} implies that $\forall k,\exists \alpha_k\in\cc,\bra{\overline{\tau}}_{R_AR_B}\ket{\Pi_k}=\alpha_k\ket{V^\dag}$.
First, we obtain
\begin{eqnarray}
    \left(\ptr{2}{S_m}\otimes\idop^{(2)}\right)\ket{\overline{\tau}}\ket{V^\dag}&=&\sum_k\overline{\alpha}_k\left(\ptr{2}{\ket{\Pi_k}\bra{V^\dag}}\otimes\idop^{(2)}\right)\ket{V^\dag}\\
    &=&\sum_k\overline{\alpha}_k V^\dag V\ket{\Pi_k}=\sum_k\overline{\alpha}_k \ket{\Pi_k}
\end{eqnarray}
On the other hand,
\begin{eqnarray}
     \left(\ptr{2}{S_m}\otimes\idop^{(2)}\right)\ket{\overline{\tau}}\ket{V^\dag}&=&\ket{\overline{\tau}}\otimes\left(\ptr{2}{\ketbra{V^\dag}}\otimes\idop^{(2)}\right)\ket{V^\dag}\\
    &=&\ket{\overline{\tau}}\otimes\left(\idop^{(1)}\otimes V^\dag V\right)\ket{V^\dag}=\ket{\overline{\tau}} \ket{V^\dag}.
\end{eqnarray}
This implies $\ket{\overline{\tau}}\ket{V^\dag}\in \range{S_m}$.
In the following, we show that $\ptr{2}{S_m}\ket{\overline{\tau}}=\ket{\overline{\tau}}\otimes\ptr{2}{E_m}$.

Let $\ptr{2}{S_n}=\ketbra{\overline{\tau}}\otimes A_n+\ket{\overline{\tau}}\otimes B_n^\dag+ \bra{\overline{\tau}}\otimes B_n+C_n$, where $(\bra{\overline{\tau}}\otimes\idop^{(1)})B_n=C_n(\ket{\overline{\tau}}\otimes\idop^{(1)})=(\bra{\overline{\tau}}\otimes\idop^{(1)})C_n=0$. Eq.~\eqref{eq:detcond} implies that
 $A_m=\ptr{2}{E_m}=\ptr{2}{\ketbra{V^\dag}}=(V V^\dag)^T$ and $\sum_{n\neq m}A_n=\sum_{n\neq m}\ptr{2}{E_n}=\idop-(V V^\dag)^T$ are orthogonal projectors.
 Since $\ptr{2}{S_n}\geq0$ for all $n$, $\range{B_n^\dag}\subseteq \range{A_n}$ for all $n$.
 This is because $\forall S\geq0,\bra{\phi}S\ket{\phi}=0\Rightarrow\bra{\phi}S=0$ and $\bra{\overline{\tau}\phi}\ptr{2}{S_n}\ket{\overline{\tau}\phi}=\bra{\phi}A_n\ket{\phi}=0$ for any $\ket{\phi}$ that is orthogonal to $ \range{A_n}$.
Since $A_m$ is a projector whose range is orthogonal to $\range{A_n}$ for all $n\neq m$, $A_m\sum_nB_n^\dag=B_m^\dag$.
On the other hand, Eq.~\eqref{eq:detcond2} implies
\begin{eqnarray}
    0&\leq&\idop-\sum_n\ptr{2}{S_n}\\
    &=&(\idop-\ketbra{\overline{\tau}})\otimes \idop^{(1)}-\ket{\overline{\tau}}\otimes\left(\sum_nB_n^\dag\right)-\bra{\overline{\tau}}\otimes\left(\sum_nB_n\right)-\left(\sum_nC_n\right).\nonumber\\
\end{eqnarray}
Thus, $\sum_nB_n^\dag=0$ since $\bra{\overline{\tau}\phi}\left(\idop-\sum_n\ptr{2}{S_n}\right)\ket{\overline{\tau}\phi}=0$ for any $\ket{\phi}\in\hh_1$.
Therefore, $B_m^\dag=A_m\sum_nB_n^\dag=0$. This completes the proof.
\end{proof}

We can also prove the following corollary.
\begin{corollary}
\label{cor:veri}
    For a local implementation of the optimal quantum state verification of $\ket{\phi}\in\hh_{\hat{A}}\otimes\hh_{\hat{B}}$, i.e., a local implementation of an instrument $\{\mathcal{E}_{\texttt{accept}},\mathcal{E}_{\texttt{reject}}\}$ defined by $\mathcal{E}_{\texttt{accept}}(\rho)=\tr{\phi\rho}$ and $\mathcal{E}_{\texttt{reject}}(\rho)=\tr{(\idop-\phi)\rho}$,
   the resource entangled state $\ket{\tau}\in\hh_{R_A}\otimes\hh_{R_B}$ must be maximally entangled if $\Schrank{{R_A}:{R_B}}{\ket{\tau}}=\Schrank{{\hat{A}}:\hat{B}}{\ket{\phi}}$.
\end{corollary}
\begin{proof}
 By letting an isometry $V$ be $V=\ket{\phi}$, we find that $\ketbra{V^\dag}=\overline{\phi}$ is the Choi operator of $\mathcal{E}_{\texttt{accept}}$ and $\Schrank{\hat{A}:\hat{B}}{\overline{\phi}}=\Schrank{{\hat{A}}:{\hat{B}}}{\ket{\phi}}=\Schrank{{R_A}:{R_B}}{\ket{\tau}}$.
 Applying Theorem~\ref{thm:EntforQI} completes the proof.
\end{proof}

Note that Yu et al.~\cite{YDY14} have shown that a two-qubit maximally entangled state is sufficient for implementing the optimal quantum verification of any state $\ket{\phi}\in\cd\otimes\cd$ for any dimension $d$ by using PPT measurements. Thus, this corollary has revealed a significant disparity in power between the separable and PPT measurements.
Moreover, they posed an open problem asking whether a $d$-dimensional maximally entangled state is always required for (deterministically) distinguishing a $d$-dimensional maximally entangled state $\phi_d^+$ and its orthogonal complement $(\idop-\phi_d^+)/(d^2-1)$ by using separable POVMs~\cite{YDY14}.
Corollary~\ref{cor:veri} solves this open problem affirmatively as follows.
Assume that $\ket{\phi}\in\hh_{\hat{A}}\otimes\hh_{\hat{B}}$ satisfies $\Schrank{{\hat{A}}:{\hat{B}}}{\ket{\phi}}=d$ and consider discrimination of $\phi$ and $(\idop-\phi)/(\dim\hh_{\hat{A}}\dim\hh_{\hat{B}}-1)$ by using separable POVMs assisted by an entangled state $\ket{\tau}\in\cd\otimes\cd$. 
If the states are deterministically distinguishable, we can implement the instrument for verifying a target state $\ket{\phi}$ as defined in the corollary.
By applying the corollary, we conclude that $\ket{\tau}$ must be maximally entangled.
(Note that the Schmidt rank of $\ket{\tau}$ must be $d$ from Theorem~\ref{thm:EntforQI}.)

\section{Discussion and open problems}
We have introduced an algebraic-geometric framework for general cone optimization and applied it to range-constrained SEP optimization problems arising in fundamental tasks in entanglement manipulation, including distillation, local state discrimination, and local implementations of unitary channels and measurements. By incorporating constraints derived from the minimum algebraic covering of the feasible region, we constructed an improved DPS hierarchy that yields nontrivial restrictions already at the first level in the form of the MFLC.
The framework is applicable to other optimization problems of a broader class of entanglement manipulation~\cite{BBKW09,C08,DFXY09,YDY12,YDY14,BCJRWY15,BHN16,BHN18,ZZLW24,PPDG25,KDS21}, as well as scenarios with limited quantum memory~\cite{OZNPQ24}. 

Two directions are particularly important for further development. First, an effective and scalable method for computing (minimum) algebraic coverings is needed; tools from computational algebraic geometry, such as Gr\"obner basis, provide a natural starting point. Second, extending the framework beyond the zero-error regime would require combining it with sum-of-squares hierarchies and semialgebraic relaxations.
Progress along these lines would move toward a general and practical theory of algebraically constrained quantum optimization.

\section*{Acknowledgments}
S.A. is greatly indebted to NTT Institute for Fundamental Mathematics team members for their comprehensive lecture about mathematics.
S.A. was partially supported by JST PRESTO Grant no.JPMJPR2111, JST Moonshot R\&D MILLENNIA Program (Grant no.JPMJMS2061), JPMXS0120319794, and CREST (Japan Science and Technology Agency) Grant no.JPMJCR2113.
J.M. was partially supported by JSPS KAKENHI Grant no.JP23K21643.
H.O. was partially supported by KAKENHI Grant no.JP20K03644 and JP25K07036.

\section*{Competing interests}
The authors declare no competing interests.

\section*{Data availability}
Numerical results together with instructions on how to reproduce them, are available online at \url{https://github.com/akibue/DPS-based-on-MFLE}.

\section*{Author contributions}
All authors contributed to the verification and analysis of the results. S.A. and J.M. contributed to the development of the algebraic covering framework and the cone optimization framework. S.A. additionally contributed to the demonstration of its applications.

\bibliography{references}

\section{Appendix}

\subsection{Quantum de Finetti theorem for prime extendable state}
\label{appendix:deFinetti}
In this section, we derive a quantum de Finetti theorem for prime $(k,\mathbb{A})$-tendable states $\rho\in\dop{\cd}$ by exploiting the quantum de Finetti theorem for $(k,\mathbb{I}_i)$-tendable states.

\noindent\textbf{Proposition} \ref{prop:prime-deFinetti}
\textit{
Let $\mathbb{A}$ be an algebraic set in $\pspace^{d-1}$ and $\{\mathbb{I}_i\}_{i\in I}$ be its irredundant irreducible components.
Suppose that, for fixed $k$ and for each $i\in I$, there exists a constant $f(\mathbb{I}_i)$ such that
\begin{equation}
    \min_{\sigma\in\dop{\mathbb{I}_i}}
    \lpnorm{1}{\rho-\sigma}
    \leq f(\mathbb{I}_i)
\end{equation}
holds for any $(k,\mathbb{I}_i)$-tendable state $\rho$,
where we define
$\dop{\mathbb{B}}:=\pos{\mathbb{B}}\cap\dop{\cd}$ for an algebraic set $\mathbb{B}$.
Then, any prime $(k,\mathbb{A})$-tendable state $\rho$ satisfies
\begin{equation}
    \min_{\sigma\in\dop{\mathbb{A}}}
    \lpnorm{1}{\rho-\sigma}
    \leq
    \max_{i\in I} f(\mathbb{I}_i).
\end{equation}
}
\begin{proof}
    Let $\{Q^k_i\in\pos{\otimes_{n=1}^k\cd}\}_{i\in I}$ be a prime $(k,\mathbb{A})$-tension of $\rho$. Define
    \[
        \rho_i:=\ptr{k-1}{Q^k_i}.
    \]
    Then, we have $\rho=\sum_{i}\rho_i$. 
    For each $i\in I$ with $\rho_i\neq 0$, the normalized state $\frac{1}{\tr{\rho_i}}\rho_i$ is $(k,\mathbb{I}_i)$-tendable. Hence, by assumption, there exists $\sigma_i\in\dop{\mathbb{I}_i}$
\[ 
    \lpnorm{1}{\frac{1}{\tr{\rho_i}}\rho_i-\sigma_i}\leq f(k,\mathbb{I}_i)\quad \Leftrightarrow\quad 
    \lpnorm{1}{\rho_i-\tr{\rho_i}\sigma_i}\leq \tr{\rho_i}f(k,\mathbb{I}_i).
\]
Now set 
\[  
    \sigma:=\sum_{i\in I}\tr{\rho_i}\sigma_i.
\]
Since $\sum_i\tr{\rho_i}=\tr{\rho}=1$ and $\dop{\mathbb{I}_i}\subseteq\dop{\mathbb{A}}$ for every $i\in I$, we have $\sigma\in\dop{\mathbb{A}}$.
Therefore, by the triangle inequality,
\[ 
    \lpnorm{1}{\rho-\sigma}\leq \sum_{i\in I}\lpnorm{1}{\rho_i-\tr{\rho_i}\sigma_i}\leq\sum_{i\in I}\tr{\rho_i}f(k,\mathbb{I}_i)\leq \max_if(k,\mathbb{I}_i).
\]
This completes the proof.
\end{proof}

\subsubsection{Example: biseparable states}
\label{subsec:example-biseparable-definetti}

We illustrate Proposition~\ref{prop:prime-deFinetti} with the example of
biseparable states. Let
\[
    \mathcal{H}:=(\mathbb{C}^{d})^{\otimes m}.
\]
The algebraic set of pure biseparable states is given by
\[
    \mathbb{A}_{\mathrm{bisep}}
    =
    \bigcup_{\emptyset\neq S\subsetneq [m]}
    \mathbb{X}_{\{S,S^c\}},
\]
where $\mathbb{X}_{\{S,S^c\}}$ denotes the Segre variety associated
with the bipartition $S|S^c$. Each $\mathbb{X}_{\{S,S^c\}}$ is irreducible.

By the quantum de Finetti theorem for bipartite separability due to Chiribella~\cite[Theorem~2]{C10}, every
$(k,\mathbb{X}_{\{S,S^c\}})$-tendable state $\rho$ satisfies
\[
    \min_{\sigma\in\dop{\mathbb{X}_{\{S,S^c\}}}}
    \|\rho-\sigma\|_1
    \leq
    \frac{
        2d^{\min\{|S|,m-|S|\}}
    }{k+d^{\min\{|S|,m-|S|\}}}.
\]
Applying Proposition~\ref{prop:prime-deFinetti} to the irreducible components
$\mathbb{X}_{\{S,S^c\}}$ of $\mathbb{A}_{\mathrm{bisep}}$, we obtain that every
prime $(k,\mathbb{A}_{\mathrm{bisep}})$-tendable state $\rho$ satisfies
\[
    \min_{\sigma\in\dop{\mathbb{A}_{\mathrm{bisep}}}}
    \|\rho-\sigma\|_1
    \leq
    \max_{\emptyset\neq S\subsetneq [m]}\frac{
        2d^{\min\{|S|,m-|S|\}}
    }{k+d^{\min\{|S|,m-|S|\}}}
    \leq
    \frac{
        2d^{\lfloor m/2\rfloor}
    }{k+d^{\lfloor m/2\rfloor}}.
\]

The same argument applies straightforwardly to other multipartite
entanglement structures described as finite unions of Segre varieties,
such as states with bounded entanglement depth and $t$-producible states.

\subsection{Entanglement cost of non-local unitary channels}
\label{sec:unitary}

Unitary channels characterize gate operations in quantum computing and the time evolution in quantum simulations. Thus, local implementations of these channels are crucial for designing distributed quantum computations~\cite{MMNI08,WMFSAPDHM23,MMDJAA24}. While the quantum teleportation protocol enables the implementation of non-local channels by consuming entanglement, more efficient protocols exist that require less entanglement~\cite{EJPP00,SKTM11,SM16,CY16,CBWRAGFDJS18}. Consequently, one of the fundamental questions in this area is determining the minimum amount of entanglement required for a local implementation~\cite{SG11,SKTM11,STM11,SM16}.
Note that the experimental demonstration of distributed realization for non-local unitary channels has recently been accomplished~\cite{DLWDTHMR21,LHZZXMOLLZLG24}.

In this subsection, we investigate the success probability of implementing a non-local unitary channel $\mathcal{U}(\rho)=U\rho U^\dag$ by using bipartite SEP channels with a resource state $\ket{\tau}$.
By using the general optimization problem given in Eq.~\eqref{eq:SEPoptformat}, the success probability can be formulated as
\begin{equation}
\label{eq:SEPoptunitary}
p(\mathcal{U},\tau)=\max\left\{ \frac{\tr{S\overline{\tau}}}{d_Ad_B}:
\begin{array}{l}
S\in \SEP{\hat{\mathcal{A}}:\hat{\mathcal{B}}},\range{S}\subseteq\hat{\ww},     \\
\idop-\ptr{2}{S}\in \SEP{\hh_{A_1}\otimes\hh_{R_A}:\hh_{R_B}\otimes\hh_{B_1}}
\end{array}
\right\},
\end{equation}
where $\dim\hh_{A_1}=\dim\hh_{A_2}=d_A$, $\dim\hh_{B_1}=\dim\hh_{B_2}=d_B$, $
\hat{\ww}:=\{\ket{\Xi}\in\hat{\mathcal{A}}\otimes\hat{\mathcal{B}}:\bra{\overline{\tau}}_{R_AR_B}\ket{\Xi}\in\vspan{\{\ket{U}\}}\}
$, and $\ket{U}=\sum_i\ket{i}_{1}\otimes(U\ket{i})_{2}\in\hh_1\otimes\hh_2\simeq\hh_{\hat{A}}\otimes\hh_{\hat{B}}$.

We assume that $ \Schrank{\hat{A}:{\hat{B}}}{\ket{U}}= \Schrank{{R_A}:{R_B}}{\ket{\tau}}=d$.
Accordingly, we can let $\ket{\overline{\tau}}=\idop^{(R_A)}\otimes L_1^{(R_B)}\ketUMES{R_AR_B}{d}$ and
$\ket{U}=V_A\otimes(V_BL_2)\ketUMES{AB}{d}$,
where $L_1$ and $L_2$ are invertible operators, and $V_A$ (or $V_B$) is an isometry from $\hh_A$ (or $\hh_B$) into $\hh_{\hat{A}}$ (or $\hh_{\hat{B}}$). 
Using this representation, we can confirm that $\hat{\ww}$ is an extended and twisted canonical subspace.
By letting $\hat{\ww}^\circ=\{\ket{\Xi}\in\hat{\mathcal{A}}\otimes\hat{\mathcal{B}}:\bra{\overline{\tau}}_{R_AR_B}\ket{\Xi}=0\}$ and using Eq.~\eqref{eq:MFLCofETCS},
we can show that the MFLC of $\Segre{\hat{\mathcal{A}}:\hat{\mathcal{B}}}\cap\left(\hat{\ww}\setminus\hat{\ww}^\circ\right)$ is
$
 \pp=\left(V_A\otimes\idop^{(R_A)}\otimes\left(L_1^\dag\right)^{-1}\otimes\left(V_BL_2\right)\right)\CMFLC{d}
$.
By using the general optimization problem given in Eq.~\eqref{eq:SEPoptformatMFLC}, we obtain
\begin{equation}
\label{eq:SEPoptunitaryMFLC}
p(\mathcal{U},\tau)=\max\left\{ \frac{\tr{S\overline{\tau}}}{d_Ad_B}:
\begin{array}{l}
S\in \SEP{\hat{\mathcal{A}}:\hat{\mathcal{B}}}\cap\pos{\pp},     \\
\idop-\ptr{2}{S}\in \SEP{\hh_{A_1}\otimes\hh_{R_A}:\hh_{R_B}\otimes\hh_{B_1}}
\end{array}
\right\}.
\end{equation}

\noindent\textbf{Numerical experiment}

Suppose that $\ket{\tau_\theta}=\cos\theta\ket{00}+\sin\theta\ket{11}$ ($\theta\in(0,\pi/4]$) and 
the target unitary is $U=\ketbra{0}_A\otimes\idop_B+\ketbra{1}_A\otimes u_B$, where $u_B=\ketbra{0}+e^{i\phi}\ketbra{1}$ and $e^{i\phi}\neq 1$.
It is known that any two-qubit non-local controlled unitary channel is locally unitarily equivalent to $U$.
In this case, we obtain
\begin{eqnarray}
        L_1&=&\cos\theta\ketbra{0}+\sin\theta\ketbra{1},\ V_A=V_B=\ket{00}\bra{0}+\ket{11}\bra{1},\\
        L_2&=&\ketbra{0}+\ket{1}\bra{0}+\ket{0}\bra{1}+e^{i\phi}\ketbra{1}.
\end{eqnarray}

It is known that such a controlled unitary channel can be exactly implemented by LOCC with a Bell pair~\cite{EJPP00}. That is, $p(U,\tau_{\frac{\pi}{4}})=1$.
However, for general $\theta$, $p(U,\tau_{\theta})$ is unknown.

We numerically solved the DPS hierarchy of Eq.~\eqref{eq:SEPoptunitary} and Eq.~\eqref{eq:SEPoptunitaryMFLC} and obtained upper bounds on $p(\mathcal{U},\tau_\theta)$, as shown in Fig.~\ref{fig:Entcost} (b) (see the details in Supplementary Information~\ref{appendix:unitary}).
Note that we also computed its lower bound based on an algorithm shown in Supplementary Information~\ref{appendix:enet} with randomly sampled $\epsilon$-nets $\left\{\ket{\Pi_x}\in\Segre{\hat{\mathcal{A}}:\hat{\mathcal{B}}}\cap(\hat{\ww}\setminus\hat{\ww}^\circ)\right\}_{x=1}^{3500}$ and $\left\{\phi_x\in\puredop{\hh_1\otimes\hh_{R_A}}\right\}_{x=1}^{330}$. 
Here, we can see that the additional constraint resulting from the MFLC improves the upper bound.
In particular, it numerically demonstrates that a maximally entangled state is required to implement a non-local unitary operator deterministically~\cite{SG11} although the upper bound derived by the second level of the DPS hierarchy cannot.

\subsection{Entanglement cost of non-local measurement}
\label{sec:POVM}
The non-local measurement is an important primitive in multipartite quantum information processing tasks, such as quantum network sensing~\cite{MA24} and data-hiding~\cite{DLT02}. Additionally, implementing non-local measurements is necessary when transitioning from a monolithic quantum computer to a distributed architecture.
The entanglement cost of implementing non-local measurements describes the quantum communication cost or the security of the data-hiding protocols, and it has been extensively studied~\cite{C08,BBKW09,BCJRWY15,BHN16,BHN18}.

In this subsection, we investigate the success probability of implementing a rank-1 POVM described by an instrument $\{\mathcal{E}_m\}_m$ defined by $\mathcal{E}_m(\rho)=\bra{M_m}\rho\ket{M_m}$ by using a bipartite SEP channel with a resource state $\ket{\tau}$.
Since the instrument does not have an output system, we let $\hh_{\hat{A}}=\hh_{A_1}$ and $\hh_{\hat{B}}=\hh_{B_1}$ in Fig.~\ref{fig:Hspaces}.
By using the general optimization problem in Eq.~\eqref{eq:SEPoptformat}, the success probability can be expressed as
\begin{equation}
\label{eq:SEPoptPOVM}
p(\{\mathcal{E}_m\}_m,\tau)=\max\left\{\min_{m} \frac{\tr{S_m\tau}}{\lpnorm{2}{\ket{M_m}}^2}:
\begin{array}{l}
\forall m, S_m\in \SEP{\hat{\mathcal{A}}:\hat{\mathcal{B}}},\range{S_m}\subseteq\hat{\ww}_m,     \\
\idop-\sum_{m}S_m\in \SEP{\hat{\mathcal{A}}:\hat{\mathcal{B}}}
\end{array}
\right\},
\end{equation}
where
$
 \hat{\ww}_m=\{\ket{\Xi}\in\hat{\mathcal{A}}\otimes\hat{\mathcal{B}}:\bra{\tau}_{R_AR_B}\ket{\Xi}\in\vspan{\{\ket{M_m}\}}\}
$.
Note that we use the complex conjugation of $S_m$ from Eq.~\eqref{eq:SEPoptformat} in Eq.~\eqref{eq:SEPoptPOVM} to eliminate the complex conjugate for $\ket{\tau}$ and $\ket{M_m}$, and $\overline{S}_m$ represents the Choi operator of each separable instrument.

We assume that $\Schrank{{\hat{A}}:{\hat{B}}}{\ket{M_m}}= \Schrank{{R_A}:{R_B}}{\ket{\tau}}=d$ for all $m$.
Accordingly, we can let $\ket{\tau}=\idop^{(R_A)}\otimes L^{(R_B)}\ketUMES{R_AR_B}{d}$ and
$\ket{M_m}=V_A\otimes (V_BL_m)\ketUMES{AB}{d}$, 
where $L$ and $L_m$ are invertible operators, and $V_A$ (or $V_B$) is an isometry from $\hh_A$ (or $\hh_B$) into $\hh_{\hat{A}}$ (or $\hh_{\hat{B}}$). 
Using this representation, we can confirm that $\hat{\ww}_m$ is an extended and twisted canonical subspace.
By letting
$\hat{\ww}^\circ=\{\ket{\Xi}\in\hat{\mathcal{A}}\otimes\hat{\mathcal{B}}:\bra{\tau}_{R_AR_B}\ket{\Xi}=0\}$ and using Eq.~\eqref{eq:MFLCofETCS}, we can show that the MFLC of $\Segre{\hat{\mathcal{A}}:\hat{\mathcal{B}}}\cap(\hat{\ww}_m\setminus\hat{\ww}^\circ)$ is
$
    \pp_{m}=\left(V_A\otimes\idop^{(R_A)}\otimes \left(L^{\dag}\right)^{-1}\otimes (V_BL_m)\right)\CMFLC{d}
$.
By using the general optimization problem in Eq.~\eqref{eq:SEPoptformatMFLC}, we obtain
\begin{equation}
\label{eq:SEPoptPOVMMFLC}
p(\{\mathcal{E}_m\}_m,\tau)=\max\left\{\min_{m} \frac{\tr{S_m\tau}}{\lpnorm{2}{\ket{M_m}}^2}:
\begin{array}{l}
\forall m, S_m\in \SEP{\hat{\mathcal{A}}:\hat{\mathcal{B}}}\cap\pos{\pp_m},     \\
\idop-\sum_{m}S_m\in \SEP{\hat{\mathcal{A}}:\hat{\mathcal{B}}}
\end{array}
\right\}.
\end{equation}

\noindent
\textbf{Numerical experiment}

Here, we consider an entanglement-assisted implementation of a projection-valued measurement $\{\ketbra{M_m}\in\puredop{\hh_A\otimes\hh_B}\}_{m=1}^4$, defined by
\begin{eqnarray}
\label{eq:SJM}
\ket{M_m}=\frac{\sqrt{3}+1}{2\sqrt{2}}\ket{\eta_m}\ket{\overline{\eta_m}}-\frac{\sqrt{3}-1}{2\sqrt{2}}(\sigma_Y\otimes\sigma_Y)\ket{\overline{\eta_m}}\ket{\eta_m},
\end{eqnarray}
where $\{\ket{\eta_m}\in\cdim{2}\}_{m=1}^4$ is a set of states proportional to the single-qubit symmetric and informationally complete (SIC) POVM.
Note that $\{\ketbra{M_m}\}_{m=1}^4$ is known as a symmetric joint POVM (SJM) or elegant joint measurement, which plays an important role in the study of quantum nonlocality~\cite{G19}, tomography~\cite{CZ21}, network sensing~\cite{MA24}, and the localization cost~\cite{PPDG25}.
We assume that an entangled state $\ket{\tau_\theta}=\cos\theta\ket{00}+\sin\theta\ket{11}\in\hh_{R_A}\otimes\hh_{R_B}$ is shared between Alice and Bob ($\theta\in(0,\frac{\pi}{4}]$).
In this case, we obtain
\begin{eqnarray}
\label{eq:SJMops1}
    L&=&\cos\theta\ketbra{0}+\sin\theta\ketbra{1},\ V_A=V_B=\idop\\
\label{eq:SJMops2}
    L_1&=&\frac{\sqrt{3}+1}{2\sqrt{2}}\ketbra{0}-\frac{\sqrt{3}-1}{2\sqrt{2}}\ketbra{1},\\
    L_2&=&\frac{\sqrt{3}-1}{2\sqrt{6}}\ketbra{0}+\frac{1}{\sqrt{3}}\ket{1}\bra{0}+\frac{1}{\sqrt{3}}\ket{0}\bra{1}+\frac{\sqrt{3}+1}{2\sqrt{6}}\ketbra{1},\\
    L_3&=&\frac{\sqrt{3}-1}{2\sqrt{6}}\ketbra{0}-\frac{\zeta}{\sqrt{3}}\ket{1}\bra{0}+\frac{\zeta^2}{\sqrt{3}}\ket{0}\bra{1}+\frac{\sqrt{3}+1}{2\sqrt{6}}\ketbra{1},\\
    L_4&=&\frac{\sqrt{3}-1}{2\sqrt{6}}\ketbra{0}+\frac{\zeta^2}{\sqrt{3}}\ket{1}\bra{0}-\frac{\zeta}{\sqrt{3}}\ket{0}\bra{1}+\frac{\sqrt{3}+1}{2\sqrt{6}}\ketbra{1},
\end{eqnarray}
where $\zeta$ is a non-real root of $\zeta^3=-1$.
 
Fig.~\ref{fig:Entcost} (c) compares upper bounds on $p(\{\mathcal{E}_m\}_m,\tau)$ by using the DPS hierarchy of Eq.~\eqref{eq:SEPoptPOVM} and Eq.~\eqref{eq:SEPoptPOVMMFLC}.
 Note that we combine the first and second levels of DPS hierarchy in the computation of Eq.~\eqref{eq:SEPoptPOVMMFLC} to improve the upper bound  (see the details in Supplementary Information~\ref{appendix:PVM}).
We also computed its lower bound based on an algorithm shown in Supplementary Information~\ref{appendix:enet} with randomly sampled $\epsilon$-nets $\left\{\ket{\Pi_x}\in\Segre{\hat{\mathcal{A}}:\hat{\mathcal{B}}}\cap(\hat{\ww}\setminus\hat{\ww}^\circ)\right\}_{x=1}^{1000}$ and $\left\{\phi_x\in\puredop{\hh_1\otimes\hh_{R_A}}\right\}_{x=1}^{300}$. 
We can see that the additional constraint resulting from the MFLC improves the approximation. 

Some authors have shown that the minimum average concurrence of the SJM is $\frac{1}{2}$~\cite{MA24}.
While this implies that $\theta\geq\frac{\pi}{12}$ ($\Leftrightarrow\sin^2\theta\gtrapprox 0.067$) is necessary to implement the SJM deterministically, we have conjectured the bound is not tight.
Indeed, our numerical experiment demonstrates that a maximally entangled state $(\theta=\frac{\pi}{4})$ is necessary for a deterministic implementation.
However, it remains an open question whether less than $1$-ebit entanglement is sufficient when a higher-Schmidt-rank resource state is allowed.

\subsection{Entanglement cost of the state verification}
\label{sec:verification}
In this subsection, we investigate the maximum $q\in(0,1]$ such that a POVM described by an instrument $\{\mathcal{E}_{\texttt{accept}},\mathcal{E}_{\texttt{reject}}\}$ defined by $\mathcal{E}_{\texttt{accept}}(\rho)=q\tr{\phi\rho}$ and $\mathcal{E}_{\texttt{reject}}(\rho)=\tr{(\idop-q\phi)\rho}$ is deterministically implementable by a bipartite SEP channel with a resource state $\ket{\tau}$.
Since this POVM does not have an output system, we let $\hh_{\hat{A}}=\hh_{A_1}$ and $\hh_{\hat{B}}=\hh_{B_1}$ in Fig.~\ref{fig:Hspaces}.
Note that one can determine whether a given state $\rho$ is a target state $\phi$ or far from it using this POVM~\cite{PLM18,KR21} at a certain confidence level.
Multiple copies of $\rho$ are required to increase this confidence level.
We call the instrument with $q=1$ an optimal verification measurement, as it requires the fewest copies~\cite{PLM18}.

By modifying the general optimization problem given in Eq.~\eqref{eq:SEPoptformat}, the maximum $q$ can be formulated as
\begin{equation}
\label{eq:SEPoptveri}
q(\phi,\tau)=\max\left\{ \tr{S\tau}:
\begin{array}{l}
S\in \SEP{\hat{\mathcal{A}}:\hat{\mathcal{B}}},\range{S}\subseteq\hat{\ww},     \\
\idop-S\in \SEP{\hat{\mathcal{A}}:\hat{\mathcal{B}}}
\end{array}
\right\},
\end{equation}
where $
 \hat{\ww}=\{\ket{\Xi}\in\hat{\mathcal{A}}\otimes\hat{\mathcal{B}}:\bra{\tau}_{R_AR_B}\ket{\Xi}\in\vspan{\{\ket{\phi}\}}\}
$.
Note that we use the complex conjugation of $S_m$ from Eq.~\eqref{eq:SEPoptformat} in Eq.~\eqref{eq:SEPoptveri} to eliminate the complex conjugate for $\ket{\tau}$ and $\ket{\phi}$. $\overline{S}$ and $\idop-\overline{S}$ correspond to the Choi operator of $\mathcal{S}_{\texttt{accept}}$ and $\mathcal{S}_{\texttt{reject}}$, where $\{\mathcal{S}_{\texttt{accept}},\mathcal{S}_{\texttt{reject}}\}$ forms a separable instrument that deterministically realizes $\{\mathcal{E}_{\texttt{accept}},\mathcal{E}_{\texttt{reject}}\}$ with the assistance of $\tau$.

We assume that $\Schrank{{\hat{A}}:{\hat{B}}}{\ket{\phi}}= \Schrank{{R_A}:{R_B}}{\ket{\tau}}=d$.
Accordingly, we can let $\ket{\tau}=\idop^{(R_A)}\otimes L^{(R_B)}\ketUMES{R_AR_B}{d}$ and
$\ket{\phi}=V_A\otimes (V_BL_1)\ketUMES{AB}{d}$, 
where $L$ and $L_1$ are invertible operators, and $V_A$ (or $V_B$) is an isometry from $\hh_A$ (or $\hh_B$) into $\hh_{\hat{A}}$ (or $\hh_{\hat{B}}$). 
Using this representation, we can confirm that $\hat{\ww}$ is an extended and twisted canonical subspace.
By letting
$\hat{\ww}^\circ=\{\ket{\Xi}\in\hat{\mathcal{A}}\otimes\hat{\mathcal{B}}:\bra{\tau}_{R_AR_B}\ket{\Xi}=0\}$ and using Eq.~\eqref{eq:MFLCofETCS}, we can show that the MFLC of $\Segre{\hat{\mathcal{A}}:\hat{\mathcal{B}}}\cap(\hat{\ww}\setminus\hat{\ww}^\circ)$ is
$
    \pp=\left(V_A\otimes\idop^{(R_A)}\otimes \left(L^{\dag}\right)^{-1}\otimes (V_BL_1)\right)\CMFLC{d}
$.
By using the general optimization problem given in Eq.~\eqref{eq:SEPoptformatMFLC}, we obtain
\begin{equation}
\label{eq:SEPoptveriMFLC}
q(\phi,\tau)=\max\left\{\tr{S\tau}:
\begin{array}{l}
S\in \SEP{\hat{\mathcal{A}}:\hat{\mathcal{B}}}\cap\pos{\pp},     \\
\idop-S\in \SEP{\hat{\mathcal{A}}:\hat{\mathcal{B}}}
\end{array}
\right\}.
\end{equation}

\noindent
\textbf{Numerical experiment}

Here, we consider the case where the target state $\ket{\phi}$ is the first state $\ket{M_1}$ of the SJM defined in Eq.~\eqref{eq:SJM} and the resource state is given by $\ket{\tau_\theta}=\cos\theta\ket{00}+\sin\theta\ket{11}$. In this setting, $L$, $V_A$, $V_B$, and $L_1$ are given in Eq.~\eqref{eq:SJMops1} and Eq.~\eqref{eq:SJMops2} .

Fig.~\ref{fig:Entcost} (d) compares the upper bounds on $q(\phi,\tau)$ by using the DPS hierarchy in Eq.~\eqref{eq:SEPoptveri} and Eq.~\eqref{eq:SEPoptveriMFLC} (see the details in Supplementary Information~\ref{appendix:verification}). 
Note that we also computed the lower bound on $q(\phi,\tau)$ based on an algorithm shown in Supplementary Information~\ref{appendix:enet} with randomly sampled $\epsilon$-nets $\left\{\ket{\Pi_x}\in\Segre{\hat{\mathcal{A}}:\hat{\mathcal{B}}}\cap(\hat{\ww}\setminus\hat{\ww}^\circ)\right\}_{x=1}^{400}$ and $\left\{\phi_x\in\puredop{\hh_{\hat{A}}\otimes\hh_{R_A}}\right\}_{x=1}^{340}$. 
The numerical results indicate that the additional constraint resulting from the MFLC effectively determines the trade-off curve between $q(\phi,\tau)$ and the strength $\theta$ of the entanglement, which is not achievable through an even higher level of the DPS hierarchy without the MFLC constraint.
The numerical results also indicate that a maximally entangled state ($\theta=\frac{\pi}{4}$) is necessary for the optimal verification measurement ($q=1$).
This observation is analytically proven in Section~\ref{sec:maxent}.
The reduction of the execution time can be understood by considering the difference in the number of parameters in $S$ in Eq.~\eqref{eq:SEPoptveri} and Eq.~\eqref{eq:SEPoptveriMFLC} ($\dim\hat{\ww}=13$ and $\dim\pp=10$).

\subsection{Entanglement distillation}
\label{sec:distillation}
A maximally entangled state is a valuable resource for distributed quantum information processing. However, in practice, it must be distilled from a noisy entangled state $\tau$. This process, known as entanglement distillation, has been extensively researched for decades. The central challenge is determining how resourceful pure entangled states can be distilled from the given state $\tau$. Notably, one of the major open problems in quantum information theory is determining the distillability of $\tau$ with a negative partial transpose (NPT)~\cite{HRZ22}. 
If we can show the existence of an NPT state $\tau$ that is not distillable under a superset of the set of LOCC channels, we can resolve the problem. To pursue this approach, we need to examine entanglement distillation using SEP channels, as any NPT state is distillable under PPT channels~\cite{EVWW01} and any entangled state is distillable under dually non-entangling operations~\cite{LB24}.

In this subsection, we investigate the success probability of distilling a pure entangled state $\ket{\psi_\theta}=\cos\theta\ket{00}+\sin\theta\ket{11}\in\hh_{A}\otimes\hh_{B}$ from a single mixed state $\tau=\sum_{i=1}^3q_i\tau_i\in\dop{\hh_{R_A}\otimes\hh_{R_B}}$ under SEP channels, where $\theta\in(0,\frac{\pi}{4}]$, $\forall q_i>0$, and
\begin{equation}
\label{eq:CDstates}
    \ket{\tau_1}=\frac{1}{\sqrt{2}}(\ket{01}+e^{i\theta_1}\ket{10}),
    \ket{\tau_2}=\frac{1}{\sqrt{2}}(\ket{02}+e^{i\theta_2}\ket{20}),
    \ket{\tau_3}=\frac{1}{\sqrt{2}}(\ket{12}+e^{i\theta_3}\ket{21}).
\end{equation}
Distillable entanglement of $\tau$ under the PPT operations has been studied~\cite{WD16,WD17}. 
However, it remains an open problem to demonstrate any gap in distillable entanglement between PPT and SEP channels.
In Supplementary Information~\ref{appendix:distillation}, we construct a SEP channel distilling $\psi_\theta$ from $\tau$ with the success probability $\min\left\{1,\frac{1}{2\sin2\theta}\right\}$ for any $\theta_i$ and $q_i$ by modifying the previous result~\cite[Theorem2 (b)]{CD09}.
By using the general optimization problem given in Eq.~\eqref{eq:SEPoptformat} and letting $\hh_{\hat{A}}=\hh_{A_2}=\hh_{A}$ and $\hh_{\hat{B}}=\hh_{B_2}=\hh_{B}$, the success probability can be formulated as
\begin{equation}
\label{eq:SEPoptdist}
p(\psi_\theta,\tau)=\max\left\{ \tr{S\overline{\tau}}:
\begin{array}{l}
S\in \SEP{\mathcal{A}:\mathcal{B}},\range{S}\subseteq\ww,     \\
\idop-\ptr{AB}{S}\in \SEP{\hh_{R_A}:\hh_{R_B}}
\end{array}
\right\},
\end{equation}
where $\mathcal{A}=\hh_A\otimes\hh_{R_A}$, $\mathcal{B}=\hh_{R_B}\otimes\hh_B$, and
$\ww:=\{\ket{\Xi}\in\mathcal{A}\otimes\mathcal{B}:\forall i,\bra{\overline{\tau}_i}_{R_AR_B}\ket{\Xi}\in\vspan{\ket{\psi_\theta}}\}$.

By letting $\ww^\circ:=\{\ket{\Xi}\in\mathcal{A}\otimes\mathcal{B}:\overline{\tau}^{(R_AR_B)}\ket{\Xi}=0\}$, we calculate the MFLC of $\Segre{\mathcal{A}:\mathcal{B}}\cap(\ww\setminus\ww^\circ)$ in the following proposition.

\begin{proposition}
\label{prop:MFLCdist}
 	Suppose that $\theta_i=\pi$ for all $i$ in the definition of $\ket{\tau_i}$ (see {\rm Eq}.~\eqref{eq:CDstates}).
    The MFLC of $\Segre{\mathcal{A}:\mathcal{B}}\cap(\ww\setminus\ww^\circ)$ is given by
    \begin{eqnarray}
        \pp&=& (\idop^{(A)}\otimes\idop^{(R_A)}\otimes(L\sigma_Y)^{(B)}\otimes\idop^{(R_B)})\bigvee_{n=1}^2(\hh_A\otimes\hh_{R_A}),
    \end{eqnarray}
    where $\sigma_Y$ is the Pauli Y operator, $\ket{\psi_\theta}=(\idop^{(A)}\otimes L^{(B)})\ketUMES{AB}{2}$, and we regard the symmetric subspace $\bigvee_{n=1}^2(\hh_A\otimes\hh_{R_A})$ as being embedded in $\hh_A\otimes\hh_{R_A}\otimes\hh_B\otimes\hh_{R_B}$ by the isomorphism $\hh_B \otimes \hh_{R_B} \simeq \hh_A \otimes \hh_{R_A}$.
\end{proposition}
A proof is given in Supplementary Information~\ref{appendix:entMFLC}.
By using Proposition~\ref{prop:MFLCdist} and the general optimization problem given in Eq.~\eqref{eq:SEPoptformatMFLC}, we obtain

\begin{equation}
\label{eq:SEPoptdistMFLC}
p(\psi_\theta,\tau)=\max\left\{ \tr{S\overline{\tau}}:
\begin{array}{l}
S\in \SEP{\mathcal{A}:\mathcal{B}}\cap\pos{\pp},     \\
\idop-\ptr{AB}{S}\in \SEP{\hh_{R_A}:\hh_{R_B}}
\end{array}
\right\}.
\end{equation}

\noindent
\textbf{Numerical experiment}

 Here, we numerically solved the DPS hierarchy of Eqs.~\eqref{eq:SEPoptdist} and \eqref{eq:SEPoptdistMFLC} for $\theta_i=\pi$ and $q_i=\frac{1}{3}$ (see details in Supplementary Information~\ref{appendix:distillationSDP}).
The numerical result in Fig.~\ref{fig:Entcost} (a) indicates that the constraints coming from the MFLC reveal the optimality of the distillation protocol shown in Supplementary Information~\ref{appendix:distillation} in the sense that it attains the maximum success probability, or equivalently, it distills maximum entanglement under a certain success probability.
The reduction of the execution time can be understood by considering the difference in the number of parameters in $S$ in Eqs.~\eqref{eq:SEPoptdist} and \eqref{eq:SEPoptdistMFLC} ($\dim\ww=27$ and $\dim\pp=10$).

\newpage
\appendix
\backmatter
\newgeometry{top=2cm,bottom=2cm,left=1cm,right=1cm}
\bmhead{Supplementary Information}

\section{Preliminaries of Algebraic geometry}
\label{sec:alggmt}
Here, let us briefly introduce notation and concepts of algebraic geometry. Readers can find a more comprehensive introduction in~\cite{RHBook,EKBook}.
We denote the set of non-negative integers by $\nn=\{0,1,2,\cdots\}$.

\subsection{Projective space}
\begin{definition}
    For $d\in\nn$, a projective $d$-space $\pspace^d$ is the set of equivalent classes $[x_0:\cdots:x_d]$ of vectors $(x_0,\cdots,x_d)\in\cdim{d+1}\setminus\{0\}$ under the equivalence relation given by $(x_0,\cdots,x_d)\sim(y_0,\cdots,y_d)$ if $(x_0,\cdots,x_d)\propto(y_0,\cdots,y_d)$.
\end{definition}

\begin{definition}
    A (linear) subspace $\vv$ in the projective space $\pspace^{d-1}$ is defined as the zero set of a family of homogeneous polynomials in $\cc[x_0,\cdots,x_{d-1}]$ of degree at most $1$.
\end{definition}
For example, $\emptyset$ and $\pspace^{d-1}$ are subspaces. 
We can verify that for a non-empty subspace $(\emptyset\neq)\vv\subseteq\pspacedim{d-1}$, there exists a rank-$d'(\geq1)$ matrix $V:\cdim{d'}\rightarrow\cdim{d}$ such that $\vv=V\pspacedim{d'-1}$.
Note that the intersection of a family of subspaces is a subspace.

\begin{definition}
    The (linear) span $\vspan{\mathbb{E}}$ of a subset $\mathbb{E}\subseteq\pspace^{d-1}$ is the minimum subspace $\vv$ containing $\mathbb{E}$.
\end{definition}

An alternative definition of the span is given as follows:
For a subset $\mathbb{E}\subseteq\pspace^{d-1}$, $\vspan{\mathbb{E}}$ is the zero set of a family of homogeneous polynomials in $\cc[x_0,\cdots,x_{d-1}]$ of degree at most $1$ that vanishes on $\mathbb{E}$.

Let $p:\cd\setminus\{0\}\rightarrow\pspace^{d-1}$ be the canonical projection from the ambient complex vector space into a projective space. We can verify that
$p^{-1}(\vspan{\mathbb{E}})\cup\{0\}=\vspan{\{\ket{\phi}:\ket{\phi}\in p^{-1}(\mathbb{E})\}}$.

While there is no standard notion of a tensor product of projective spaces, we use the tensor product notation 
\[
[x_0:x_1:\cdots:x_{d-1}]\otimes[y_0:y_1:\cdots:y_{d'-1}]=[x_0y_0:x_0y_1:\cdots:x_1y_0:x_1y_1:\cdots:x_{d-1}y_{d'-1}]
\]
to denote the image of the Segre embedding from $\pspace^{d-1}\times\pspace^{d'-1}$ into $\pspace^{dd'-1}$.

\subsection{Algebraic set in projective space}

\begin{definition}
    A (commutative) ring $S$ is called a graded ring if $S=\oplus_{n\in\nn}S_n\left(=\left\{\sum_{n\in N}f_n:f_n\in S_n,|N|<\infty\right\}\right)$, where each $S_n$ is an abelian group and $S_n\cdot S_m\subseteq S_{m+n}$ for any $m,n\in\nn$. An element of $S_n$ is called a homogeneous element of degree $n$.
\end{definition}
For example, the polynomial ring $\cc[x_0,x_1,\cdots,x_{d-1}]$ in (finite) $d$ variables over the field $\cc$ is a graded ring, where $S_n$ is the set consisting of $0$ and homogeneous polynomials with degree $n$. Note that $S_n$ is isomorphic to the set of linear functionals $\bra{f}:\vee_n\mathbb{C}^d\rightarrow\mathbb{C}$ via the isomorphism $f(\ket{x})=\bra{f}(\ket{x}^{\otimes n})$, where $f\in S_n$ and $\ket{x}=(x_0,x_1,\cdots,x_{d-1})^T$ \cite{HNB24}.

\begin{definition}
    An ideal $\mathfrak{a}\subseteq S$ of a graded ring $S=\oplus_{n\in\nn}S_n$ is a homogeneous ideal if $\mathfrak{a}=\oplus_{n\in\nn}(\mathfrak{a}\cap S_n)$.
\end{definition}

Since $\mathfrak{a}\supseteq\oplus_{n\in\nn}(\mathfrak{a}\cap S_n)$ holds for any ideal $\mathfrak{a}$, the condition that $\mathfrak{a}$ be homogeneous can equivalently be stated as follows: for any $f\in\mathfrak{a}$ and any $n\in\nn$, the homogeneous element of $f$ with degree $n$ also belongs to $\mathfrak{a}$.
The following facts are known:
\begin{itemize}
    \item An ideal $\mathfrak{a}$ is homogeneous if and only if it is generated by a set of homogeneous elements, i.e., $\mathfrak{a}=\langle\{f_\alpha\}_{\alpha\in A}\rangle$, where $f_\alpha\in S_{n_\alpha}$.
    \item If $\mathfrak{a}$ and $\mathfrak{b}$ are homogeneous ideals, $\mathfrak{a}+\mathfrak{b}$, $\mathfrak{a}\mathfrak{b}$, $\mathfrak{a}\cap\mathfrak{b}$, and $\sqrt{\mathfrak{a}}$ are homogeneous ideals.
\end{itemize}

\begin{definition}
The zero set of a family $T\subseteq\cc[x_0,\cdots,x_{d-1}]$ of homogeneous polynomials is defined by $Z(T):=\{x\in\pspace^{d-1}:\forall f\in T,f(x)=0\}$.
For a homogeneous ideal $\mathfrak{a}$, we define $Z(\mathfrak{a}):=Z(T)$, where $T$ is the set of homogeneous elements in $\mathfrak{a}$.
\end{definition}

\begin{definition}
 $\mathbb{E}\subseteq\pspace^{d-1}$ is an algebraic set if there exists a family $T$ of homogeneous polynomials such that $\mathbb{E}=Z(T)$.
\end{definition}

\begin{definition}
 The Zariski topology on $\pspace^{d-1}$ is defined by taking the closed sets to be the algebraic sets.
\end{definition}

\begin{definition}
 A nonempty subset $\mathbb{E}$ of a topological space $X$ is irreducible if it cannot be decomposed as a union $\mathbb{E}=\mathbb{E}_1\cup\mathbb{E}_2$ of two proper subsets, each one of which is closed in $\mathbb{E}$.
\end{definition}

\begin{definition}
    An irreducible algebraic set in $\pspace^{d-1}$ with respect to the Zariski topology is called a projective variety.
\end{definition}

\begin{definition}
    For a subset $\mathbb{E}$ of $\pspace^{d-1}$, 
    the homogeneous ideal $I(\mathbb{E})$ of $\mathbb{E}$ is defined by $\langle\{f\in \cup_{n\in\nn} S_n:\forall \ket{x}\in\mathbb{E},f(\ket{x})=0\}\rangle$, i.e., the ideal generated by homogeneous polynomials that vanishes on $\mathbb{E}$.
\end{definition}

The following facts are known:
\begin{itemize}
    \item An algebraic set $\mathbb{E}\subseteq\pspace^{d-1}$ is irreducible if and only if $I(\mathbb{E})$ is a prime ideal \cite[Exercise 2.4]{RHBook}.
    \item An algebraic set $\mathbb{E}\subseteq\pspace^{d-1}$ can be uniquely decomposed into irreducible components; i.e., there exists a unique finite family $\{\mathbb{I}_k\}_{k\in K}$ of projective varieties such that $\mathbb{E}=\cup_{k\in K}\mathbb{I}_k$ and $\mathbb{I}_k\not\subseteq\mathbb{I}_{k'}$ for any $k\neq k'$ \cite[Exercise 2.5]{RHBook}. 
    \item Any nonempty subspace in $\pspace^{d-1}$ is a projective variety.
    \item For a subset $\mathbb{E}$ of $\pspace^{d-1}$,  $Z(I(\mathbb{E}))=\overline{\mathbb{E}}$ \cite[Exercise 2.3]{RHBook}.
    \item A subset $\mathbb{E}$ of a topological space $X$ is irreducible if and only if its closure $\overline{\mathbb{E}}$ is irreducible~\cite[Example 1.1.4]{RHBook}.
    \item The set $\{\otimes_{n=1}^N\ket{\phi_n}:\ket{\phi_n}\in\pspace^{d_n-1}\}=p(\Segre{\cdim{d_1}:\cdots:\cdim{d_N}}\setminus\{0\})$ is a projective variety in $\pspacedim{d_1\cdots d_N-1}$ \cite[Exercise 2.14]{RHBook}.
 \item Any nonempty open set $\mathbb{V}$ in an irreducible set $\mathbb{E}$ is irreducible~\cite[Example 1.1.3]{RHBook}.
\end{itemize}

The following fact is also known (in a different notation)~\cite{HNB24}. For completeness, we provide a proof.

\begin{proposition}
    \label{prop:ZIm}
    For a subset $\mathbb{E}\subseteq\pspace^{d-1}$ and positive integer $m$, $\ket{x}\in Z(I_m(\mathbb{E}))$ if and only if $\ket{x}^{\otimes m}\in\vee_m\mathbb{E}:=\vspan{\{\ket{x}^{\otimes m}:\ket{x}\in\mathbb{E}\}}$.
\end{proposition}
In the proof, we use the following lemma.
\begin{lemma}
    For a subset $\mathbb{E}$ of $\pspace^{d-1}$ and $m\in\nn$, $Z(I_m(\mathbb{E}))=Z(\langle I(\mathbb{E})\cap S_m\rangle)$.
\end{lemma}
\begin{proof}
    Since $I_m(\mathbb{E})\supseteq I(\mathbb{E})\cap S_m$, we find $Z(I_m(\mathbb{E}))\subseteq Z(\langle I(\mathbb{E})\cap S_m\rangle)$. To complete the proof, we show that $\forall\ket{x}\in Z(\langle I(\mathbb{E})\cap S_m\rangle), \forall f\in I_m(\mathbb{E}),f(\ket{x})=0$.
    This is equivalent to $\forall\ket{x}\in Z(\langle I(\mathbb{E})\cap S_m\rangle), \forall f\in \cup_{n=0}^{m}S_m,(\forall\ket{y}\in\mathbb{E},f(\ket{y})=0)\Rightarrow f(\ket{x})=0$. For all $\ket{x}\in Z(\langle I(\mathbb{E})\cap S_m\rangle)$ and for all $f\in S_n$ such that $0\leq n\leq m$ and $\forall\ket{y}\in\mathbb{E},f(\ket{y})=0$, there exists $g\in S_{m-n}$ such that $g(\ket{x})\neq0$. Since $fg\in I(\mathbb{E})\cap S_m$, $f(\ket{x})g(\ket{x})=0$. Thus, $f(\ket{x})=0$.
\end{proof}

\begin{proof}[Proof of Proposition \ref{prop:ZIm}]
    \begin{eqnarray}
        &&\ket{x}\in Z(I_m(\mathbb{E}))=Z(\langle I(\mathbb{E})\cap S_m\rangle)\\
        &\Leftrightarrow&\forall f\in I(\mathbb{E})\cap S_m,f(\ket{x})=0\\
        &\Leftrightarrow&\forall f\in S_m,(\forall \ket{x}\in\mathbb{E},f(\ket{x})=0)\Rightarrow f(\ket{x})=0\\
        &\Leftrightarrow&\forall\ket{f}\in\vee_m\cd,(\forall\ket{x}\in\mathbb{E},\bra{f}(\ket{x}^{\otimes m})=0)\Rightarrow\bra{f}(\ket{x}^{\otimes m})=0\\
        &\Leftrightarrow&\forall\ket{f}\in\cdim{md},(\forall\ket{x}\in\mathbb{E},\bra{f}(\ket{x}^{\otimes m})=0)\Rightarrow\bra{f}(\ket{x}^{\otimes m})=0\\
        &\Leftrightarrow&\ket{x}^{\otimes m}\in Z(I_1(\{\ket{x}^{\otimes m}:\ket{x}\in\mathbb{E}\}))=\vspan{\{\ket{x}^{\otimes m}:\ket{x}\in\mathbb{E}\}}.
    \end{eqnarray}

\end{proof}

\section{Application of MFLCs in two qubits to unambiguous local state discrimination}
\label{appendix:Koashi}
Unambiguous state discrimination tries to distinguish without error quantum states that are not necessarily orthogonal. At a glance, this contradicts the nature of quantum mechanics. However, it is possible by allowing an "I don't know" outcome~\cite{C98,D88,P88,I87,HB04,MHD10,A04}. Since the non-orthogonal quantum states are at the heart of quantum cryptography, the possibility of unambiguous state discrimination is used in quantum cryptographic protocols~\cite{EHPP94}.

In~\cite{KTYI08}, Koashi et al. consider an unambiguous local-state discrimination of
\begin{equation}
    \hat{\rho}_0=\ketbra{00},\ \hat{\rho}_1=\frac{1}{2}\left(\ketbra{++}+\ketbra{--}\right),
\end{equation}
where $\ket{\pm}=\frac{1}{\sqrt{2}}(\ket{0}\pm\ket{1})$ under LOCC and SEP channels.
To generalize the scenario of SEP channels, we examine unambiguous local discrimination of general two-qubit mixed states $\rho_0$ and $\rho_1$ by using a separable instrument.
Formally, we construct a positive operator-valued measure (POVM) $\{M_0,M_1,M_2\}\subseteq\SEP{\cdim{2}:\cdim{2}}$ such that
\begin{equation}
  \label{eq:USDcond}
    \tr{\rho_0M_1}=\tr{\rho_1M_0}=0,\ \sum_{m=0}^2M_m=\idop.
\end{equation}
By following~\cite{KTYI08}, we focus on the success probability $\gamma_m:=\tr{\rho_mM_m}$ of guessing $\rho_m$ and analyze the maximum $P^{(sep)}_{opt}(\gamma_0)$ of $\gamma_1$ when $\gamma_0$ is given.
Formally, $P^{(sep)}_{opt}(\gamma_0)$ is the maximum value of $\gamma_1$ when $\tr{\rho_0M_0}=\gamma_0$ and Eqs.~\eqref{eq:USDcond} are satisfied.

Eqs.~\eqref{eq:USDcond} imply that
\begin{eqnarray}
    \range{M_1}\subseteq\vv_0,
    \range{M_0}\subseteq\vv_1,
\end{eqnarray}
where $\vv_m$ is the orthogonal complement of $\range{\rho_m}$.
This implies that $M_0$ (or $M_1$) is a convex combination of $\ketbra{\Pi}$, where $\ket{\Pi}$ is contained in the MFLC of $\Segre{\cdim{2}:\cdim{2}}\cap\vv_1$ (or $\Segre{\cdim{2}:\cdim{2}}\cap\vv_0$).
Combined with Propositions~\ref{prop:MFLCof2Q}, this fact makes it simpler to calculate $P^{(sep)}_{opt}(\gamma_0)$.
For example, it is known that $P^{(sep)}_{opt}(\gamma_0)$ can be computed by using an SDP since $\SEP{\cdim{2}:\cdim{2}}=\PPT{\cdim{2}:\cdim{2}}$~\cite{HHH96}.
We can obtain a simpler SDP by incorporating the constraints resulting from the MFLCs.
Below, we demonstrate this simplification by using a specific class of $\rho_0$ and $\rho_1$.

Here, we consider
\begin{eqnarray}
    \rho_0&=&\ketbra{00},\\
    \rho_1&=&a\ketbra{++}+c\ketbra{--}+b\ket{++}\bra{--}+b\ket{--}\bra{++},
\end{eqnarray}
where $a\in(0,1)$, $c=1-a$ and $b^2< ac$.
Note that $\range{\rho_1}=\vspan{\{\ket{++},\ket{--}\}}$ under these conditions.
We can show that $\vv_0=\vspan{\{\ket{01},\ket{10},\ket{11}\}}$ and $\vv_1=\vspan{\{\ket{+-},\ket{-+}\}}$ through a straightforward calculation.
By following the calculation in the proof of Proposition~\ref{prop:MFLCof2Q}, we obtain their irreducible components, which coincide with their MFLCs:
\begin{eqnarray}
    \mathbb{S}\cap\vv_0&=& (\ket{1}\otimes\cdim{2})\cup(\cdim{2}\otimes\ket{1})\\
    \mathbb{S}\cap\vv_1&=& \vspan{\{\ket{+-}\}}\cup\vspan{\{\ket{-+}\}}.    
\end{eqnarray}

Observe that the values $\tr{\rho_nM_m}$ and $\sum_{m=0}^2M_m$ do not change if we replace $M_m$ by $\hat{M}_m=\frac{1}{4}(M_m+\overline{M}_m+P(M_m+\overline{M}_m)P)$ with the swap operator $P=\sum_{i,j}\ket{ij}\bra{ji}$ since $\rho_1$, $\rho_2$ and $\idop$ are invariant under swap and complex-conjugation.
By considering the MFLCs and the invariance of $\hat{M}_m$ under swap and complex conjugation, we can let
\begin{eqnarray}
\label{eq:M0}
    \hat{M}_0&=&p(\ketbra{+-}+\ketbra{-+})\\
    \label{eq:M1}
    \hat{M}_1&=&\ketbra{1}\otimes S+S\otimes\ketbra{1}+q\ketbra{1}\otimes\ketbra{1},
\end{eqnarray}
where $p\geq0$, $q\in\rr$, $S^T=S$, $S\in\pos{\cdim{2}}$, and $S+q\ketbra{1}\in\pos{\cdim{2}}$.
Since $\pos{\hh}$ is convex, $S,S+q\ketbra{1}\in\pos{\cdim{2}}$ implies $S+\frac{q}{2}\ketbra{1}\in\pos{\cdim{2}}$.
Thus, we can let
\begin{eqnarray}
    \label{eq:M1p}
    \hat{M}_1&=&\ketbra{1}\otimes S+S\otimes\ketbra{1},
\end{eqnarray}
where $S^T=S$ and $S\in\pos{\cdim{2}}$ without loss of generality.

By using these parameterizations, we can represent the success probability $\gamma_n$ of guessing $\rho_n$ as
\begin{eqnarray}
    \label{eq:gamma1}
    \gamma_0=\tr{\rho_0\hat{M}_0}=\frac{p}{2},\ \ 
    \gamma_1=\tr{\rho_1\hat{M}_1}=\tr{S\sigma},
\end{eqnarray}
where $\sigma=a\ketbra{+}+c\ketbra{-}+b\ket{+}\bra{-}+b\ket{-}\bra{+}$.
Thus, $P^{(sep)}_{opt}(\gamma_0)$ can be formulated as the following optimization problem:
\begin{eqnarray}
	\label{eq:koashiprob}
    P^{(sep)}_{opt}(\gamma_0)&=&\max\tr{S\sigma}\\
	\label{eq:CPcond}
    &s.t.&S\geq0, S^T=S\\
    \label{eq:TPcond}
    &&2\gamma_0(\ketbra{+-}+\ketbra{-+})\nonumber\\
    &&+\ketbra{1}\otimes S+S\otimes\ketbra{1}\leq\idop.
\end{eqnarray}
Note that Eq.~\eqref{eq:TPcond} is imposed under the condition $\hat{M}_0+\hat{M}_1\leq\idop$, which guarantees the existence of $\hat{M}_2$ such that $\hat{M}_2=\idop-\hat{M}_0-\hat{M}_1\in\pos{\cdim{4}}$ and $\sum_{m=0}^2\hat{M}_m=\idop$. We do not explicitly impose $\hat{M}_2\in\SEP{\cdim{2}:\cdim{2}}$ in the optimization problem. However, this condition is satisfied since 
$\hat{M}_2^{T_1}=\idop-\hat{M}_0-\hat{M}_1=\hat{M}_2\in\pos{\cdim{4}}$ and $\SEP{\cdim{2}:\cdim{2}}=\PPT{\cdim{2}:\cdim{2}}$, where $T_1$ represents partial transposition on the first qubit.

\begin{figure}[ht]
    \centering
    \includegraphics[width=6cm]{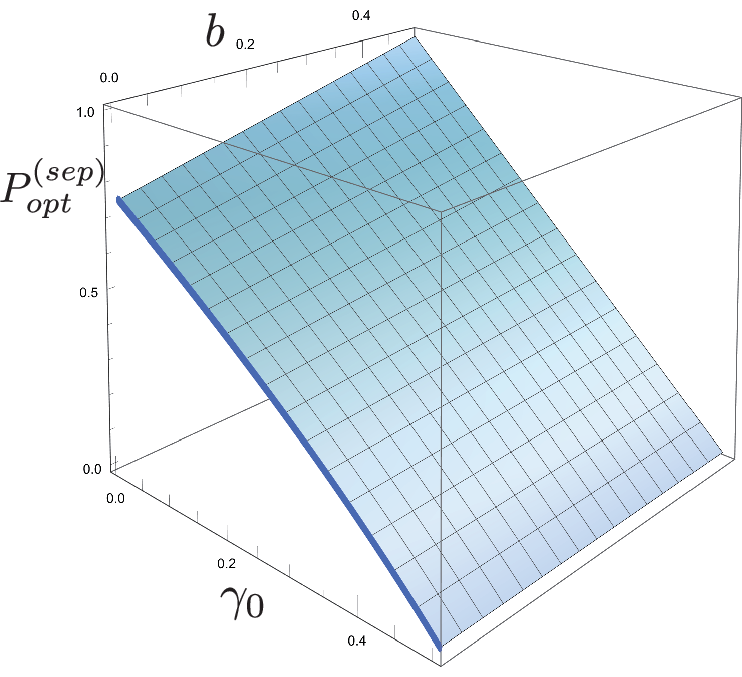}
    \caption{Plot of $P^{(sep)}_{opt}(\gamma_0)$ solved by an SDP represented by Eqs.~\eqref{eq:koashiprob}--\eqref{eq:TPcond} for $a=c=\frac{1}{2}$. For the case of $b=0$, the solution of the SDP coincides with the analytical curve $1-\gamma_0-(4(1-\gamma_0))^{-1}$ derived in~\cite{KTYI08}, depicted by the thick blue curve.}
    \label{fig:Koashi}
\end{figure}

It is important to note that this optimization problem, defined in Eqs.~\eqref{eq:koashiprob}--\eqref{eq:TPcond}, is an SDP without any condition resulting from the DPS hierarchy. This illustrates that the MFLC is advantageous for optimization over $\mathbf{SEP}$, independently of the DPS hierarchy.
We can ensure the validity of the optimization problem by plotting its numerical solutions (Fig.~\ref{fig:Koashi}).

\section{Proofs for MFLCs of canonical subspace}
\label{sec:MFLCCS}
\begin{figure}[ht]
    \centering
    \includegraphics[width=5.5cm]{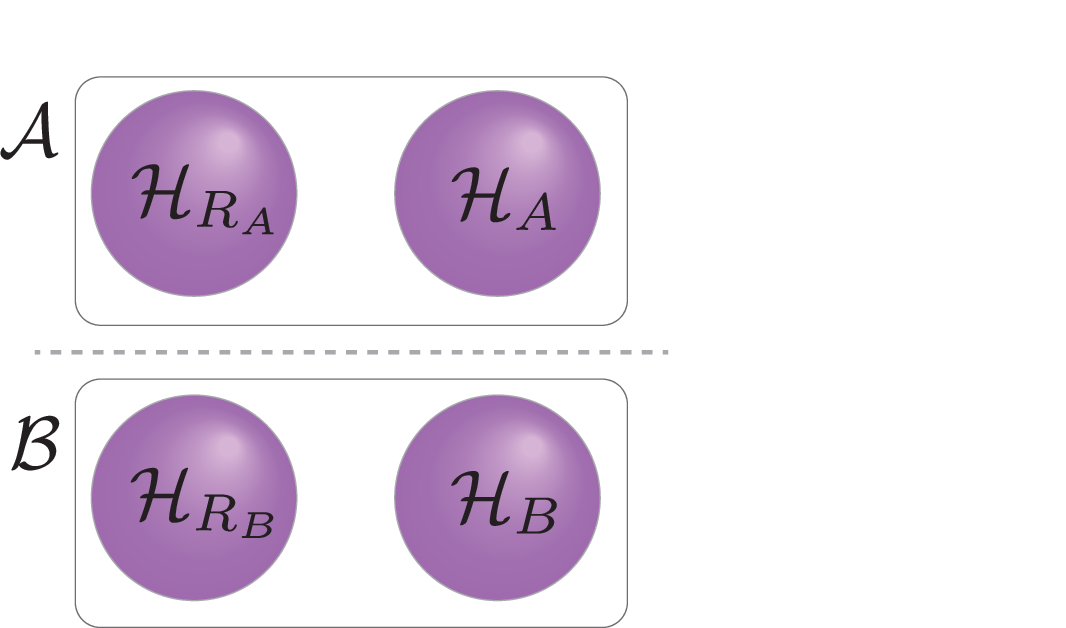}
    \caption{Partitioning of the composite Hilbert spaces where the canonical subspace is defined. We consider the product vectors between $\mathcal{A}$ and $\mathcal{B}$.}
    \label{fig:Hspacessimple}
\end{figure}
Let $\mathcal{A}=\hh_A\otimes\hh_{R_A}$, $\mathcal{B}=\hh_{R_B}\otimes\hh_B$, and $\dim\hh_A=\dim\hh_{R_A}=\dim\hh_B=\dim\hh_{R_B}=d$ (see Fig.~\ref{fig:Hspacessimple}).
A canonical subspace $\Csubspace{d}$ and a related subspace $\CVsubspace{d}$ is defined as
\begin{eqnarray}
    \Csubspace{d}&:=&\{\ket{\Xi}\in\mathcal{A}\otimes\mathcal{B}:\braUMES{R_AR_B}{d}\ket{\Xi}\in\vspan{\{\ketUMES{AB}{d}\}}\},\\
        \CVsubspace{d}&:=&\{\ket{\Xi}\in\mathcal{A}\otimes\mathcal{B}:\braUMES{R_AR_B}{d}\ket{\Xi}=0\}.    
\end{eqnarray}

\begin{proposition}
\label{prop:firstCS}
 $\CMFLC{d}\cup\CVsubspace{d}$ forms an FLC of $\Segre{\mathcal{A}:\mathcal{B}}\cap\Csubspace{d}$, where
 \begin{eqnarray}
        \CMFLC{d}&:=&\Csubspace{d}\cap\CDsubspace{d},\\
        \CDsubspace{d}&:=&\{\ket{\Xi}\in\mathcal{A}\otimes\mathcal{B}:\braUMES{AB}{d}\ket{\Xi}\in\vspan{\{\ketUMES{R_AR_B}{d}\}}\}.
\end{eqnarray}
\end{proposition}
\begin{proof}
First, through a straightforward calculation, we can show
\begin{eqnarray}
\label{eq:pvsub1}
      \Segre{\mathcal{A}:\mathcal{B}}\cap\Csubspace{d}&=&\{\ket{A}\ket{B}:[A][B]^T\in\{0,\idop\}\}\\
      &=&\{\ket{A}\ket{B}:[A][B]^T=0\}\cup\{\ket{A}\ket{B}:[A][B]^T=\idop\},
\end{eqnarray}
where $\ket{A}_{AR_A}=(A\otimes\idop^{(R_A)})\ketUMES{R_AR_A}{d}$ and $\ket{B}_{R_BB}=(\idop^{(R_B)}\otimes B)\ketUMES{R_BR_B}{d}$ for $A\in\linop{\hh_{R_A}:\hh_A}$ and $B\in\linop{\hh_{R_B}:\hh_B}$, and $[A]$ and $[B]$ are matrix representations of $A$ and $B$ with respect to the computational basis defining the maximally entangled states.
This can be done by observing $\braUMES{R_AR_B}{d}(\ket{A}\ket{B})\in\{0,\ketUMES{AB}{d}\}\Leftrightarrow [A][B]^T=\{0,\idop\}$.
Note that we regard each vector as an element of a projective space.

Since $\braUMES{AB}{d}(\ket{A}\ket{B})\in\{0,\ketUMES{R_AR_B}{d}\}\Leftrightarrow [B]^T[A]=\{0,\idop\}$ and $[A][B]^T=\idop\Leftrightarrow [B]^T[A]=\idop$, we find that
\begin{eqnarray}
\{\ket{A}\ket{B}:[A][B]^T=I\}&\subseteq&\Csubspace{d}\cap\CDsubspace{d}=\CMFLC{d},\\
\{\ket{A}\ket{B}:[A][B]^T=I\}&\cap&\CVsubspace{d}=\emptyset,\\
\{\ket{A}\ket{B}:[A][B]^T=0\}&\subseteq&\CVsubspace{d},\\
\{\ket{A}\ket{B}:[A][B]^T=0\}&\not\subseteq&\CDsubspace{d}.
\end{eqnarray}
This completes the proof.
\end{proof}

\begin{proposition}
\label{prop:secondCS}
$\mathbb{V}:=\left(\Segre{\mathcal{A}:\mathcal{B}}\cap\Csubspace{d}\right)\setminus\CVsubspace{d}$ is irreducible.
\end{proposition}
\begin{proof}
In the proof of Proposition \ref{prop:firstCS}, we obtained $\Segre{\mathcal{A}:\mathcal{B}}\cap\Csubspace{d}=\{\ket{A}\ket{B}:[A][B]^T=I\}\cup\{\ket{A}\ket{B}:[A][B]^T=0\}$, $\{\ket{A}\ket{B}:[A][B]^T=I\}\cap\CVsubspace{d}=\emptyset$, and $\{\ket{A}\ket{B}:[A][B]^T=0\}\subseteq\CVsubspace{d}$. These imply that
\begin{equation}
\mathbb{V}=\{\ket{A}\ket{B}:[A][B]^T=I\}=\{\ket{A}\ket{B}:[B]={\rm adj}([A])^T,\det([A])\neq0\},
\end{equation}
where ${\rm adj}([A])$ is the adjugate matrix of $[A]$, and we used $[A]{\rm adj}([A])=\det([A])\idop$.
Since $\mathbb{D}:=\{\ket{A}:\det([A])\neq0\}\subset\pspacedim{\dim\mathcal{A}-1}$ is an open set in an irreducible set $\pspacedim{\dim\mathcal{A}-1}$, $\mathbb{D}$ is also irreducible.
Since $f:\mathbb{D}\rightarrow\mathbb{V}$ defined by $f(\ket{A}):=\ket{A}\ket{B(A)}$ with $[B(A)]={\rm adj}([A])^T$ satisfies the condition of Proposition \ref{prop:mintest}, we find that $\mathbb{V}=f(\mathbb{D})$ is irreducible.
\end{proof}

\begin{proposition}
\label{prop:thirdCS}
Let $\mathbb{V}:=\left(\Segre{\mathcal{A}:\mathcal{B}}\cap\Csubspace{d}\right)\setminus\CVsubspace{d}$ and $\CMFLC{d}$ be defined as Proposition \ref{prop:firstCS}. Then, $\CMFLC{d}\subseteq\vspan{\mathbb{V}}$.
\end{proposition}
\begin{proof}
Since we obtained $\mathbb{V}=\{\ket{A}\ket{B}:[A][B]^T=I\}$ in the proof of Proposition \ref{prop:secondCS}, we find that
\begin{equation}
\mathbb{E}:=\{\ket{A}\ket{B}:[A]=U,[B]=\overline{U},U{\rm\ is\ a\ unitary\ matrix}\}\subseteq\mathbb{V}.
\end{equation}
In the following, we prove that $\vspan{\mathbb{E}}=\CMFLC{d}(\subseteq\vspan{\mathbb{V}})$, which completes the proof.
To do that, we show that
\begin{equation}
\label{eq:spanMFLC}
    \range{\int dU \ketbra{A}\otimes\ketbra{B}}=\CMFLC{d},
\end{equation}
where we regard vectors and subspaces as residing in a complex vector space, set $[A]=\overline{[B]}=U$, and compute the integral with respect to the Haar measure.
From now on, we will use the matrix representation in the calculation.
\begin{eqnarray}
    &&\int dU S(\ketbra{A}\otimes\ketbra{B})S^\dag\\
    &=&\int dUS\left( \sum_{ijkl}(U\ket{i}\bra{k}U^\dag)\otimes \ket{i}\bra{k}\otimes\ket{j}\bra{l}\otimes( \overline{U}\ket{j}\bra{l}\overline{U}^\dag)\right)S^\dag\\
    &=&\int dU \sum_{ijkl}\ket{ij}\bra{kl} \otimes\left((U\otimes \overline{U})\ket{ij}\bra{kl}(U\otimes \overline{U})^\dag\right)\\
    \label{eq:pvspan_analysis1}
    &=&\sum_{ijkl} \ket{ij}\bra{kl} \otimes\left(\int dU(U\otimes U)\ket{il}\bra{kj}(U\otimes U)^\dag\right)^{T_2},
\end{eqnarray}
where $S=\sum_{ijk}\ket{ijk}\bra{kij}\otimes\idop$ is a permutation operator and $T_2$ represents the partial transpose acting on the second system.

For a $d^2$ by $d^2$ matrix $X$ and $d$-dimensional unitary matrix $U$, it is known that $Y:=\int dU(U\otimes U)X(U\otimes U)^\dag$ can be decomposed as $Y=\alpha\idop+\beta P$, where $P=\sum_{ij}\ket{ij}\bra{ji}$ is the swap matrix~\cite[Theorem 7.15]{WBook}. Since $\tr{Y}=\tr{X}=\alpha d^2+\beta d$ and  $\tr{PY}=\tr{PX}=\alpha d+\beta d^2$, we obtain
\begin{equation}
    \int dU(U\otimes U)X(U\otimes U)^\dag=\frac{d\tr{X}-\tr{PX}}{d(d^2-1)}\idop+
    \frac{d\tr{PX}-\tr{X}}{d(d^2-1)}P.
\end{equation}
By using this equation, we can proceed as follows.
\begin{eqnarray}
    Eq.~\eqref{eq:pvspan_analysis1}&=&\frac{d\idop-\ketbra{I_d}}{d(d^2-1)}\otimes\idop+
    \frac{d\ketbra{I_{d}}-\idop}{d(d^2-1)}\otimes P^{T_2}\\
    &=&\frac{1}{d(d^2-1)}\big(d\idop\otimes\idop+d\ketbra{I_{d}}\otimes\ketbra{I_{d}}-\ketbra{I_{d}}\otimes\idop-\idop\otimes\ketbra{I_{d}}\big)\nonumber\\\\
    &=&\phi^+_{d}\otimes\phi^+_{d}+\frac{1}{d^2-1}(\idop-\phi^+_{d})\otimes(\idop-\phi^+_{d}).\nonumber\\
\end{eqnarray}
This proves Eq.~\eqref{eq:spanMFLC}.
\end{proof}

\subsection{Twisted canonical subspace}
\label{sec:MFLCTCS}
Let $\ket{\tau}=\idop^{(R_A)}\otimes L_1^{(R_B)}\ketUMES{R_AR_B}{d}$ and $\ket{L_2}=\idop^{(A)}\otimes L_2^{(B)}\ketUMES{AB}{d}$ with full rank operators $L_1\in\linop{\hh_{R_B}}$ and $L_2\in\linop{\hh_B}$.
We consider a subspace $\ww$ defined by

\begin{eqnarray}
    \ww&:=&\{\ket{\Xi}\in\mathcal{A}\otimes\mathcal{B}:\bra{\tau}_{R_AR_B}\ket{\Xi}\in\vspan{\{\ket{L_2}\}}\}\\
    &=&\{\ket{\Xi}\in\mathcal{A}\otimes\mathcal{B}:\braUMES{R_AR_B}{d}L_1^\dag\otimes L_2^{-1}\ket{\Xi}\in \vspan{\{\ketUMES{AB}{d}\}}\}\\
    &=&\left(L_1^{\dag}\right)^{-1}\otimes L_2\{\ket{\Xi}\in\mathcal{A}\otimes\mathcal{B}:\braUMES{R_AR_B}{d}\ket{\Xi}\in\vspan{\{\ketUMES{AB}{d}\}}\}\\
    &=&\left(\left(L_1^{\dag}\right)^{-1}\otimes L_2\right)\Csubspace{d}.
\end{eqnarray}

We also define a subspace $\ww^\circ$ as follows:
\begin{eqnarray}
        \ww^\circ&:=&\{\ket{\Xi}\in\mathcal{A}\otimes\mathcal{B}:\bra{\tau}_{R_AR_B}\ket{\Xi}=0\}\\
        &=&\left(L_1^{\dag}\right)^{-1}\otimes L_2\{\ket{\Xi}\in\mathcal{A}\otimes\mathcal{B}:\braUMES{R_AR_B}{d}\ket{\Xi}=0\}\\
        &=&\left(\left(L_1^{\dag}\right)^{-1}\otimes L_2\right)\CVsubspace{d}.
\end{eqnarray}

Since 
\begin{eqnarray}
\Segre{\mathcal{A}:\mathcal{B}}\cap\ww&=&\left(\left(L_1^{\dag}\right)^{-1}\otimes L_2\right)\left(\Segre{\mathcal{A}:\mathcal{B}}\cap\Csubspace{d}\right),\\
\left(\Segre{\mathcal{A}:\mathcal{B}}\cap\ww\right)\setminus\ww^\circ&=&\left(\left(L_1^{\dag}\right)^{-1}\otimes L_2\right)\left(\left(\Segre{\mathcal{A}:\mathcal{B}}\cap\Csubspace{d}\right)\setminus\ww^\circ\right),
\end{eqnarray}
the three propositions shown in this section remain valid if we replace $\Csubspace{d}$, $\CVsubspace{d}$ and $\CMFLC{d}$ with $\ww$, $\ww^\circ$ and $\left(\idop^{(AR_A)}\otimes\left( L_1^{\dag}\right)^{-1}\otimes L_2\right)\CMFLC{d}$, respectively.

\subsection{Extended canonical subspace}
\label{appendix:MFLCexCS}
\begin{figure}[ht]
    \centering
    \includegraphics[width=5.5cm]{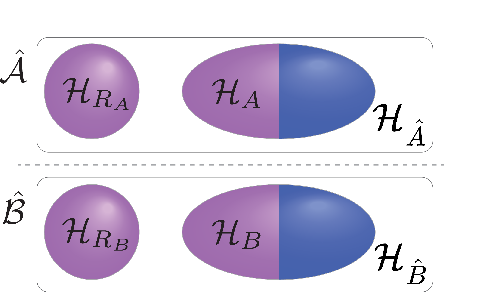}
    \caption{Hilbert spaces where the extended canonical subspace is defined. We consider the product vectors between $\hat{\mathcal{A}}$ and $\hat{\mathcal{B}}$.}
    \label{fig:Hspacesext}
\end{figure}
Suppose that the Hilbert spaces $\hh_A$ and $\hh_B$ are embedded in an extended Hilbert space, $\hh_A\subseteq\hh_{\hat{A}}$ and $\hh_B\subseteq\hh_{\hat{B}}$ (see Fig.~\ref{fig:Hspacesext}), and define two subspaces
\begin{eqnarray}
\hat{\vv}&:=&\{\ket{\Xi}\in\hat{\mathcal{A}}\otimes\hat{\mathcal{B}}:\braUMES{R_AR_B}{d}\ket{\Xi}\in\vspan{\{\ketUMES{AB}{d}\}}\},\\
\hat{\vv}^\circ&:=&\{\ket{\Xi}\in\hat{\mathcal{A}}\otimes\hat{\mathcal{B}}:\braUMES{R_AR_B}{d}\ket{\Xi}=0\},
\end{eqnarray}
where $\hat{\mathcal{A}}=\hh_{\hat{A}}\otimes\hh_{R_A}$, $\hat{\mathcal{B}}=\hh_{R_B}\otimes\hh_{\hat{B}}$.

Since we can verify that
\begin{eqnarray}
	\Segre{\hat{\mathcal{A}}:\hat{\mathcal{B}}}\cap\hat{\vv}&=&\{\ket{\hat{A}}\ket{\hat{B}}:[\hat{A}][\hat{B}]^T\in\{0,\idop_d\oplus0\}\}\\
	\label{eq:extfirstCS1}
      &=&\{\ket{\hat{A}}\ket{\hat{B}}:[\hat{A}][\hat{B}]^T=0\}\cup\{\ket{\hat{A}}\ket{\hat{B}}:[\hat{A}][\hat{B}]^T=\idop_d\oplus0\},
\end{eqnarray}
where $\ket{\hat{A}}_{AR_A}=(\hat{A}\otimes\idop^{(R_A)})\ketUMES{R_AR_A}{d}$ and $\ket{\hat{B}}_{R_BB}=(\idop^{(R_B)}\otimes \hat{B})\ketUMES{R_BR_B}{d}$ for $\hat{A}\in\linop{\hh_{R_A}:\hh_A}$ and $\hat{B}\in\linop{\hh_{R_B}:\hh_B}$, and $[\hat{A}]$ and $[\hat{B}]$ are matrix representations of $\hat{A}$ and $\hat{B}$ with respect to the computational basis defining the maximally entangled states and an orthonormal basis of the orthogonal complement $\hh_A^\bot$ (or $\hh_B^\bot$) of $\hh_A$ (or $\hh_B$) within $\hh_{\hat{A}}$ (or $\hh_{\hat{B}}$).
Here, $\idop_d$ represents the $d$ by $d$ identity matrix, which acts on a subspace $\hh_A$ (or $\hh_B$) in $\hh_{\hat{A}}$ (or $\hh_{\hat{B}}$).
Since $[\hat{A}][\hat{B}]^T=\idop_d\oplus 0$ implies that $[\hat{A}]$ and $[\hat{B}]$ can be decomposed as
    \begin{equation}
    \label{eq:embedLO}
        [\hat{A}]=\begin{pmatrix}
            [A]\\0
        \end{pmatrix},\ \ 
        [\hat{B}]=\begin{pmatrix}
            [B]\\0
        \end{pmatrix}
    \end{equation}
    by using $d$ by $d$ matrices $[A]$ and $[B]$ satisfying $[A][B]^T=\idop$, we obtain
\begin{equation}
	\label{eq:extfirstCS2}
	\{\ket{\hat{A}}\ket{\hat{B}}:[\hat{A}][\hat{B}]^T=\idop_d\oplus0\}=(V_A\otimes\idop^{(R_AR_B)}\otimes V_B)\{\ket{A}\ket{B}:[A][B]^T=\idop\},
\end{equation}
where $V_A:\hh_A\rightarrow\hh_{\hat{A}}$ and $V_B:\hh_B\rightarrow\hh_{\hat{B}}$ are isometry operators that can be represented by $V_A=V_B=\sum_{i=0}^{d-1}\ketbra{i}$ with the computational basis $\{\ket{i}\}_{i=0}^{d-1}$ of $\hh_A$ or $\hh_B$ defining the maximally entangled state in $\hh_A\otimes\hh_B$.

By using Eq.~\eqref{eq:extfirstCS1} and Eq.~\eqref{eq:extfirstCS2}, we can verify that Proposition \ref{prop:firstCS} remains valid if we replace $\Csubspace{d}$, $\CVsubspace{d}$ and $\CMFLC{d}$ with $\hat{\vv}$, $\hat{\vv}^\circ$ and $(V_A\otimes\idop^{(R_AR_B)}\otimes V_B)\CMFLC{d}$, respectively.
Since
\begin{eqnarray}
\hat{\mathbb{V}}&:=&\left(\Segre{\hat{\mathcal{A}}:\hat{\mathcal{B}}}\cap\hat{\vv}\right)\setminus\hat{\vv}^\circ\\
&=&(V_A\otimes\idop^{(R_AR_B)}\otimes V_B)\{\ket{A}\ket{B}:[A][B]^T=\idop\},
\end{eqnarray}
we find that Proposition \ref{prop:secondCS} and Proposition \ref{prop:thirdCS} remain valid if we replace $\mathbb{V}$ and $\CMFLC{d}$ with $\hat{\mathbb{V}}$ and $(V_A\otimes\idop^{(R_AR_B)}\otimes V_B)\CMFLC{d}$, respectively.

\subsection{Extended and twisted canonical subspace}
\label{sec:MFLCexttw}
Let $\ket{\tau}=\idop^{(R_A)}\otimes L_1^{(R_B)}\ketUMES{R_AR_B}{d}$ and $\ket{L_2}=\idop^{(A)}\otimes L_2^{(B)}\ketUMES{AB}{d}$ with full-rank operators $L_1^{(R_B)}\in\linop{\hh_{R_B}}$ and $L_2^{(B)}\in\linop{\hh_B}$.

Consider two Hilbert spaces $\hh_A$ and $\hh_B$ that are embedded in an extended Hilbert space as $\hh_A\subseteq\hh_{\hat{A}}$ and $\hh_B\subseteq\hh_{\hat{B}}$, and define two subspaces
\begin{eqnarray}
\hat{\ww}&:=&\{\ket{\Xi}\in\hat{\mathcal{A}}\otimes\hat{\mathcal{B}}:\bra{\tau}_{R_AR_B}\ket{\Xi}\in\vspan{\{\ket{L_2}\}}\},\\
\hat{\ww}^\circ&:=&\{\ket{\Xi}\in\hat{\mathcal{A}}\otimes\hat{\mathcal{B}}:\bra{\tau}_{R_AR_B}\ket{\Xi}=0\},
\end{eqnarray}
where $\hat{\mathcal{A}}=\hh_{\hat{A}}\otimes\hh_{R_A}$, $\hat{\mathcal{B}}=\hh_{R_B}\otimes\hh_{\hat{B}}$.
By using a similar calculation to those in the previous cases, we can show that the MFLC of $\Segre{\hat{\mathcal{A}}:\hat{\mathcal{B}}}\cap(\hat{\ww}\setminus\hat{\ww}^\circ)$ is
\begin{equation}
    \pp:=\left(V_A\otimes\idop^{(R_A)}\otimes\left(L_1^\dag\right)^{-1}\otimes\left(V_BL_2\right)\right)\CMFLC{d}.
\end{equation}

\section{Independence of the success probability from the input state}
\label{appendix:constsuccessprob}
In general, we say a family of CP maps $\{ \mathcal{E}'_m : \linop{\mathcal{H}_1} \rightarrow \linop{\mathcal{H}_2} \}_{m \in \Sigma}$ probabilistically implements an instrument $\{ \mathcal{E}_m : \linop{\mathcal{H}_1} \rightarrow \linop{\mathcal{H}_2} \}_{m \in \Sigma}$ if there is a function $p:\mathbf{D}(\mathcal{H}_1) \rightarrow [0,1]$ such that
\begin{equation}
    \mathcal{E}'_m(\rho) = p(\rho) \mathcal{E}_m (\rho), \qquad (\forall \rho \in \mathbf{D}(\mathcal{H}_1), ~\forall m \in \Sigma).
\end{equation}
In this appendix, we show that if the instrument $\{ \mathcal{E}_m \}$ is taken from the list in Table~\ref{table:Nop}, then the above $p$ must be a constant function for any probabilistic implementation $\{ \mathcal{E}'_m \}$ of $\{ \mathcal{E}_m \}$ (The precise statement will be given in Theorem~\ref{thm:constant_instrument}).

In the case of entanglement-assisted implementation of quantum operations considered in the main text, the separable instrument $\{ \mathcal{S}_m : \linop{(\mathcal{H}_{A_1} \otimes \mathcal{H}_{R_A}) \otimes (\mathcal{H}_{B_1} \otimes \mathcal{H}_{R_B})} \rightarrow \linop{\mathcal{H}_2} \}_{m \in \Sigma \cup \{ \texttt{fail} \}}$ induces a family of CP maps $\{ \mathcal{E}'_m : \linop{\mathcal{H}_1} \rightarrow \linop{\mathcal{H}_2} \}_{m \in \Sigma}$ defined by
\begin{equation}
    \mathcal{E}'_m (A) := \mathcal{S}_m(A \otimes \tau). \qquad (\forall A \in \linop{\mathcal{H}_1} , \forall m \in \Sigma)
\end{equation}
(Note that the index $\texttt{fail}$ no longer appears in the family.) The argument in this section does not depend on the form of $\{ \mathcal{E}'_m \}$, whether it is given by the entanglement-assisted form or not.

\begin{lemma}\label{lem:constant}
    Let $\mathcal{E} : \linop{\mathcal{H}_1} \rightarrow \linop{\mathcal{H}_2} $ be a CPTP map and $\mathcal{E}': \linop{\mathcal{H}_1} \rightarrow \linop{\mathcal{H}_2}$ be its probabilistic implementation, that is, a function $p:\mathbf{D}(\mathcal{H}_1) \mapsto [0,1]$ exists and satisfies $\mathcal{E}'(\rho) = p(\rho) \mathcal{E}(\rho)$ for any $\rho$. If there is a state $\rho_\ast \in \mathbf{D}(\mathcal{H}_1)$ such that $\mathcal{E}(\rho) \neq \mathcal{E}(\rho_\ast)$ whenever $\rho \neq \rho_\ast$, then $p$ is a constant function.
\end{lemma}
\begin{proof}
 Let $\rho \in \dop{\mathcal{H}_1}$ be a state distinct from $\rho_\ast$. We have
\begin{equation}
    \mathcal{E}'(\rho + \rho_\ast) = \mathcal{E}'(\rho) + \mathcal{E}'(\rho_\ast) = p(\rho) \mathcal{E}(\rho) + p(\rho_\ast) \mathcal{E}(\rho_\ast),
\end{equation}
from the linearity of $\mathcal{E}'$. On the other hand, we also have
\begin{align}
    \mathcal{E}'(\rho + \rho_\ast) &= 2 \mathcal{E}' \left( \frac{\rho + \rho_\ast}{2} \right) = 2p \left( \frac{\rho + \rho_\ast}{2} \right) \mathcal{E} \left( \frac{\rho + \rho_\ast}{2} \right) \\
    &= p \left( \frac{\rho + \rho_\ast}{2} \right) \mathcal{E}(\rho) + p \left( \frac{\rho + \rho_\ast}{2} \right) \mathcal{E}(\rho_\ast),
\end{align}
by the linearity of $\mathcal{E}$. By equating these two expressions, we arrive at
\begin{equation}
   \left( p(\rho) - p \left( \frac{\rho+\rho_\ast}{2} \right) \right) \mathcal{E}(\rho) =   -\left( p(\rho_\ast) - p \left( \frac{\rho+\rho_\ast}{2} \right) \right) \mathcal{E}(\rho_\ast).
\end{equation}
Since $\mathcal{E}$ is a trace preserving map, by taking the trace of both sides, we have
\begin{equation}
   p(\rho) - p \left( \frac{\rho+\rho_\ast}{2} \right)  =    -p(\rho_\ast) + p \left( \frac{\rho+\rho_\ast}{2} \right).
\end{equation}
Because $\mathcal{E}(\rho)$ and $\mathcal{E}(\rho_\ast)$ are different by assumption, both sides of the above equality must be zero, which implies
\begin{equation}
    p(\rho) = p \left( \frac{\rho + \rho_\ast}{2}  \right) = p(\rho_\ast).
\end{equation}
Since this holds for any state $\rho$ ($\neq \rho_\ast$), the function $p$ takes the constant value $p(\rho_\ast)$.
\end{proof}

\begin{lemma}\label{lem:list}
    Let $\{ \mathcal{E}_m : \linop{\mathcal{H}_1} \rightarrow \linop{\mathcal{H}_2} \}_{m \in \Sigma}$ be any instrument from the list in Table~\ref{table:Nop}. Define a linear map $\mathcal{E}:\linop{\mathcal{H}_1} \rightarrow \mathcal{L} (\mathcal{H}_2 \otimes \mathbb{C}^\Sigma)$ by
    \begin{equation}
    \label{eq:conditional_operation}    \mathcal{E}(\rho) = \sum_m\mathcal{E}_m (\rho) \otimes \ketbra{m}.
    \end{equation}
Then there exists a state $\rho_\ast \in \mathbf{D}(\mathcal{H}_1)$ such that $\mathcal{E}(\rho ) \neq \mathcal{E}(\rho_\ast )$ whenever $\rho \neq \rho_\ast$.
\end{lemma}
\begin{proof}
The instruments in the list in Table~\ref{table:Nop} that have an input state are unitary channels, rank-1 POVMs, and verification of pure states.
\begin{description}
    \item[unitary channel] For unitary channels, $\Sigma$ is a singleton, and $\mathcal{E}(\rho) = U \rho U^\dagger$. Any state in $\mathbf{D}(\mathcal{H}_1)$ can play the role of $\rho_\ast$ since unitary channels are bijective.
    \item[rank-1 POVM]  Let $\{ \ketbra{M_m} \}_{m =1,\ldots,n}$ be a rank-1 POVM. The CPTP map \eqref{eq:conditional_operation} is defined by $\mathcal{E}(\rho) = \sum_m \bra{M_m} \rho \ket{M_m} \ketbra{m}$. In this case, we can take, e.g., $\rho_\ast = \ketbra{M_1}/\braket{M_1}{M_1}$. This is the unique state that makes $\bra{1} \mathcal{E}(\rho_\ast) \ket{1} = \bra{M_1} \rho_\ast \ket{M_1}$ equal to its maximum value $\braket{M_1}{M_1}$, so we have $\bra{1} \mathcal{E}(\rho_\ast) \ket{1} > \bra{1} \mathcal{E}(\rho) \ket{1}$ and hence $\mathcal{E}(\rho) \neq \mathcal{E}(\rho_\ast)$ whenever $\rho \neq \rho_\ast$. 
  \item[verification of pure state] The pure state verification is described by a POVM $\{ M_\texttt{accept} := q\phi,~M_\texttt{reject} := \idop - q \phi \}$, with some $q \in [0,1]$ and pure state $\phi \in \mathbf{D}(\mathcal{H}_1)$. The CPTP map \eqref{eq:conditional_operation} is defined by $\mathcal{E}(\rho) = q \tr{\phi \rho} \ketbra{a} + (1-q \tr{\phi \rho})\ketbra{r}$. Since $\bra{a} \mathcal{E}(\rho) \ket{a} = q \tr{\phi \rho}$ reaches its maximum value $q$ if and only if $\rho = \phi$, we have $\mathcal{E}(\rho) \neq \mathcal{E}(\phi)$ whenever $\rho \neq \phi$. 
\end{description} 
\end{proof}

\begin{theorem}\label{thm:constant_instrument}
    Let $\{ \mathcal{E}_m : \linop{\mathcal{H}_1} \rightarrow \linop{\mathcal{H}_2} \}_{m \in \Sigma}$ be any instrument from the list in Table 1. If $\{ \mathcal{E}'_m : \linop{\mathcal{H}_1} \rightarrow \linop{\mathcal{H}_2} \}_{m \in \Sigma}$ probabilistically implements $\{ \mathcal{E}_m : \linop{\mathcal{H}_1} \rightarrow \linop{\mathcal{H}_2} \}_{m \in \Sigma}$ in the sense that a function $p:\mathbf{D}(\mathcal{H}_1) \rightarrow [0,1]$ exists and satisfies $\mathcal{E}'_m(\rho) = p(\rho) \mathcal{E}_m (\rho)$ for all $m \in \Sigma$, then $p$ must be a constant function.
\end{theorem}
\begin{proof}
 Define linear maps $\mathcal{E},\mathcal{E}':\linop{\mathcal{H}_1} \rightarrow \mathcal{L}(\mathcal{H}_2  \otimes \mathbb{C}^\Sigma )$ by Eq.~\eqref{eq:conditional_operation} and by
 \begin{equation}
        \mathcal{E}'(\rho) := \sum_m \mathcal{E}'_m (\rho) \otimes \ketbra{m}
 \end{equation}
respectively. $\mathcal{E}$ is a CPTP map. From $\mathcal{E}'_m(\rho) = p(\rho) \mathcal{E}_m (\rho)$ ($\forall m \in \Sigma$) we obtain
\begin{equation}
    \mathcal{E}'(\rho) = \sum_m p(\rho) \mathcal{E}_m (\rho) \otimes \ketbra{m} = p(\rho) \sum_m \mathcal{E}_m (\rho) \otimes \ketbra{m} = p(\rho) \mathcal{E}(\rho).
\end{equation}
From Lemma~\ref{lem:list}, there exists a state $\rho_\ast$ such that $\mathcal{E}(\rho ) \neq \mathcal{E}(\rho_\ast )$ whenever $\rho \neq \rho_\ast$. So Lemma~\ref{lem:constant} applies to the CPTP map $\mathcal{E}$ and its probabilistic implementation $\mathcal{E}'$, and implies that $p$ is a constant function.

\end{proof}

\section{Extraction of range constraint}
\label{appendix:range_constraint}
\begin{lemma}
    \label{lemma:range_constraint}
    Let $S\in\pos{\hh\otimes\hh_R}$, $\tau\in\pos{\hh_R}$ and $E\in\pos{\hh}$.
    If there exists $p\in\rr$ such that $\ptr{R}{S\tau}=pE$, then $\range{S}\subseteq\ww$, where
    \begin{equation}
        \ww:=\{\ket{\Xi}\in\hh\otimes\hh_R:\forall\ket{\eta}\in\range{\tau},\braket{\eta}{\Xi}\in\range{E}\}.
    \end{equation}
    Moreover, if ${\rm rank}(E)=1$, the converse holds.
\end{lemma}
In the proof of this lemma, we use the following auxiliary lemma. Although this fact is standard in matrix analysis, we include a proof for completeness.
\begin{lemma}
\label{lemma:POSspan}
    $\range{\sum_{i\in I}\ketbra{\Theta_i}}=\vspan{\{\ket{\Theta_i}\}_{i\in I}}$ for any finite set $\{\ket{\Theta_i}\}_{i\in I}\subseteq\hh$ of vectors.
\end{lemma}
\begin{proof}
    Since $\range{\sum_{i\in I}\ketbra{\Theta_i}}\subseteq\vspan{\{\ket{\Theta_i}\}_{i\in I}}$ is trivial, we show the converse by contradiction.
    Assume that $\range{\sum_{i\in I}\ketbra{\Theta_i}}\subsetneq\vspan{\{\ket{\Theta_i}\}_{i\in I}}$. Then, there exists a unit vector $\ket{\phi}\in\hh$ such that $\ket{\phi}\in\vspan{\{\ket{\Theta_i}\}_{i\in I}}$ and $\bra{\phi}\sum_{i\in I}\ketbra{\Theta_i}\ket{\phi}=0$.
    Since the second condition implies $\forall i,\braket{\phi}{\Theta_i}=0$, this contradicts the first condition. This completes the proof.
\end{proof}
\begin{proof}[Proof of Lemma~\ref{lemma:range_constraint}]
    Let $S$ and $\tau$ be diagonalized as $S=\sum_x\ketbra{\Xi_x}$ and $\tau=\sum_yp_y\ketbra{\eta_y}$ ($p_y>0$), respectively.
    Since $pE=\ptr{R}{S\tau}=\sum_{x,y}p_y\braket{\eta_y}{\Xi_x}\braket{\Xi_x}{\eta_y}$, we obtain
    \begin{equation}
        \vspan{\{\braket{\eta_y}{\Xi_x}\}_{x,y}}=\range{pE}\subseteq\range{E}
    \end{equation}
    by using Lemma~\ref{lemma:POSspan} with $\ket{\Theta_i}=\braket{\eta_y}{\Xi_x}(\in\hh)$.
    This implies that $\braket{\eta}{\Xi_x}\in\range{E}$ for any $x$ and $\ket{\eta}\in\range{\tau}$. This proves that $\range{S}\subseteq\ww$.

    Conversely, $\range{S}\subseteq\ww$ implies that $\braket{\eta_y}{\Xi_x}\in\range{E}$ for all $x$ and $y$. Since $\ptr{R}{S\tau}=\sum_{x,y}p_y\braket{\eta_y}{\Xi_x}\braket{\Xi_x}{\eta_y}$, we obtain $\exists p\in\rr,\ptr{R}{S\tau}=pE$ if ${\rm rank}(E)=1$.
\end{proof}

\section{Computing a lower bound based on $\epsilon$-net}
\label{appendix:enet}
Here, we provide an algorithm to obtain a lower bound on Eq.~\eqref{eq:SEPoptformat}.
First, we show that for any finite set $\left\{\ket{\Pi_x^{(m)}}\in\mathbb{E}_m\right\}_x$ of product vectors,
\begin{equation}
\label{eq:SEPoptformatnet}
{\rm Eq.}~\eqref{eq:SEPoptformat}\geq\max\left\{\min_{m\in\Sigma} \frac{\tr{S_m\overline{\tau}}}{\lpnorm{2}{\ket{E_m}}^2(1+\delta)}:
\begin{array}{l}
\forall m\in\Sigma,x,p_{x}^{(m)}\geq0,\\
\forall m\in\Sigma,S_m=\sum_xp_{x}^{(m)}\ketbra{\Pi_x^{(m)}},\\
\Delta=\idop-\sum_{m\in\Sigma}\ptr{2}{S_m},\\
 \min_{S\in \SEP{\hh_{A_1}\otimes\hh_{R_A}:\hh_{R_B}\otimes\hh_{B_1}}}\lpnorm{1}{\Delta-S}\leq \delta
\end{array}
\right\},
\end{equation}
where $\lpnorm{p}{X}:=\tr{(XX^\dag)^\frac{p}{2}}^\frac{1}{p}$ is the Schatten $p$-norm.
This is because we can show $\frac{1}{1+\delta}\left\{S_m\right\}_{m\in\Sigma}$ is a feasible solution of the optimization problem given in the right-hand side of Eq.~\eqref{eq:SEPoptformat} when $\{p_x^{(m)},S_m,\Delta,\delta\}$ is the one given in the right-hand side of Eq.~\eqref{eq:SEPoptformatnet} as follows:
\begin{itemize}
 \item Since $\mathbb{E}_m\subseteq\Segre{\hat{\mathcal{A}}:\hat{\mathcal{B}}}\cap\ww_m$, we can verify $\frac{1}{1+\delta}S_m\in\SEP{\hat{\mathcal{A}}:\hat{\mathcal{B}}}$ and $\range{\frac{1}{1+\delta}S_m}\subseteq\ww_m$.
 \item Let $S^*\in\SEP{\hh_{A_1}\otimes\hh_{R_A}:\hh_{R_B}\otimes\hh_{B_1}}$ achieve the minimum, i.e.,   $\lpnorm{1}{\Delta-S^*}=\min_{S}\lpnorm{1}{\Delta-S}$ in Eq.~\eqref{eq:SEPoptformatnet}. Since $\lpnorm{1}{\Delta-S^*}\leq\delta$ implies that $\delta\idop+(\Delta-S^*)$ is an element of the separable cone~\cite{GB02}, $I-\sum_m\frac{1}{1+\delta}\ptr{2}{S_m}=\frac{1}{1+\delta}\left(\delta I+\Delta-S^*+S^*\right)$ is also an element of the separable cone.
\end{itemize}

 Next, we can verify, by definition, that for any finite subsets $\{\phi_x\in\puredop{\hh_{A_1}\otimes\hh_{R_A}}\}_x$ and $\{B_x\in\pos{\hh_{R_B}\otimes\hh_{B_1}}\}_x$, 
\begin{equation}
\min_{S\in \SEP{\hh_{A_1}\otimes\hh_{R_A}:\hh_{R_B}\otimes\hh_{B_1}}}\lpnorm{1}{\Delta-S}\leq\lpnorm{1}{\Delta-\sum_{x}\phi_x\otimes B_x}. 
\end{equation}
Moreover, since $\lpnorm{1}{X}=\min_{P\geq0,P\geq X}2\tr{P}-\tr{X}$ for any Hermitian operator $X$, 
\begin{equation}
\lpnorm{1}{\Delta-\sum_{x}\phi_x\otimes B_x}\leq2\tr{P}+\sum_{x}\tr{ B_x}-\tr{\Delta}
\end{equation}
for any $P\geq0$ such that $P+\sum_{x}\phi_x\otimes B_x\geq\Delta$.

Thus, we obtain the following lower bound:
\begin{equation}
\label{eq:SEPoptformatnetjoined}
{\rm Eq.}~\eqref{eq:SEPoptformat}\geq\max\left\{\min_{m\in\Sigma} \frac{\tr{S_m\overline{\tau}}}{\lpnorm{2}{\ket{E_m}}^2(1+\delta)}:
\begin{array}{l}
\forall m\in\Sigma,x,p_{x}^{(m)}\geq0,\\
\forall m\in\Sigma,S_m=\sum_xp_{x}^{(m)}\ketbra{\Pi_x^{(m)}},\\
\Delta=\idop-\sum_{m\in\Sigma}\ptr{2}{S_m},\\
 \delta=2\tr{P}+\sum_{x}\tr{ B_x}-\tr{\Delta},\\
 P\geq0,P+\sum_{x}\phi_x\otimes B_x\geq\Delta, B_x\geq0
\end{array}
\right\}.
\end{equation}
This is because $\{p_x^{(m)},S_m,\Delta,\delta\}$ is a feasible solution of the optimization problem given in the right-hand0side of Eq.~\eqref{eq:SEPoptformatnet} when $\{p_x^{(m)},S_m,\Delta,\delta,P, B_x\}$ is the one given in the right-hand side of Eq.~\eqref{eq:SEPoptformatnetjoined}.
Note that the right-hand side converges to Eq.~\eqref{eq:SEPoptformat} if we use finer $\epsilon$-nets $\left\{\ket{\Pi_x^{(m)}}\right\}_x$ of $\mathbb{E}_m$ and $\{\phi_x\}_x$ of $\puredop{\hh_{A_1}\otimes\hh_{R_A}}$.
However, the right-hand side cannot be computed by an SDP directly since the target function is not linear.

Alternatively, our algorithm solves the following SDP
\begin{equation}
\label{eq:SEPoptformatnetSDP}
\max\left\{\min_{m\in\Sigma} \frac{\tr{S_m\overline{\tau}}}{\lpnorm{2}{\ket{E_m}}^2}-\delta:
\begin{array}{l}
\forall m\in\Sigma,x,p_{x}^{(m)}\geq0,\\
\forall m\in\Sigma,S_m=\sum_xp_{x}^{(m)}\ketbra{\Pi_x^{(m)}},\\
\Delta=\idop-\sum_{m\in\Sigma}\ptr{2}{S_m},\\
 \delta=2\tr{P}+\sum_{x}\tr{ B_x}-\tr{\Delta},\\
 P\geq0,P+\sum_{x}\phi_x\otimes B_x\geq\Delta, B_x\geq0
\end{array}
\right\},
\end{equation}
and compute $r(\tau):=\min_{m\in\Sigma} \frac{\tr{S^*_m\overline{\tau}}}{\lpnorm{2}{\ket{E_m}}^2(1+\delta^*)}$ by using $S^*_m$ and $\delta^*$ attaining the maximum of Eq.~\eqref{eq:SEPoptformatnetSDP}. We can find that $r(\tau)$ is a lower bound on the right-hand side of Eq.~\eqref{eq:SEPoptformatnetjoined}.

By using lower bounds $\{r(\tau_\lambda)\}_\lambda$ for finite resource states $\{\tau_\lambda\}_\lambda$, we can obtain lower bounds $r(\tau)$ for any $\tau$ as follows:
Assume we can transform $\tau$ into an ensemble $\{(p_\lambda,\tau_\lambda)\}_\lambda$ by using an LOCC instrument $\{\mathcal{L}_\lambda\}_\lambda$, i.e., $\mathcal{L}_\lambda(\tau)=p_\lambda\tau_\lambda$ for all $\lambda$.
Let $\{\mathcal{S}^{(\lambda)}_m\}_m$ be a separable instrument satisfying $\mathcal{S}^{(\lambda)}_m(\rho\otimes\tau_\lambda)=p(\{\mathcal{E}_m\}_m,\tau_\lambda)\mathcal{E}_m(\rho)$ for all $\lambda$, $\rho$, and $m\in\Sigma$.
Then, we can verify that $\{\mathcal{S}_m=\sum_\lambda\mathcal{S}^{(\lambda)}_m\circ\mathcal{L}_\lambda\}_m$ is a separable instrument and satisfies
\begin{equation}
 \mathcal{S}_m(\rho\otimes\tau)=\sum_\lambda p_\lambda\mathcal{S}^{(\lambda)}_m(\rho\otimes\tau_\lambda)=\sum_\lambda p_\lambda p(\{\mathcal{E}_m\}_m,\tau_\lambda)\mathcal{E}_m(\rho)
\end{equation}
for all $\rho$ and $m\in\Sigma$. 
Thus, $p(\{\mathcal{E}_m\}_m,\tau)\geq\sum_\lambda p_\lambda p(\{\mathcal{E}_m\}_m,\tau_\lambda)\geq\sum_\lambda p_\lambda r(\tau_\lambda)$.

Accordingly, we can show the following proposition.
\begin{proposition}
\label{prop:concavity_tradeoff}
 Let $\ket{\tau(s)}=\sqrt{1-s}\ket{00}+\sqrt{s}\ket{11}$, where $s\in[0,\frac{1}{2}]$. Then, $f(s)=p(\{\mathcal{E}_m\}_m,\tau(s))$ is concave, where $p(\{\mathcal{E}_m\}_m,\tau)$ is defined in Eq.~\eqref{eq:SEPoptformat}.
\end{proposition}
\begin{proof}
 For any $s_1,s_2\in[0,\frac{1}{2}]$ and $p\in[0,1]$, Theorem 1 in~\cite{DM99} implies that  $\tau(ps_1+(1-p)s_2)$ can be transformed into $\{(p,\tau(s_1)),(1-p,\tau(s_2))\}$ by using an LOCC instrument. Thus,
\begin{eqnarray}
 f\left(ps_1+(1-p)s_2\right)&=&p\left(\{\mathcal{E}_m\}_m,\tau\left(ps_1+(1-p)s_2\right)\right)\\
 &\geq& pp(\{\mathcal{E}_m\}_m,\tau(s_1))+(1-p)p(\{\mathcal{E}_m\}_m,\tau(s_2))\\
 &=&pf(s_1)+(1-p)f(s_2).
\end{eqnarray}
 This completes the proof.
\end{proof}

Utilizing this proposition, we take the convex hull of the set $\{r(\tau_x)\}_x$, which are numerically obtained lower bounds for finite resource states $\{\tau_x\}_x$, to serve as a lower bound on $p(\{\mathcal{E}_m\}_m,\tau)$ presented in Fig.~\ref{fig:Entcost}.

\section{SDPs in numerical experiments}
Here, we summarize the SDPs used in the numerical experiments.
We wrote the SDPs using Python and utilized the PICOS ~\cite{PICOS} and QICS~\cite{QICS} packages to solve them.

\subsection{Non-local unitary channels}
\label{appendix:unitary}
In Fig.~\ref{fig:Entcost} (b), we compute three upper bounds on the success probability $p(\mathcal{U},\tau)$ to implement nonlocal unitary channel $\mathcal{U}$ by SEP channels with a resource state $\ket{\tau}$, given in Eq.~\eqref{eq:SEPoptunitary} and Eq.~\eqref{eq:SEPoptunitaryMFLC}. Each upper bound is computed by solving the following SDPs:
\begin{itemize}
 \item \textbf{PPT + MFLC}:
 \begin{eqnarray}
 &&\max \frac{\tr{S\overline{\tau}_\theta}}{d_Ad_B}\\
&s.t.&S\in\PPT{\hat{\mathcal{A}}:\hat{\mathcal{B}}},\range{S}\subseteq\pp\\
&&\idop-\ptr{2}{S}\in\PPT{\hh_{A_1}\otimes\hh_{R_A}:\hh_{R_B}\otimes\hh_{B_1}},
\end{eqnarray}
where $\pp$ is defined in Section~\ref{sec:unitary}.

 \item \textbf{PPT (DPS 1st Lv.)}:
 \begin{eqnarray}
 &&\max \frac{\tr{S\overline{\tau}_\theta}}{d_Ad_B}\\
&s.t.&S\in\PPT{\hat{\mathcal{A}}:\hat{\mathcal{B}}},\range{S}\subseteq\hat{\ww}\\
&&\idop-\ptr{2}{S}\in\PPT{\hh_{A_1}\otimes\hh_{R_A}:\hh_{R_B}\otimes\hh_{B_1}},
\end{eqnarray}
where $\hat{\ww}$ is defined in subsection~\ref{sec:unitary}.

 \item \textbf{DPS 2nd Lv.}:
 \begin{eqnarray}
 &&\max \frac{\tr{S\overline{\tau}_\theta}}{d_Ad_B}\\
&s.t.&S_{ext}\in\PPT{\hat{\mathcal{A}}:\hat{\mathcal{A}}':\hat{\mathcal{B}}},\range{S_{ext}}\subseteq\vee_{n=1}^2\hat{\mathcal{A}}\otimes\hat{\mathcal{B}},\\
&&S=\ptr{\hat{\mathcal{A}'}}{S_{ext}},\range{S}\subseteq\hat{\ww},\\
\label{eq:extraconstraint}
&&\range{S}\subseteq\range{V_A}\otimes \hh_{R_A}\otimes\hh_{R_B}\otimes\range{V_B},\\
&&R\in\PPT{\hh_{A_1}\otimes\hh_{R_A}:\hh_{A_1'}\otimes\hh_{R_A'}:\hh_{R_B}\otimes\hh_{B_1}},\\
&&\range{R}\subseteq\vee_{n=1}^2(\hh_{A_1}\otimes\hh_{R_A})\otimes(\hh_{R_B}\otimes\hh_{B_1}),\\
&&\idop-\ptr{2}{S}=\ptr{R_A' A_1'}{R},
\end{eqnarray}
where $\hat{\ww}$ is defined in subsection~\ref{sec:unitary} and we consider $\vee_{n=1}^2\hat{\mathcal{A}}$ and $\vee_{n=1}^2(\hh_{A_1}\otimes\hh_{R_A})$ are embedded in $\hat{\mathcal{A}}\otimes\hat{\mathcal{A}}'$ and $(\hh_{A_1}\otimes\hh_{R_A})\otimes(\hh_{A_1'}\otimes\hh_{R_A'})$, respectively.
Note that the original second level of the DPS hierarchy does not impose Eq.~\eqref{eq:extraconstraint}. We impose Eq.~\eqref{eq:extraconstraint} since we can assume $\range{S}\subseteq\pp(\subset\range{V_A}\otimes \hh_{R_A}\otimes\hh_{R_B}\otimes\range{V_B})$. Thus, this optimization problem can be regarded as a second level of the DPS hierarchy partially strengthened by the MFLC. This modification significantly reduces the size of the SDP.

\end{itemize}

\subsection{Non-local measurement}
\label{appendix:PVM}
In Fig.~\ref{fig:Entcost} (c), we compute three upper bounds on the success probability to implement the SJM by SEP channels with a resource state $\ket{\tau}$, given in Eq.~\eqref{eq:SEPoptPOVM} and Eq.~\eqref{eq:SEPoptPOVMMFLC}. Each upper bound is computed by solving the following SDPs:
\begin{itemize}
 \item \textbf{PPT* + MFLC}:
 \begin{eqnarray}
 &&\max \min_m\tr{S_m\tau_\theta}\\
&s.t.&\forall m,S_m\in\PPT{\hat{\mathcal{A}}:\hat{\mathcal{B}}},\range{S_m}\subseteq\pp_m,\\
&&R\in\PPT{\hat{\mathcal{A}}:\hat{\mathcal{A}}':\hat{\mathcal{B}}},
\range{R}\subseteq\vee_{n=1}^2\hat{\mathcal{A}}\otimes\hat{\mathcal{B}},\\
&&\idop-\sum_mS_m=\ptr{\hat{\mathcal{A}}'}{R},
\end{eqnarray}
where $\pp_m$ is defined in Section~\ref{sec:POVM} and we consider $\vee_{n=1}^2\hat{\mathcal{A}}$ is embedded in $\hat{\mathcal{A}}\otimes\hat{\mathcal{A}}'$.
Note that we partially used a condition resulting from the second level of the DPS hierarchy to improve the upper bound.

 \item \textbf{PPT (DPS 1st Lv.)}:
 \begin{eqnarray}
 &&\max \min_m\tr{S_m\tau_\theta}\\
&s.t.&\forall m,S_m\in\PPT{\hat{\mathcal{A}}:\hat{\mathcal{B}}},\range{S_m}\subseteq\hat{\ww}_m,\\
&&\idop-\sum_mS_m\in\PPT{\hat{\mathcal{A}}:\hat{\mathcal{B}}},
\end{eqnarray}
where $\hat{\ww}_m$ is defined in Section~\ref{sec:POVM}.

 \item \textbf{DPS 2nd Lv.}:
 \begin{eqnarray}
 &&\max \min_m\tr{S_m\tau_\theta}\\
&s.t.&\forall m,S_{ext,m}\in\PPT{\hat{\mathcal{A}}:\hat{\mathcal{A}}':\hat{\mathcal{B}}},\\
&&\forall m,\range{S_{ext,m}}\subseteq\vee_{n=1}^2\hat{\mathcal{A}}\otimes\hat{\mathcal{B}},\\
&&\forall m,S_m=\ptr{\hat{\mathcal{A}'}}{S_{ext,m}},\range{S_m}\subseteq\hat{\ww}_m,\\
&&R\in\PPT{\hat{\mathcal{A}}:\hat{\mathcal{A}}':\hat{\mathcal{B}}},
\range{R}\subseteq\vee_{n=1}^2\hat{\mathcal{A}}\otimes\hat{\mathcal{B}},\\
&&\idop-\sum_mS_m=\ptr{\hat{\mathcal{A}}'}{R},
\end{eqnarray}
where $\hat{\ww}_m$ is defined in Section~\ref{sec:POVM} and we consider $\vee_{n=1}^2\hat{\mathcal{A}}$ are embedded in $\hat{\mathcal{A}}\otimes\hat{\mathcal{A}}'$.

\end{itemize}

\subsection{State verification}
\label{appendix:verification}
In Fig.~\ref{fig:Entcost} (d), we compute three upper bounds on the maximum parameter $q(\phi,\tau)$ to deterministically implement a state verification of $\phi$ by SEP channels with a resource state $\ket{\tau}$, given in Eq.~\eqref{eq:SEPoptveri} and Eq.~\eqref{eq:SEPoptveriMFLC}. Each upper bound is computed by solving the following SDPs:
\begin{itemize}
 \item \textbf{PPT + MFLC}:
 \begin{eqnarray}
 &&\max \tr{S\tau_\theta}\\
&s.t.&S\in\PPT{\hat{\mathcal{A}}:\hat{\mathcal{B}}},\range{S}\subseteq\pp,\\
&&\idop-S\in\PPT{\hat{\mathcal{A}}:\hat{\mathcal{B}}},
\end{eqnarray}
where $\pp$ is defined in Section~\ref{sec:verification}.

 \item \textbf{PPT (DPS 1st Lv.)}:
 \begin{eqnarray}
 &&\max \tr{S\tau_\theta}\\
&s.t.&S\in\PPT{\hat{\mathcal{A}}:\hat{\mathcal{B}}},\range{S}\subseteq\hat{\ww},\\
&&\idop-S\in\PPT{\hat{\mathcal{A}}:\hat{\mathcal{B}}},
\end{eqnarray}
where $\hat{\ww}$ is defined in Section~\ref{sec:verification}.

 \item \textbf{DPS 2nd Lv.}:
 \begin{eqnarray}
 &&\max \tr{S\tau_\theta}\\
&s.t.&S_{ext}\in\PPT{\hat{\mathcal{A}}:\hat{\mathcal{A}}':\hat{\mathcal{B}}},\range{S_{ext}}\subseteq\vee_{n=1}^2\hat{\mathcal{A}}\otimes\hat{\mathcal{B}},\\
&&S=\ptr{\hat{\mathcal{A}'}}{S_{ext}},\range{S}\subseteq\hat{\ww},\\
&&R\in\PPT{\hat{\mathcal{A}}:\hat{\mathcal{A}}':\hat{\mathcal{B}}},\\
&&\idop-S=\ptr{\hat{\mathcal{A}}'}{R},
\end{eqnarray}
where $\hat{\ww}$ is defined in Section~\ref{sec:verification} and we consider $\vee_{n=1}^2\hat{\mathcal{A}}$ are embedded in $\hat{\mathcal{A}}\otimes\hat{\mathcal{A}}'$.

\end{itemize}

\subsection{Entanglement distillation}
\label{appendix:distillationSDP}
In Fig.~\ref{fig:Entcost} (a), we compute three upper bounds on the success probability $p(\psi_\theta,\tau)$ to distill a pure entangled state $\psi_\theta$ from a mixed state $\tau$ by SEP channels, given in Eq.~\eqref{eq:SEPoptdist} and Eq.~\eqref{eq:SEPoptdistMFLC}. Each upper bound is computed by solving the following SDPs:
\begin{itemize}
 \item \textbf{PPT + MFLC}:
 \begin{eqnarray}
 &&\max \tr{S\overline{\tau}}\\
&s.t.&S\in\PPT{\mathcal{A}:\mathcal{B}},\range{S}\subseteq\pp,\\
&&\idop-\ptr{AB}{S}\in\PPT{\hh_{R_A}:\hh_{R_B}},
\end{eqnarray}
where $\pp$ is defined in Proposition~\ref{prop:MFLCdist}.

 \item \textbf{PPT (DPS 1st Lv.)}:
 \begin{eqnarray}
 &&\max \tr{S\overline{\tau}}\\
&s.t.&S\in\PPT{\mathcal{A}:\mathcal{B}},\range{S}\subseteq\ww,\\
&&\idop-\ptr{AB}{S}\in\PPT{\hh_{R_A}:\hh_{R_B}},
\end{eqnarray}
where $\ww$ is defined in Section~\ref{sec:distillation}.

 \item \textbf{PPT (DPS 2nd Lv.)}:
 \begin{eqnarray}
 &&\max \tr{S\overline{\tau}}\\
&s.t.&S_{ext}\in\PPT{\mathcal{A}:\mathcal{A}':\mathcal{B}},\range{S_{ext}}\subseteq\vee_{n=1}^2\mathcal{A}\otimes\mathcal{B},\\
&&S=\ptr{\mathcal{A}'}{S_{ext}},\range{S}\subseteq\ww,\\
&&\idop-\ptr{AB}{S}\in\PPT{\hh_{R_A}:\hh_{R_B}},
\end{eqnarray}
where $\ww$ is defined in Section~\ref{sec:distillation} and we consider $\vee_{n=1}^2\mathcal{A}$ are embedded in $\mathcal{A}\otimes\mathcal{A}'$.

\end{itemize}

\section{Distillation protocol}
\subsection{From a mixed state $\tau$ in $\dop{\cdim{3}\otimes\cdim{3}}$ into a pure state $\psi_\theta$}
\label{appendix:distillation}
Here, we construct a SEP channel $\mathcal{S}:\linop{\hh_{R_A}\otimes\hh_{R_B}}\rightarrow\linop{\hh_A\otimes\hh_B}$ for distilling a pure entangled state $\ket{\psi_\theta}=\cos\theta\ket{00}+\sin\theta\ket{11}\in\hh_{A}\otimes\hh_{B}$ from a mixed state $\tau=\sum_{i=1}^3q_i\tau_i\in\dop{\hh_{R_A}\otimes\hh_{R_B}}$, where $\theta\in(0,\frac{\pi}{4}]$, $\forall q_i>0$, and
\begin{equation}
    \ket{\tau_1}=\frac{1}{\sqrt{2}}(\ket{01}+e^{i\theta_1}\ket{10}),
    \ket{\tau_2}=\frac{1}{\sqrt{2}}(\ket{02}+e^{i\theta_2}\ket{20}),
    \ket{\tau_3}=\frac{1}{\sqrt{2}}(\ket{12}+e^{i\theta_3}\ket{21}).
\end{equation}
We can find a Kraus representation $\mathcal{S}(\rho)=\sum_{i,j}E_{i,j}\rho E_{i,j}^\dag$ for the distillation by modifying the proof of Theorem 2 (b) in~\cite{CD09}.
\begin{eqnarray}
 E_{1,1}&=&\sqrt{p}(\sqrt{\cos\theta}\ket{0}_A\bra{0}_{R_A}+\sqrt{\sin\theta}\ket{1}_A\bra{1}_{R_A})\nonumber\\
 &&\otimes(\sqrt{\cos\theta}\ket{0}_B\bra{1}_{R_B}+e^{-i\theta_1}\sqrt{\sin\theta}\ket{1}_B\bra{0}_{R_B})\\
 E_{1,2}&=&\sqrt{p}(\sqrt{\cos\theta}\ket{0}_A\bra{1}_{R_A}+\sqrt{\sin\theta}\ket{1}_A\bra{0}_{R_A})\nonumber\\
 &&\otimes(e^{-i\theta_1}\sqrt{\cos\theta}\ket{0}_B\bra{0}_{R_B}+\sqrt{\sin\theta}\ket{1}_B\bra{1}_{R_B})\\
 E_{2,1}&=&\sqrt{p}(\sqrt{\cos\theta}\ket{0}_A\bra{0}_{R_A}+\sqrt{\sin\theta}\ket{1}_A\bra{2}_{R_A})\nonumber\\
 &&\otimes(\sqrt{\cos\theta}\ket{0}_B\bra{2}_{R_B}+e^{-i\theta_2}\sqrt{\sin\theta}\ket{1}_B\bra{0}_{R_B})\\
 E_{2,2}&=&\sqrt{p}(\sqrt{\cos\theta}\ket{0}_A\bra{2}_{R_A}+\sqrt{\sin\theta}\ket{1}_A\bra{0}_{R_A})\nonumber\\
 &&\otimes(e^{-i\theta_2}\sqrt{\cos\theta}\ket{0}_B\bra{0}_{R_B}+\sqrt{\sin\theta}\ket{1}_B\bra{2}_{R_B})\\
 E_{3,1}&=&\sqrt{p}(\sqrt{\cos\theta}\ket{0}_A\bra{1}_{R_A}+\sqrt{\sin\theta}\ket{1}_A\bra{2}_{R_A})\nonumber\\
 &&\otimes(\sqrt{\cos\theta}\ket{0}_B\bra{2}_{R_B}+e^{-i\theta_3}\sqrt{\sin\theta}\ket{1}_B\bra{1}_{R_B})\\
 E_{3,2}&=&\sqrt{p}(\sqrt{\cos\theta}\ket{0}_A\bra{2}_{R_A}+\sqrt{\sin\theta}\ket{1}_A\bra{1}_{R_A})\nonumber\\
 &&\otimes(e^{-i\theta_3}\sqrt{\cos\theta}\ket{0}_B\bra{1}_{R_B}+\sqrt{\sin\theta}\ket{1}_B\bra{2}_{R_B}).
\end{eqnarray}
Through a straightforward calculation, we can show that $ E_{i,1}\ket{\tau_i}= E_{i,2}\ket{\tau_i}=\frac{\sqrt{p}}{\sqrt{2}}\ket{\psi_\theta}$ for all $i$ and $ E_{i,1}\ket{\tau_j}= E_{i,2}\ket{\tau_j}=0$ for $i\neq j$.
This implies $\mathcal{S}(\tau)=p\psi_\theta$. Thus, we can obtain the maximum success probability $p$ of the distillation by maximizing $p$ under the constraint $\idop- \sum_{i,j}E_{i,j}^\dag E_{i,j}\in\SEP{\cdim{3}:\cdim{3}}$. By a straightforward calculation, we obtain
\begin{eqnarray}
 E_{1,1}^\dag E_{1,1}&=&p(\cos\theta\ketbra{0}+\sin\theta\ketbra{1})\otimes(\sin\theta\ketbra{0}+\cos\theta\ketbra{1})\\
 E_{1,2}^\dag E_{1,2}&=&p(\sin\theta\ketbra{0}+\cos\theta\ketbra{1})\otimes(\cos\theta\ketbra{0}+\sin\theta\ketbra{1})\\
 E_{2,1}^\dag E_{2,1}&=&p(\cos\theta\ketbra{0}+\sin\theta\ketbra{2})\otimes(\sin\theta\ketbra{0}+\cos\theta\ketbra{2})\\
 E_{2,2}^\dag E_{2,2}&=&p(\sin\theta\ketbra{0}+\cos\theta\ketbra{2})\otimes(\cos\theta\ketbra{0}+\sin\theta\ketbra{2})\\
 E_{3,1}^\dag E_{3,1}&=&p(\cos\theta\ketbra{1}+\sin\theta\ketbra{2})\otimes(\sin\theta\ketbra{1}+\cos\theta\ketbra{2})\\
 E_{3,2}^\dag E_{3,2}&=&p(\sin\theta\ketbra{1}+\cos\theta\ketbra{2})\otimes(\cos\theta\ketbra{1}+\sin\theta\ketbra{2}).
\end{eqnarray}
This implies
\begin{equation}
  \sum_{i,j}E_{i,j}^\dag E_{i,j}=p \times{\rm diag}(2\sin2\theta,1,1,1,2\sin2\theta,1,1,1,2\sin2\theta).
\end{equation}
Therefore, the maximum success probability of this protocol is $p=\min\left\{1,\frac{1}{2\sin2\theta}\right\}$.

\section{Proof for Proposition~\ref{prop:MFLCdist}}
\label{appendix:entMFLC}

Before presenting the proof, we show the following lemma.
\begin{lemma}
\label{lemma:MFLCdist}
 The MFLC of $\Segre{\mathcal{A}:\mathcal{B}}\cap(\vv\setminus\vv^\circ)$ is $\vee_{n=1}^2(\hh_{A}\otimes\hh_{R_A})$, 
 where $\mathcal{A}=\hh_{A}\otimes\hh_{R_A}$, $\mathcal{B}=\hh_{R_B}\otimes\hh_B$, $\dim\hh_A=\dim\hh_B=2$, $\dim\hh_{R_A}=\dim\hh_{R_B}=d$,
\begin{eqnarray}
 \vv&=&\left\{\ket{\Xi}\in\mathcal{A}\otimes\mathcal{B}:\forall \ket{\psi}\in\wedge_2\cd,\bra{\psi}_{R_AR_B}\ket{\Xi}\in\vspan{\{\ket{01}-\ket{10}\}}\right\},\\
 \vv^\circ&=&\{\ket{\Xi}\in\mathcal{A}\otimes\mathcal{B}:\forall \ket{\psi}\in\wedge_2\cd,\bra{\psi}_{R_AR_B}\ket{\Xi}=0\},
\end{eqnarray}
 and we regard the symmetric subspace $\vee_{n=1}^2(\hh_{A}\otimes\hh_{R_A})$ as being embedded in $\hh_A\otimes\hh_{R_A}\otimes\hh_B\otimes\hh_{R_B}$ by the isomorphism $\hh_B\otimes\hh_{R_B}\simeq\hh_{A}\otimes\hh_{R_A}$.
\end{lemma}
\begin{proof}
For any $\ket{A}\ket{B}\in\Segre{\mathcal{A}:\mathcal{B}}\cap\vv$, it holds that for any $0\leq i<j\leq d-1$,
\begin{equation}
\label{eq:antSYMcond1}
 \exists\alpha_{ij}\in\cc,\ \bra{i}_{R_A}\ket{A}\bra{j}_{R_B}\ket{B}- \bra{j}_{R_A}\ket{A}\bra{i}_{R_B}\ket{B}=\alpha_{ij}(\ket{01}_{AB}-\ket{10}_{AB}).
\end{equation}
By defining,
$[A_{ij}]= \begin{pmatrix}
 (\bra{i}_{R_A}\bra{0}_A)\ket{A}& (\bra{i}_{R_A}\bra{1}_A)\ket{A}\\
  (\bra{j}_{R_A}\bra{0}_A)\ket{A}& (\bra{j}_{R_A}\bra{1}_A)\ket{A}
\end{pmatrix}$, 
$[B_{ij}]= \begin{pmatrix}
 (\bra{i}_{R_B}\bra{0}_B)\ket{B}& (\bra{i}_{R_B}\bra{1}_B)\ket{B}\\
  (\bra{j}_{R_B}\bra{0}_B)\ket{B}& (\bra{j}_{R_B}\bra{1}_B)\ket{B}
\end{pmatrix}$, Eq.~\eqref{eq:antSYMcond1} is equivalent to
\begin{equation}
\exists\alpha_{ij}\in\cc,\ 
 [A_{ij}]^T
\begin{pmatrix}
 0&1\\
 -1&0
\end{pmatrix}
[B_{ij}]=\alpha_{ij}\begin{pmatrix}
 0&1\\
 -1&0
\end{pmatrix}.
\end{equation}
By straightforward calculation, we find that Eq.~\eqref{eq:antSYMcond1} is equivalent to either of the following cases:
\begin{enumerate}
    \item $\exists\alpha_{ij}\in\cc^\times,\ [B_{ij}]=\alpha_{ij}[A_{ij}]\wedge\det(A)\neq0$
    \item $[A_{ij}]=0$
    \item $\exists
    \begin{pmatrix}
        a\\b
    \end{pmatrix}\in\cdim{2}\setminus\{0\},
    \exists\begin{pmatrix}
        c\\d
    \end{pmatrix}\in\cdim{2},\ 
    [A_{ij}]=
    \begin{pmatrix}
        0&0\\ a& b
    \end{pmatrix},
        [B_{ij}]=
    \begin{pmatrix}
        0&0\\ c& d
    \end{pmatrix}$
    \item $\exists \beta\in\cc,\exists
    \begin{pmatrix}
        a\\b
    \end{pmatrix}\in\cdim{2}\setminus\{0\},
    \exists\begin{pmatrix}
        c\\d
    \end{pmatrix}\in\cdim{2},\ 
    [A_{ij}]=
    \begin{pmatrix}
        a&b\\\beta a&\beta b
    \end{pmatrix},
        [B_{ij}]=
    \begin{pmatrix}
        c&d\\\beta c&\beta d
    \end{pmatrix}$
\end{enumerate}
Note that except for the first case, $[A_{ij}]^T
\begin{pmatrix}
 0&1\\
 -1&0
\end{pmatrix}
[B_{ij}]=0$ holds.

Thus, for any $\ket{A}\ket{B}\in\Segre{\mathcal{A}:\mathcal{B}}\cap(\vv\setminus\vv^\circ)$, there exists $i$ and $j$ such that the first case holds.
We let
\begin{equation}
    [A_{ij}]= \begin{pmatrix}
 (\bra{i}_{R_A}\bra{0}_A)\ket{A}& (\bra{i}_{R_A}\bra{1}_A)\ket{A}\\
  (\bra{j}_{R_A}\bra{0}_A)\ket{A}& (\bra{j}_{R_A}\bra{1}_A)\ket{A}
\end{pmatrix}=
\begin{pmatrix}
    a&b\\c&d
\end{pmatrix},\ 
[B_{ij}]= \begin{pmatrix}
 (\bra{i}_{R_B}\bra{0}_B)\ket{B}& (\bra{i}_{R_B}\bra{1}_B)\ket{B}\\
  (\bra{j}_{R_B}\bra{0}_B)\ket{B}& (\bra{j}_{R_B}\bra{1}_B)\ket{B}
\end{pmatrix}=\alpha_{ij}[A_{ij}],
\end{equation}
where $\alpha_{ij}\in\cc^\times$ and $\det\begin{pmatrix}
    a&b\\c&d
\end{pmatrix}\neq0$.
Observe that for all integers $k$ such that $i<k\leq d-1$, the matrices $[A_{ik}]$ and $[B_{ik}]$, associated with the pair $(i,k)$ satisfy either the first or the fourth case. This implies $[B_{ik}]=\alpha_{ij}[A_{ik}]$.
On the other hand, observe that for all integers $k$ such that $0\leq k<j$, the matrices $[A_{kj}]$ and $[B_{kj}]$, associated with the pair $(k,j)$, satisfy either the first, the third, or the fourth case. This also implies $[B_{kj}]=\alpha_{ij}[A_{kj}]$.
Therefore, we obtain $\exists\alpha\in\cc^{\times},\ket{B}=\alpha\ket{A}$. This completes the proof.

\end{proof}

\begin{proof}[Proof of Proposition~\ref{prop:MFLCdist}]
First, by following the argument in the twisted canonical subspace in Section~\ref{sec:MFLCTCS}, we can show that
\begin{eqnarray}
 \ww&=&\left\{\ket{\Xi}\in\mathcal{A}\otimes\mathcal{B}:\forall i,\bra{\overline{\tau}_i}_{R_AR_B}(\sigma_YL^{-1})^{(B)}\ket{\Xi}\in\vspan{\{ \ket{01}-\ket{10}\}}\right\}\\
 &=& (L\sigma_Y)^{(B)}\vv,\\
 \ww^\circ&=& (L\sigma_Y)^{(B)}\vv^\circ,
\end{eqnarray}
where
\begin{eqnarray}
 \vv&=&\left\{\ket{\Xi}\in\mathcal{A}\otimes\mathcal{B}:\forall i,\bra{\overline{\tau}_i}_{R_AR_B}\ket{\Xi}\in\vspan{\{\ket{01}-\ket{10}\}}\right\},\\
 \vv^\circ&=&\{\ket{\Xi}\in\mathcal{A}\otimes\mathcal{B}:\forall i,\bra{\overline{\tau}_i}_{R_AR_B}\ket{\Xi}=0\}.
\end{eqnarray}
This implies that the MFLC of $\Segre{\mathcal{A}:\mathcal{B}}\cap(\ww\setminus\ww^\circ)$ is $(L\sigma_Y)^{(B)}\pp'$, where $\pp'$ is the MFLC of $\Segre{\mathcal{A}:\mathcal{B}}\cap(\vv\setminus\vv^\circ)$.
Applying Lemma~\ref{lemma:MFLCdist} completes the proof.
\end{proof}



\end{document}